\documentclass[AMA,Times1COL]{WileyNJDv5} 

\articletype{RESEARCH ARTICLE}%

\received{Date Month Year}
\revised{Date Month Year}
\accepted{Date Month Year}
\journal{Journal}
\volume{00}
\copyyear{2023}
\startpage{1}

\usepackage[flushleft]{threeparttable}
\usepackage{hyperref}
\hypersetup{colorlinks,linkcolor={blue},citecolor={blue},urlcolor={blue}} 

\begin{document}

\title{Evaluation and comparison of covariate balance metrics in studies with time-dependent confounding}

\author[1,2]{David Adenyo}

\author[1,2]{Jason R. Guertin}

\author[1]{Bernard Candas}

\author[2,3]{Caroline Sirois}

\author[1,2]{Denis Talbot}

\authormark{DAVID \textsc{et al.}}
\titlemark{Evaluation and comparison of covariate balance metrics in studies with time-dependent confounding}

\address[1]{\orgdiv{Département de médecine sociale et préventive}, \orgname{Université Laval}, \orgaddress{\state{Québec}, \country{Canada}}}

\address[2]{\orgdiv{Centre de recherche du CHU de Québec}, \orgname{Université Laval}, \orgaddress{\state{Québec}, \country{Canada}}}

\address[3]{\orgdiv{Faculté de pharmacie}, \orgname{Université Laval}, \orgaddress{\state{Québec}, \country{Canada}}}

\corres{Denis Talbot, Département de médecine sociale et préventive, Université Laval. \email{denis.talbot@fmed.ulaval.ca}}


\abstract[Abstract]{Marginal structural models have been increasingly used by analysts in recent years to account for confounding bias in studies with time-varying treatments. The parameters of these models are often estimated using inverse probability of treatment weighting. To ensure that the estimated weights adequately control confounding, it is possible to check for residual imbalance between treatment groups in the weighted data. Several balance metrics have been developed and compared in the cross-sectional case but have not yet been evaluated and compared in longitudinal studies with time-varying treatment. We have first extended the definition of several balance metrics to the case of a time-varying treatment, with or without censoring. We then compared the performance of these balance metrics in a simulation study by assessing the strength of the association between their estimated level of imbalance and bias. We found that the Mahalanobis balance performed best. Finally, the method was illustrated for estimating the cumulative effect of statins exposure over one year on the risk of cardiovascular disease or death in people aged 65 and over in population-wide administrative data. This illustration confirms the feasibility of employing our proposed metrics in large databases with multiple time-points.}

\keywords{covariate balance, time-dependent confounding, inverse probability of treatment weighting, balance metric, time-varying covariates}

\jnlcitation{\cname{%
\author{Adenyo D.},
\author{Guertin J R},
\author{Candas B}, 
\author{Sirois C}, and
\author{Talbot D}}.
\ctitle{Evaluation and comparison of covariate balance metrics in studies with time-dependent confounding.} \cjournal{\it Statistics in Medicine.} \cvol{2024;00(00):1--18}.}

\maketitle

%

\section{Introduction}\label{sec1}

Marginal structural models (MSMs) are an increasingly popular approach for estimating the effect of a time-varying treatment or exposure using observational data \cite{a,ba}. Unlike traditional approaches, such as propensity score matching or covariate adjustment in an outcome model, MSMs can appropriately deal with covariates that are both confounders for the effect of a treatment at a given time-point and intermediate variables in the causal pathway between a previous treatment and the outcome \cite{a,ba}. The most common approach for estimating the parameters of an MSM is to adjust for measured confounders with an inverse probability of treatment weighting (IPTW) estimator \cite{a,ba}. When using this approach, each individual is assigned a weight that corresponds to the inverse of the probability of their observed treatment at each time-point, conditional on their past treatment and covariate history. Intuitively, IPTW creates a pseudo-population where the treatment at each time-point is independent of previously measured confounders, thus mimicking a sequential randomized trial with respect to these confounders \cite{a,ba}.  

One condition to obtain valid causal effect estimates when using this IPTW estimator is the absence of residual systematic differences in the observed baseline and time-varying covariates between treatment groups in the weighted data \cite{d,e}. In settings where the effect of a treatment at a single time-point is of interest, various methods, or metrics, have been proposed to assess if treatment groups are balanced. Notably, several authors have developed methods for evaluating imbalance on one covariate at a time in the weighted sample, including the standardized mean difference (SMD), the Kolmogorov-Smirnov (KS) test statistic and graphical comparisons of the distribution of continuous variables \cite{l,a2,m}. In addition, other methods such as the L\'{e}vy distance (LD) and the non-parametric overlap coefficient (OVL) have been developed in the context of matching on the propensity score\cite{m}. Various other metrics have also been developed to assess the balance on several covariates simultaneously, including the Mahalanobis balance (MHB) \cite{m,n}, the C-statistic of the propensity score model \cite{m,p}, the $L_1$ balance metric or the $L_1$ median \cite{m,r}. One potential advantage of such global metrics is their ability to take into account the correlations between the covariates and to assess how the joint distribution of the covariates is imbalanced between treatment groups. Recently \textit{Franklin et al}\cite{m} proposed two new measures in addition to the previous metrics in the context of matching on the propensity score, namely the post-matching C-statistic as well as the general weighted difference (GWD). Using a simulation study, \textit{Franklin et al}\cite{m} compared the performance of various metrics when using a matching estimator and concluded that the post-matching C-statistic, SMD, and GWD performed best with regard to their association with bias.

Although there have been significant methodological developments regarding balance checking methods, there are still important gaps in knowledge, specifically in the longitudinal-MSM setting. For example, while \textit{Franklin et al} \cite{m} have indicated how each of the above-mentioned metrics are calculated when adjusting for confounders using an IPTW estimator, the performance of the metrics was not compared in their simulation study when using this estimator. In addition, prior research mainly focused on balance metrics for a treatment measured at a single time-point. Recently, the SMD metric has been adapted to the time-varying treatment setting \cite{d,NoahGreifer}. To the best of our knowledge, the other metrics have not yet been extended to this longitudinal setting and their relative performance has not been examined in simulation studies.

The objective of this study is to evaluate and compare various metrics in the context of MSMs using IPTW in order to determine which metrics are most associated with bias in the estimation of the effect of a time-varying treatment. This paper is structured as follows. In Section \ref{sec2}, we first introduce the notation as well as the concepts underlying MSMs. We first present the case without censoring (e.g., loss-to-follow up) and subsequently extend the notation to the case with censoring. Then, we define the metrics to evaluate covariate balance in a longitudinal weighted sample. Section \ref{sec3} presents a Monte Carlo simulation study that evaluates and compares the balance metrics discussed above in the context of an MSM. In Section \ref{sec4}, we present an application to real-world data in which we study the effect of statin use for the primary prevention of cardiovascular disease among older adults. The paper concludes in Section \ref{sec5} with a discussion.

\section{Methods}\label{sec2}
\subsection{Notation and marginal structural models for uncensored data}

We first present the notation for marginal structural models for uncensored data. Consider a study with $T+1$ follow-up times and $n$ individuals sampled from a population so that the time-ordered data take the following form $\mathcal{O}_i=\{X_{i0},A_{i0},X_{i1},A_{i1},...,X_{iT},A_{iT},Y_i\}$ where $A_{it}$ represents the treatment (or exposure) variable of individual $i$ ($i=1,2,...,n$) at time $t$ ($t=0,1,... ,T$), $X_{it}$ is a vector containing the time-dependent confounders, and $Y_i$ is the outcome at the end of the study at time $T+1$. To simplify the presentation, we assume that $A_{it}$ is a binary variable where $A_{it}=1 \ (A_{it}= 0)$ indicates that individual $i$ is treated (untreated) at time $t$, but the extension to the case of a multilevel exposure is straightforward. At each time $t$, we admit that $X_{it}$ is realized before $A_{it}$ and is thus not affected by $A_{it}$. Let $\bar{A}_{it}=\{{A}_{i0},{A}_{i1},...,{A}_{it}\}$ and $\bar{X}_{it}=\{{X}_{i0},{X}_{i1},...,{X}_{it}\}$ denote the history of treatment and covariates up to time $t$ for individual $i$, respectively. We denote by $Y_{i}^{\bar{a}_{T}}$, the counterfactual outcome for individual $i$  under the full treatment history $\bar{A}_{iT}=\bar{a}_{T}$. 

The problem of estimating the effect of the treatment history can be expressed in a general way as the estimation of the parameters of a model for the counterfactual outcome expectation according to the treatment history. The following expression for an MSM is obtained: $E[Y^{\bar{a}_{T}}]=g(\beta,\bar{a}_{T})$ where $\beta$ is a vector of parameters that captures the causal effect of the treatment history on the outcome and $g$ is a function defined by the analyst.

The nonparametric identification of the parameters of an MSM from the observed data can be achieved under four key assumptions: i) consistency, which requires that, for any individual, if $\bar{A}_{T}=\bar{a}_{T}$, then $Y=Y^{\bar{a}_{T}}$; ii) sequential exchangeability, which requires that the treatment at each time-point is not confounded by unobserved factors, i.e., $Y^{\bar{a}_{T}} \coprod A_{t}|\bar{A}_{t-1},\bar{X}_{t}$; iii) positivity, which requires $0<P({A}_{t}=1|\bar{A}_{t-1}=\bar{a}_{t-1},\bar{X}_{t}=\bar{x}_{t})<1$ at every time-point $t$ for a given treatment history $\bar{a}_{t-1}$ and covariate history $\bar{x}_{t}$; iv) a unique and stable treatment value: there is no interference between subjects and there is only one version of each treatment level \cite{f,s}.

The most common approach for estimating the parameters of an MSM is to use an IPTW estimator, where each individual is assigned a weight that corresponds to the product of the inverse of their probability of receiving the treatment they received at each time-point: 
\begin{align}
	U_{i}=\prod_{t=0}^{T}\dfrac{1}{P(A_{t}=a_{it}|\bar{A}_{t-1}=\bar{a}_{i,t-1},\bar{X}_{t}=\bar{x}_{it})}. \label{eq.w}
\end{align}

As mentioned in the introduction, the IPTW creates a pseudo-population where the treatment at each time-point is independent of previously measured confounders \cite{a,ba}. However, individuals receiving an unusual treatment at a time-point conditional on their past can have extreme weights. This may increase the variance of the treatment effect estimator. To overcome this problem, it has been suggested to use stabilized weights instead of using standard weights. Stabilized weights are defined as:
\begin{align}
	SW_{i}=\prod_{t=0}^{T}\dfrac{P(A_{t}=a_{it}|\bar{A}_{t-1}=\bar{a}_{i,t-1})}{P(A_{t}=a_{it}|\bar{A}_{t-1}=\bar{a}_{i,t-1},\bar{X}_{t}=\bar{x}_{it})}. \label{eq.sw}
\end{align}
Robins \cite{a} has shown that these stabilized weights are optimal from a variance perspective. However, they do not balance the treatment groups according to prior treatments, and only balance prior covariates conditionally on previous treatment history \cite{d}. In other words, when checking balance in data weighted according to these stabilized weights, one would need to compare each treatment group at each time-point for each possible treatment history, that is $A_t \coprod \bar{X}_{t-1} | \bar{A}_{t-1} = \bar{a}_{t-1}$ for all $t= 1,...,T$ and all $\bar{a}_{t-1}$. In situations with multiple time-points, the number of balance check to make may become unmanageable. Previous approaches for checking balance in a longitudinal setting are affected by this challenge \cite{d}. As such, we recommend not using these stabilized weights to check balance. Another form of stabilized weights, the marginal stabilized weights \cite{Sall:2019aa}, could instead be considered:
\begin{align}
	W_{i}=\prod_{t=0}^{T}\dfrac{P(A_{t}=a_{it})}{P(A_{t}=a_{it}|\bar{A}_{t-1}=\bar{a}_{i,t-1},\bar{X}_{t}=\bar{x}_{it})}. \label{eq.swm}
\end{align}
These weights are expected to have a lower variance than the unstabilized weights and are theoretically expected to balance treatment groups at each time-point according to both prior treatment and prior covariates, unconditionally on previous treatment history. In other words, one only needs to verify if $A_t \coprod \bar{X}_{t-1}$ for all $t=1,...,T$ in the weighted data according to these weights. 

Various methods can be used for estimating the conditional probabilities involved in the weights (\ref{eq.w}), (\ref{eq.sw}), (\ref{eq.swm}), including parametric methods, in particular logistic regression \cite{t,u}, or non-parametric and machine learning methods such as neural networks, classification/regression trees \cite{v,w,x} as well as ensemble methods such as Super Learner \cite{y}. However, using non-parametric approaches is fraught with challenges and does not always improve the validity above parametric methods \cite{z,aa}. In the following, we focus on marginal stabilized weights and logistic regression for simplicity.

\subsection{Notation and marginal structural models for censored data}

The notation we have just presented can be extended to the case of censored data, that is, for data where the follow-up is incomplete for some subjects. Such censoring is very common in longitudinal studies. Ignoring these losses to follow-up may induce selection bias. It is possible to adjust for such selection bias when using MSMs by employing inverse probability of censoring weights \cite{ab}. More precisely, denote by $\bar{C}_{it}=\{{C}_{i0},{C}_{i1},...,{C}_{it}\}$ the right censoring history for subject $i$ up to time $t$, where ${C}_{it}$ is the variable corresponding to the loss to follow-up status at time $t$ for subject $i$ with $C_t=1$ if the subject is lost to follow-up (censored) and $C_t=0$ otherwise. We assume that once a subject becomes censored, they do not re-enter the study in the future (if $C_t = 0$, then $C_k = 0$ for all $k > t$). In this case, the time-ordered data take the following form $\mathcal{O}_i=\{C_{i0},X_{i0},A_{i0},C_{i1},X_{i1},A_{i1},...,C_{iT},X_{iT},A_{iT},Y_i\}$.

The problem of interest then becomes the estimation of the counterfactual outcome expectation under treatment history $\bar{a}_T$ and under no censoring $\bar{c}_T=0$. The MSM can be expressed in a general way as $E[Y^{\bar{a}_T,\bar{c}_T=0}] = g(\beta,\bar{a}_T)$. The identification of the parameters of this MSM can be achieved under similar assumptions as those that were presented in the previous section, by considering $C_t$ as a second time-varying treatment. More precisely, the identification requires the sequential exchangeability for $A_t$ and $C_t$, that is, $Y^{(\bar{a}_{T},\bar{c}_{T}=0)} \coprod A_{t}|\bar{A}_{t-1},\bar{C}_{t}=0,\bar{X}_{t}$ and $Y^{(\bar{a}_{T},\bar{c}_{T}=0)} \coprod C_{t}|\bar{A}_{t-1},\bar{C}_{t-1}=0,\bar{X}_{t-1}$; the joint positivity for $(A_t,C_t)$, that is, ${0<P({A}_{t}=1|\bar{A}_{t-1}=\bar{a}_{t-1},\bar{X}_{t}=\bar{x}_{t},\bar{C}_{t}=0)<1}$ and ${0<P({C}_{t}=0|\bar{A}_{t-1}=\bar{a}_{t-1},\bar{X}_{t-1}=\bar{x}_{t-1},\bar{C}_{t-1}=0)}$; and consistency.

The marginal stabilized inverse probability of censoring weights (IPCW) $W_{i}^{C}$ are constructed similarly to the IPTW by modeling the probability of not being lost to follow-up:
\begin{align}
	W_{i}^{C}=\prod_{t=0}^{T}\dfrac{P(C_{t}=0|\bar{C}_{t-1}=0)}{P(C_{t}=0|\bar{A}_{t-1}=a_{i,t-1},\bar{X}_{t-1}=\bar{x}_{it-1},\bar{C}_{t-1}=0)}. \label{eq.swm1}
\end{align}
The IPTW also needs to be modified to account for the fact that it cannot be computed for subjects that are censored:
\begin{align}
	W_{i}^{A}=\prod_{t=0}^{T}\dfrac{P(A_{t}=a_{it}|\bar{C}_{t}=0)}{P(A_{t}=a_{it}|\bar{A}_{t-1}=\bar{a}_{i,t-1},\bar{X}_{t}=\bar{x}_{it},\bar{C}_{t}=0)}. \label{eq.swm2}
\end{align}
Under the previous identifiability assumptions, the parameters of the MSM can be estimated by fitting a model $E[Y|\bar{A}, \bar{C}_T = 0] = g(\beta, \bar{A})$ weighting by the product of the IPTW and the IPCW.
We denote this product by $W^{A,C}$. These weights control for both confounding and selection biases by mimicking a sequential randomized trial without censoring relative to measured covariates. 

\subsection{Balance metrics}\label{sec22}

In this section, we extend the definition of several balance metrics to the time-varying treatment without censoring and with censoring setting. We considered eight of the ten metrics that were defined in \textit{Franklin et al} \cite{m} for the case of a treatment and covariates measured at a single time-point. We did not retain the $L_1$ metrics ($L_1$ measure and $L_1$ median) because of their inferior performance in previous simulations \cite{m}. The first five metrics we present below evaluate the balance on a single covariate at a time and the last three evaluate the balance on several covariates simultaneously. 

Recall that the covariates must be chosen to satisfy the sequential exchangeability assumption to yield consistent estimators. In the time-varying treatment without censoring setting, treatment groups at each time-point should be balanced relative to previous covariates in data weighted according to $W$, that is $A_t \coprod \bar{X}_t$ for all $t = 0, ..., T$. To verify this balancing property, we propose to compare the distribution between treatment groups at each time-point $t = 1, ..., T$ of each previous covariate $X_{t-k}$ for $k = 0,..., t$ when considering metrics that apply to a single covariate at a time. When considering metrics that can consider multiple covariates simultaneously, we propose to consider jointly all covariates that are measured at a given time-point, but to consider separately the covariates that are measured at different time-points. In other words, we propose comparing the joint distribution between treatment groups at each time-point $t = 0, ..., T$ of all the covariates $X_{t-k}$, separately for all $k = 0, ..., t$. This choice avoids considering simultaneously the repeated version of the same variables $X_{t-k}$ and $X_{t-k-1}$, which may lead to collinearity issues. To simplify the presentation, we define the metrics  considering the marginal stabilized weights $W$. 

\begin{enumerate}
	\item The \textit{absolute difference} is defined as the absolute value of the difference in the means of a given covariate between the treatment groups. This metric as well as standardized means difference have recently been developed in the longitudinal setting by \textit{Jackson et al} \cite{d}. Unlike these authors, we propose using products up to time $T$ instead of up to time $t$ in Equations (\ref{eq.w}), (\ref{eq.sw}), (\ref{eq.swm}), (\ref{eq.swm1}) and (\ref{eq.swm2}) to check the balance at time $t$. We believe the choice of using the weights with all product terms is preferable since the parameters of the MSM are actually estimated using these weights, not with the partial products. As such, we define the absolute difference for covariate $X_{t-k}$ between treatment groups at time $t$ as 
	${D_{t,k}=\lvert E[W\times I(A_{t}=1)\times X_{t-k}]-E[W\times I(A_{t}=0)\times X_{t-k}]}\rvert$, where $0\leq k\leq t$.	
	\item The \textit{standardized means difference} is the absolute difference divided by the the pooled standard deviation of the covariate between treatment groups 
	$SMD_{t,k}=D_{t,k}/\sqrt{(s^2_{1,t,k}+s^2_{0,t,k})/2}$, where $s^2_{1,t,k}$ and $s^2_{0,t,k}$ are the estimated weighted variances in the treated ($A=1$) and untreated ($A=0$) groups respectively, $0\leq k\leq t$.
	The threshold usually chosen to define the existence of a residual imbalance is 0.1 (10$\%$) \cite{p,u}.
	\item The \textit{overlap coefficient} is defined as the proportion of overlap in two density functions, calculated by finding the area under the minimum of the two curves (treated and untreated). For a binary variable, the OVL quantifies the overlap in probability densities between the two groups of treated and untreated subjects \cite{a2}. However, for a continuous variable, the OVL is estimated by the non-parametric kernel density estimation method \cite{a2,ad}. In this paper, we calculated the OVL for a continuous variable in a general way as follows :
	$ \mbox{OVL}_{t,k}=\int \underset{x_{t-k}}{min} \left( \hat{f}_{W\times I(A_{t}=1)\times X_{t-k}}(x_{t-k}),\hat{f}_{W\times I(A_{t}=0)\times X_{t-k}}(x_{t-k})\right) dx_{t-k}$ where ${0\leq k\leq t}$ and $\hat{f}_{W\times I(A_{t}=J)\times X_{t-k}|t,k}(x_{t-k})$ is the weighted density function of the covariate in the treatment group $J$. The OVL varies between $0$ (no overlap) and $1$ (optimal balance) and its value is independent of the unit of measurement \cite{m,ae}.
	\item The \textit{Kolmogorov-Smirnov} distance is defined as the maximum vertical distance of the cumulative distribution functions between the groups of treatments. In weighted data, its expression is given by $\mbox{KS}_{t,k}=\underset{x_{t-k}}{\mbox{max}}\left|\hat{F}_{W\times I(A_{t}=1)\times X_{t-k}}(x_{t-k})-\hat{F}_{W\times I(A_{t}=0)\times X_{t-k}}(x_{t-k})\right|$, where $\hat{F}_{W\times I(A_{t}=1)\times X_{t-k}}(x_{t-k})$ and $\hat{F}_{W\times I(A_{t}=0)\times X_{t-k}}(x_{t-k})$ denote the weighted empirical cumulative distribution function in the treated and untreated groups, respectively, where $0\leq k\leq t$. This metric ranges from $0$ (optimal balance) to $1$ \cite{m,ae}. 
	\item The \textit{L\'{e}vy distance} is defined as the length of the side of the largest square that can fit between two cumulative distribution weighted functions: 
	${\mbox{LV}_{t,k}=\underset{x_{t-k}}{\mbox{min}}\{\epsilon>0 : \hat{F}_{W\times I(A_{t}=0)\times X_{t-k}}(x_{t-k}-\epsilon)-\epsilon \leqslant \hat{F}_{W\times I(A_{t}=1)\times X_{t-k}|t,k}(x_{t-k}-\epsilon)} \leqslant \hat{F}_{W\times I(A_{t}=0)\times X_{t-k}|t,k}(x_{t-k}+\epsilon)+\epsilon, \ \ \forall \ \ x_{t-k}\}$, where $0\leq k\leq t$. Similarly, the L\'{e}vy distance varies from $0$ to $1$, where the value $0$ indicates optimal balance \cite{m,ae}.
	\item We propose a new version of the \textit{Mahalanobis balance}. Unlike \textit{Franklin et al}\cite{m}, we consider the pooled variance-covariance matrix of covariates in the definition of the MHB instead of the sample variance-covariance matrix of covariates. We made this choice for two reasons. Firstly, it reinforces the concordance between MHB and SMD, since MHB is simply the matrix version of SMD. Secondly, it offers the possibility of proposing a threshold for MHB since, unlike SMD, there is no established acceptable limit. Based on these connections, we propose to use $p \times 0.01$ as a threshold, where $p$ is the number of covariates (see Appendix 3 in in the Supporting information). It is defined as  ${\mbox{MHB}_{t,k}=(\bar{X}_1}-{\bar{X}_0)^{T}}\Sigma^{-{1}}(\bar{X}_1-\bar{X}_0)$, where $\bar{X}_j = E[W\times I(A_{t}=j)\times  \mathbf{X}_{t-k}]$ is the vector of weighted means of the covariates in treatment group $j$, and $\Sigma$ is the pooled within-group variance-covariance matrix of the covariates, with $0\leq k\leq t$. 
	\item We also define the \textit{post-weighting C-statistic (CS)}. It is given by the area under the receiver operating characteristic (ROC) curve which measures the ability of model estimated in the weighted sample to discriminate treated subjects from untreated subjects. Its value varies between $0.5$ (inability of the Propensity score (PS) model to discriminate treated subjects from untreated subjects after weighting) and $1$ (the worst balance after weighting) \cite{m}.
	\item \textit{The general weighted difference} is given by $GW\!D_{t,k}=\sum_{0\leqslant a \leqslant b \leqslant C} w_{ab}\left|E[W\times I(A_{t}=1)\times X_{a,t-k}X_{b,t-k}]\right.-\left.E[W\times I(A_{t}=0)\times X_{a,t-k}X_{b,t-k}]\right|$, where $C$ is the number of measured covariates, $X_{0}$ is a unit vector and $w_{ab}$ is a weight assigned to the pair of covariates $X_{a}X_{b}$, where $w_{ab}=1/s_{ab}$ if $a=0$ and $w_{ab}=0.5/s_{ab}$ otherwise,
	$s_{ab}$ the pooled intra-group standard deviation of $X_{a,t-k}X_{b,t-k}$, $0\leq k\leq t$. 
\end{enumerate}

In the time-varying treatment with censoring setting, treatment groups at each time-point should be balanced relative to previous covariates among persons who have not been censored in data weighted according to $W^{A,C}$, that is $A_t \coprod \bar{X}_t,\bar{C}_t=0$ for all $t=0,...,T$. For example for the absolute difference we have: ${D_{t,k}=\lvert E[W^{A,C}\times I(A_{t}=1)\times X_{t-k},\bar{C}_{t-k}=0]-E[W^{A,C}\times I(A_{t}=0)\times X_{t-k},\bar{C}_{t-k}=0]}\rvert$, where $0\leq k\leq t$. The definitions are analogous when considering the others metrics. The R code for implementing all balance metrics is available on Github (\url{https://github.com/detal9/LongitudinalBalanceMetrics}).

\section{Simulations}\label{sec3}

We now present the simulation study we conducted to evaluate and compare the performance of the balance metrics we presented in the previous section. We first provide a general overview of the simulation study before presenting the data-generating equations in Subsection \ref{sec32}.

\subsection{General presentation of the simulation study}

Among the simulation scenarios we considered, seven were adapted from \textit{Franklin et al} \cite{m}, and three are new scenarios. These scenarios differ regarding the type of relations between covariates and treatments (linear or nonlinear), the type of relations between covariates and outcome (linear or nonlinear), the sample size ($n = 10,000$ or $n = 1,000$) and the strength of the relations between the variables (see Section \ref{sec32} and \ref{sec33} for more details). We first elaborated the scenarios in the case of uncensored data and subsequently for censored data. To simplify the simulation study, all scenarios feature two time-points where covariates and treatment are measured, and the outcome is measured at a third time-point. 

\begin{figure*}[t]
	\centerline{\includegraphics[width=10in,width=\textwidth]{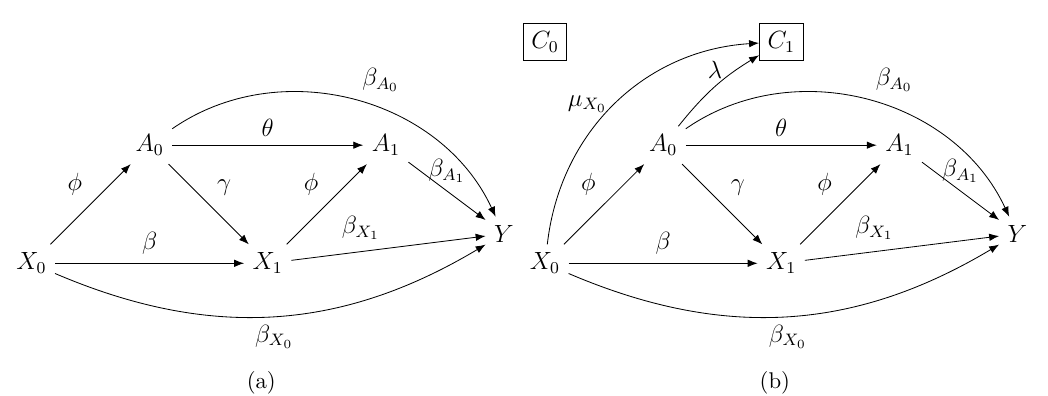}}
	\caption{Causal Directed Acyclic Graphs describing measured covariates $X_t$, exposures $A_t$, outcome $Y$ and parameters of the data-generating mechanism for an observational study where exposure is (a) not censored (b) censored, with continuing data collection among the uncensored at time $t$ i.e. $C_t=0$.\label{fig3}}
\end{figure*}

For each scenario in the time-varying treatment without censoring setting, a vector of six time-varying covariates $X_t=(L_t,M_t,N_t,O_t,P_t,Q_t), t=0,1$ and a binary time-varying treatment variable were generated (Figure \ref{fig3}). The variables $L_t$, $M_t$, $N_t$ were continuous, more precisely $L_t$ and $N_t$ were normally distributed and $M_t$ was a lognormal variable. We simulated $O_t$ and $P_t$ as binomial variables and specified that $O_t$ was strongly correlated with $L_t$ in all scenarios. Finally, $Q_t$ was simulated as an ordered categorical variable. Transformation of these variables were also considered to create non-linear relations (${T_t=\sin(L_t)}$, $R_t=M_t^{2}$, $V_t=N_t\times O_t$, $Z_t=O_t\times P_t$). Inspired by our application of interest, we generated data on $Y$ such that treatment at each time-point reduces the risk of the outcome while higher covariate values increase the risk of outcome. 
The MSM of interest was $\text{logit}(E[Y^{\bar{a}_1}]) = \beta_0 + \beta_1 a_0 + \beta_2 a_1$ and the true values of its parameters were estimated by Monte Carlo simulation of the counterfactual outcomes with a sample size of $n=100,\!000$ (see \url{https://github.com/detal9/LongitudinalBalanceMetrics} for the source R code). 

For each simulation scenario, we generated $1,\!000$ datasets of size $n$. For each dataset, we estimated the treatment probabilities $P(A_t|\bar{A}_{t-1}, \bar{X}_t)$ and $P(A_t)$ for $t=0,1$ using logistic regression models, either including the independent variables as main terms only (simple specification), or including both main terms and two-way interactions between the variables (complex specification). Using the predicted probabilities from these models, we computed five different types of weights based on $W_0=\frac{P(A_0)}{P(A_0|X_0)}$
and $W_1=\frac{P(A_1)}{P(A_1|A_{0}, X_0,X_1)}$ in order to produce datasets with varying levels of balance and bias. For the first weight, we considered $W_0\times W_1$, for the second weight we considered only $W_1$, for the third weight we considered only $W_0$, for the fourth weight we considered the product of $W_0$ truncated at the $90$th percentiles with $W_1$ and finally we considered the product of $W_0$ with $W_1$ truncated at the $90$th percentiles. We then computed each balance metric at all time-points and either for each covariates separately (D, SMD, OVL, KS, LD) or for all covariates globally (MHB, CS, GWD) in the unweighted data and in data weighted according to each of the five aforementioned weights. To calculate the CS, we re-estimated in the weighted data the propensity score at each time-point without conditioning on previous treatment to measure the ability of the covariates to distinguish between treated and untreated patients after weighting. We also estimated the bias in each of the dataset for each of the parameters of the MSM by computing the difference between the estimated odds ratio using the IPTW estimator and the true odds ratio. The bias of the crude (unweighted) model was also estimated. For metrics that measure the balance of only one covariate at a time (D, SMD, OVL, KS, and LD), we calculated the metric for each covariate at each time-point and then averaged the values over covariates, resulting in three average balance variables: average balance between $A_0$ groups according to covariates $X_0$, average balance between $A_1$ groups according to covariates $X_0$, and average balance between $A_1$ groups according to covariates $X_1$. As will be seen shortly, these averages were used to evaluate the performance of the balance metrics. For metrics that measure the balance for multiple covariates at a time, we also obtained three analogous balance variables since we considered jointly all variables measured at a given time-point, but considered separately variables measured at different time-points. To make the balance metrics comparable, we transformed the balance metrics so that $0$ indicates perfect balance and higher values indicate larger imbalances. In particular, for the $OVL$ we calculated $1-OVL$ and for $C$ we calculated $2\times(C-0.5)$.

To evaluate the performance of the eight balance metrics mentioned in Section \ref{sec22}, we first created, for each balance metric, a dataset with $6,\!000$ rows (obtained by considering the $1,\!000$ simulated datasets of each of the unweighted and five types of weights) and four columns representing the estimated bias and the average balance variables. We then fitted a separate linear regression model for each metric on these data where the dependent variable was the estimated bias and the independent variables were the balance variables, featuring both linear and quadratic terms for each independent variable. For each estimated model, we extracted the proportion of variation explained ($R^2$). We also extracted the estimated intercept ($\beta_0$) of each model, which measures the average bias in absence of imbalance in the explored scenario. As in \textit{Franklin et al}\cite{m}, we focus the interpretation around the comparison of  relative $R^2$ across metrics, noting that the $R^2$ themselves are not very informative since they are highly dependent on the simulation scenarios.

We repeated the analyses as in the previous paragraphs in the time-varying treatment with censoring setting (only for the base case and for simple propensity score) by simulating binary censoring indicators $C_t$ where $C_t=1$ censored and $C_t=0$ otherwise such that $20\%$ of individuals were censored at time $t=1$. In such case we replaced $W_0$ and $W_1$ by $W^{A,C}_0=\frac{P(A_0|C_0=0)}{P(A_0|X_0,C_0=0)}$ and $W^{A,C}_1=\frac{P(A_1|\bar{C}_1=0)}{P(A_1|A_{0}, X_0,X_1,\bar{C}_1=0)}\times \frac{P(C_1|C_0=0)}{P(C_1|A_0,X_0,C_0=0)}$ respectively.

\subsection{Data-generating equations}\label{sec32}

We now present the specific data-generating equations that we used. For each simulation scenario, $n$ independent observations were generated according to these equations. To simplify the presentation, we present the data-generating equations in a general framework where steps $2, 3, 5, 6$, and $7$ correspond to the case of uncensored data and steps $1-7$ correspond to the case of data with censoring.

\begin{enumerate}
	\item Censoring indicator at time-point 0: $C_{0}=0$ for all subjects
	\item Covariates at time-point 0: $L_{0}\sim  N(0,1)$, $M_{0}\sim  logN(0,1)$, $N_{0}\sim  N(0,10)$, ${O_{0}\sim  Bern(p=\text{expit}(\delta_{0}+2L_{0}))}$, $P_{0}\sim  Bern(p=0.2)$,
	$Q_{0}$ was simulated as an ordered categorical variable with prevalences of $50\%$, $30\%$, $10\%$, $5\%$, and $5\%$ in categories 1 to 5, respectively, ${T_{0}=\sin(L_{0})}$, $R_{0}=M_{0}^{2}$, $V_{0}=N_{0}\times O_{0}$, $Z_{0}=O_{0}\times P_{0}$
	\item Exposure at time-point 0: 
	$A_{0} \sim Bern(p_{A_{0}}=expit(\alpha_{0}+\phi_{L_{0}}L_{0}+\phi_{M_{0}}M_{0}+\phi_{N_{0}}N_{0}+\phi_{O_{0}}O_{0}+\phi_{P_{0}}P_{0}+\phi_{Q_{0}}Q_{0}+\phi_{T_{0}}T_{0}+\phi_{R_{0}}R_{0}+\phi_{V_{0}}V_{0}+\phi_{Z_{0}}Z_{0}))$
	\item Censoring indicator at time-point 1:
	$C_{1} \sim Bern(p_{C_{1}}=expit(\mu_{1}+\mu_{L_{0}}L_{0}+\mu_{M_{0}}M_{0}+\mu_{N_{0}}N_{0}+\mu_{O_{0}}O_{0}+\mu_{P_{0}}P_{0}+\mu_{Q_{0}}Q_{0}+\lambda A_{0}))$
	\item Covariates at time-point 1: $L_{1}\sim  N(\beta L_{0}+\gamma_{0} A_{0},1)$, $M_{1}\sim  logN(\beta M_{0}+\gamma_{1} A_{0},1)$, ${N_{1}\sim  N(\beta N_{0}+\gamma_{2} A_{0},10)}$, $O_{1}\sim  Bern(p=expit(\delta_{1}+\beta O_{0}+2 L_{1}+\gamma_{3} A_{0}))$, ${P_{1}\sim  Binary(p=expit(\mu_{0}+\beta P_{0}+\gamma_{4} A_{0}))}$, 
	$Q_{1}$ was simulated as an ordered categorical variable with prevalences of $40\%$, $30\%$, $20\%$, $5\%$, and $5\%$ in categories 1 to 5, $T_{1}=\sin(L_{1})$, $R_{1}=M_{1}^{2}$, $V_{1}=N_{1}\times O_{1}$, $Z_{1}=O_{1}\times P_{1}$
	\item Exposure at time-point 1: $A_{1} \sim Bern(p_{A_{1}}=expit(\alpha_{1}+\phi_{L_{1}}L_{1}+\phi_{M_{1}}M_{1}+\phi_{N_{1}}N_{1}+\phi_{O_{1}}O_{1}+\phi_{P_{1}}P_{1}+\phi_{Q_{1}}Q_{1}+\phi_{T_{1}}T_{1}+\phi_{R_{1}}R_{1}+\phi_{V_{1}}V_{1}+\phi_{Z_{1}}Z_{1}+\theta A_{0}))$
	\item Outcome: $Y \sim Bern(p_{Y}=expit(\alpha_{Y}+\sum_{j=0}^{1}
	\beta_{L_{j}}L_{j}+\sum_{j=0}^{1}\beta_{M_{j}}M_{j}+\sum_{j=0}^{1}
	\beta_{N_{j}}N_{j}+\sum_{j=0}^{1}\beta_{O_{j}}O_{j}+\sum_{j=0}^{1}
	\beta_{P_{j}}P_{j}+\sum_{j=0}^{1}\beta_{Q_{j}}Q_{j}+\sum_{j=0}^{1}
	\beta_{T_{j}}T_{j}+\sum_{j=0}^{1}\beta_{R_{j}}R_{j}+\sum_{j=0}^{1}
	\beta_{V_{j}}V_{j}+\sum_{j=0}^{1}\beta_{Z_{j}}Z_{j}+\beta_{A_{0}} A_{0}+\beta_{A_{1}} A_{1}))$
\end{enumerate}
\subsection{Description of the simulation scenarios}\label{sec33}

In this section we provide a description of the ten different scenarios we have run. We first describe the base case scenario and then describe how each scenario differs from this base case scenario. Scenarios 1, 2, 3, 7,8, 9 and 10 were based on \textit{Franklin et al} \cite{m} and the detailed for covariate parameters for simulations studies can be found in Appendix 1 in the Supporting information.

\textit{Scenario 1} (Base case): In this scenario, we used a sample size of $n=10,\!000$ and did not include the covariates $T_t$, $R_t$, $V_t$ and $Z_t$, $t=0,1$ in the treatment or outcome generating equations (i.e., we set $\phi_{T_t}=\beta_{T_t}=0, \phi_{R_t}=\beta_{R_t}=0, \phi_{V_t}=\beta_{V_t}=0, \phi_{Z_t}=\beta_{Z_t}=0$). Recall that these covariates represent non-linear terms. We then chose values for $\alpha_0$ and $\alpha_1$ such that approximately $50\%$ of individuals were treated at each time-point. In addition, we chose values for $\alpha_Y$ so that the overall outcome prevalence was approximately $20\%$. For each time-point, we simulated the variables $O_t$ and $P_t$ with prevalences of $50\%$ and $20\%$, respectively. Finally, we specified that $\beta=0$ indicating no effect of the covariates $X_0$ on the covariates $X_1$. This choice reduces the collinearity between the imbalance variables that measure the imbalance at different time-points and thus facilitates the interpretation of the simulation results. 

\textit{Scenario 2}  (Low prevalence of exposure): We chose $\alpha_0 = -3.08$, $\alpha_1 = -3.37$ and $\alpha_Y = -5.1$ such that $20\%$ of individuals were treated at each time-point and overall outcome prevalence was approximately $20\%$.

\textit{Scenario 3} (Small sample): We used a smaller sample size of $n = 1,\!000$.

Scenarios 4, 5 and 6 were designed to evaluate the performance of the metrics when the data is highly imbalanced or weakly imbalanced.

\textit{Scenario 4} (High imbalance, no confounding): All covariates are strongly associated with treatment at each time-point (larger values for the $\phi$ parameters) and had no effect on the outcome ($\beta_{X_0} = \beta_{X_1} = 0$).

\textit{Scenario 5}  (Low imbalance, moderate confounding): 
All covariates at both time-points are weakly associated with treatment, the treatment at both time-points had no effect on the outcome ($\beta_{A_0} = \beta_{A_1} = 0$) and covariates had strong relations with the outcome (large $\beta_{X_0}$ and $\beta_{X_1}$ values). 

It should be noted that the degree of bias depends on the magnitude of the effect of the covariate on the outcome in addition to the amount of imbalance. More precisely, for a given level of imbalance, a weaker bias is expected when the covariates are weakly prognostic of the outcome than when the covariates are strongly prognostic of the outcome. Because the balance metrics we considered do not incorporate information about the associations of covariates with the outcome, we expected that all metrics would perform poorly in Scenarios 4 and 5. 

\textit{Scenario 6} (High imbalance-high confounding): We chose larger values for the $\phi$ and $\beta$ parameters to create datasets with high imbalance and relatively high confounding at each time-point.

Scenarios 7 and 8 were designed to evaluate the performance of balance metrics when the outcome or treatment are a non-linear function of the covariates (e.g., body mass index).

\textit{Scenario 7} (Nonlinear outcome): We included the covariates $T$, $R$, $V$ and $Z$ in the outcome generating equation (i.e. $\phi_{T_t}=\phi_{R_t}=\phi_{V_t}=\phi_{Z_t}=0,\beta_{T_t}\neq 0, \beta_{R_t}\neq 0,\beta_{V_t}\neq 0,\beta_{Z_t}\neq 0$), $t=0,1$

\textit{Scenario 8} (Nonlinear outcome and exposure): We included the covariates $T$, $R$, $V$ and $Z$ in the outcome and treatment generating equations (i.e. all $\phi$ and $\beta$ parameters are not equal to zero). 

\textit{Scenario 9} (Redundant Covariates): 
This scenario was designed to understand the advantages of the MHB, which uses covariance between variables to avoid over-penalizing the imbalance on several highly correlated covariates. We included the covariates $T$, $R$, $V$ and $Z$ in the outcome and treatment generating equations and specified that $O_0$ and $O_1$ had no effect on the treatments or outcome by setting the coefficients to zero on all terms of the treatments or outcome generating model involving $O_0$ and $O_1$ (ie, $\phi_{O_0} = \phi_{O_1} =0,\beta_{O_0} = \beta_{O_1} =0$). As such, because $L_0$ and $O_0$, as well as $L_1$ and $O_1$, are highly correlated but only $L_0$ and $L_1$ affect the exposure and the outcome, $O_0$ and $O_1$ are redundant variables. 

\textit{Scenario 10}  (Instrumental variables):  
This scenario was designed to assess the performance of the balance metrics when there are instrumental variables present in the set of covariates to be balanced. Indeed, adjustment for the latter is known to increase the variance of the estimators and is likely to amplify the bias in the treatment effect estimate \cite{ah,ai}. We included the covariates $T$, $R$, $V$ and $Z$ in the outcome and treatment generating equations and specified that the covariates $M_0$ and $M_1$ were instrumental variables by setting to zero the coefficients related to these variables in the outcome generating equation (i.e. $\beta_{M_0}=\beta_{M_1}=\beta_{R_0}=\beta_{R_1}=0$).

\subsection{Results}
\begin{figure*}[t]
	\centerline{\includegraphics[width=10in,width=\textwidth]{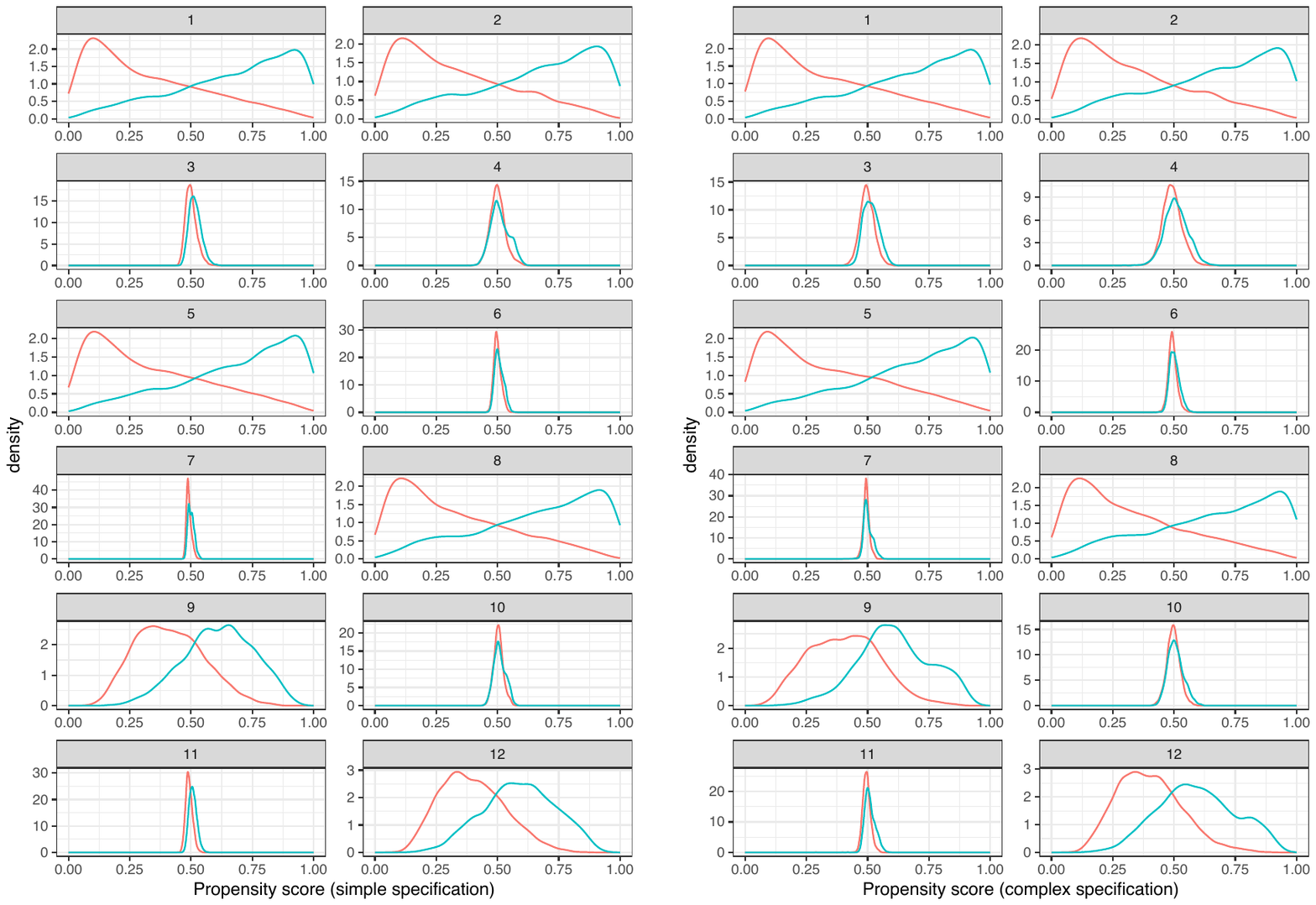}}
	\caption{Distributions of the propensity score in an example dataset of the base case simulation scenario without censoring with different types of weights. The red curve corresponds to the treated group and the blue curve to the untreated. Odd panels correspond to the distribution at $t=0$ and even panels to the distribution at $t = 1$. The first row (panels 1-2) correspond to unweighted data, the second (panels 3-4) to data weighted by $W_0\times W_1$, the third (panels 5-6) to data weighted by $W_1$ only, the fourth (panels 7-8) to data weighted according to $W_0$ only, the fifth (panels 9-10) to the product of $W_0$ truncated at the 90th percentile with $W_1$ and the last row (panels 11-12) to the product of $W_0$ with $W_1$ truncated at the 90th percentile.\label{fig4}}
\end{figure*}

Figure \ref{fig4} shows the distributions of the propensity score at each time-point for an example of one dataset of the base case scenario without censoring. The unweighted data were highly imbalanced. With both propensity score specifications (simple or complex), the balance on the propensity score improved in data weighted by $W_0\times W_1$, and the two groups had very similar distributions at each time-point. In contrast, the data weighted by $W_1$ only were highly imbalanced at time $t=0$ and well balanced at time $t=1$. The opposite result is obtained when the data are weighted by $W_0$ only. Similarly in the data weighted by the product of $W_0$ truncated at the 90th percentile with $W_1$, the data are moderately imbalanced at time $t=0$ and well balanced at time $t=1$. Finally, in the data weighted by the product of $W_0$ and $W_1$ truncated at the 90th percentile, the data are well balanced at time $t=0$ and moderately imbalanced at time $t=1$. The results for the base case with censoring are similar and are presented in Appendix 4 in the Supporting information. Likewise results for all other simulation scenarios are presented in Appendix 4 in the Supporting information.

Table \ref{tab1} presents the average bias and the average imbalance for each of the eight balance metrics for the base case scenario with uncensored and censored data. We present the results just for the simple propensity score because the two specifications for the propensity score provided nearly identical results. The objective is to determine whether the calculated balance metrics are consistent with the true balances as in the Figure \ref{fig4} and whether the calculated balances are consistent with the bias. An ideal balance metric would show high imbalance between groups $A_0$ by covariates $X_0$ or high imbalance between groups $A_1$ by covariates $X_1$ when the estimates are biased, and low (near zero) imbalance when the bias approaches zero. Apart from OVL and CS, all other calculated balance metrics are consistent with true imbalance. In addition, the effect estimates are biased in the unweighted data and in the data weighted by $W_0$ or $W_1$ only. These biases decreased in the data weighted by the other three weights. Tables for the other simulation scenarios are presented in Appendix 2 of the Supporting information.

Table \ref{tab2} and \ref{tab3} summarizes the estimated intercept and explained variation ($R^2$) for each of the eight metrics in all scenarios for the uncensored and censored data setting. Almost all imbalance metrics were strongly associated with bias. In particular for the base case (Scenario 1), the performance of D, SMD, KS, LD, MHB and GWD in terms of $R^2$ is similar, with $R^2$ ranging from $93\%$ to $94\%$. However, the intercept for MHB was closer to $0$ than that of the other metrics, indicating a better fit between MHB and bias. In contrast, OVL and CS had a consistently weaker association with bias than most other measures of balance and  OVL had intercept far from $0$, indicative that substantial bias could be present even when this metric did not detect any imbalance. The results for the other scenarios are overall in line with those of the base case scenario. Of note, in the low imbalance moderate confounding (Scenario 5) and the low sample size (Scenario 3) scenarios, all metrics were moderately less good at predicting bias, but the comparative performance of the metrics remained unchanged. In addition, all metrics were poor at predicting bias in the high imbalance no confounding (Scenario 4) scenario and in the censored data setting. 

\begin{table}[h]  \centering  \begin{threeparttable} 
		\caption{Average covariate imbalances across covariates at each time-point and average bias for the base case scenario. Averages are taken over $1,\!000$ simulated datasets in the unweighted data and in data weighted according to the five weights.\label{tab1}} 
\begin{tabular}{rrrrrrrrrrrrrrrrrr}          &       &       &       &       &       &       &       &       &       &       &       &       &       &       &       &       &  \\    \midrule    \multicolumn{18}{c}{Uncensored data} \\    \midrule          &       & \multicolumn{3}{c}{\textbf{D}} &       & \multicolumn{3}{c}{\textbf{SMD}} &       & \multicolumn{3}{c}{\textbf{OVL}} &       & \multicolumn{3}{c}{\textbf{KS}} & \multicolumn{1}{c}{\textbf{Bias}} \\\cmidrule{3-5}\cmidrule{7-9}\cmidrule{11-13}\cmidrule{15-17}    \multicolumn{2}{l}{Unweighted} & \multicolumn{1}{c}{1.44} & \multicolumn{1}{c}{0.08} & \multicolumn{1}{c}{1.42} &       & \multicolumn{1}{c}{0.32} & \multicolumn{1}{c}{0.02} & \multicolumn{1}{c}{0.29} &       & \multicolumn{1}{c}{0.42} & \multicolumn{1}{c}{0.33} & \multicolumn{1}{c}{0.41} &       & \multicolumn{1}{c}{0.14} & \multicolumn{1}{c}{0.02} & \multicolumn{1}{c}{0.12} & \multicolumn{1}{c}{0.29} \\    \multicolumn{2}{l}{$W_0\times W_1$} & \multicolumn{1}{c}{0.05} & \multicolumn{1}{c}{0.05} & \multicolumn{1}{c}{0.05} &       & \multicolumn{1}{c}{0.02} & \multicolumn{1}{c}{0.02} & \multicolumn{1}{c}{0.02} &       & \multicolumn{1}{c}{0.36} & \multicolumn{1}{c}{0.34} & \multicolumn{1}{c}{0.36} &       & \multicolumn{1}{c}{0.02} & \multicolumn{1}{c}{0.02} & \multicolumn{1}{c}{0.02} & \multicolumn{1}{c}{0.01} \\    \multicolumn{2}{l}{$W_1$} & \multicolumn{1}{c}{1.44} & \multicolumn{1}{c}{0.02} & \multicolumn{1}{c}{0.03} &       & \multicolumn{1}{c}{0.32} & \multicolumn{1}{c}{0.01} & \multicolumn{1}{c}{0.01} &       & \multicolumn{1}{c}{0.42} & \multicolumn{1}{c}{0.33} & \multicolumn{1}{c}{0.35} &       & \multicolumn{1}{c}{0.14} & \multicolumn{1}{c}{0.01} & \multicolumn{1}{c}{0.01} & \multicolumn{1}{c}{0.11} \\    \multicolumn{2}{l}{$W_0$} & \multicolumn{1}{c}{0.03} & \multicolumn{1}{c}{0.05} & \multicolumn{1}{c}{1.42} &       & \multicolumn{1}{c}{0.01} & \multicolumn{1}{c}{0.02} & \multicolumn{1}{c}{0.29} &       & \multicolumn{1}{c}{0.36} & \multicolumn{1}{c}{0.34} & \multicolumn{1}{c}{0.41} &       & \multicolumn{1}{c}{0.01} & \multicolumn{1}{c}{0.02} & \multicolumn{1}{c}{0.12} & \multicolumn{1}{c}{0.09} \\    \multicolumn{2}{l}{$W_0$tr90$\times W_1$} & \multicolumn{1}{c}{0.51} & \multicolumn{1}{c}{0.02} & \multicolumn{1}{c}{0.03} &       & \multicolumn{1}{c}{0.11} & \multicolumn{1}{c}{0.01} & \multicolumn{1}{c}{0.01} &       & \multicolumn{1}{c}{0.38} & \multicolumn{1}{c}{0.34} & \multicolumn{1}{c}{0.35} &       & \multicolumn{1}{c}{0.05} & \multicolumn{1}{c}{0.01} & \multicolumn{1}{c}{0.01} & \multicolumn{1}{c}{0.02} \\    \multicolumn{2}{l}{$W_0\times W_1$tr90} & \multicolumn{1}{c}{0.03} & \multicolumn{1}{c}{0.04} & \multicolumn{1}{c}{0.50} &       & \multicolumn{1}{c}{0.01} & \multicolumn{1}{c}{0.01} & \multicolumn{1}{c}{0.10} &       & \multicolumn{1}{c}{0.36} & \multicolumn{1}{c}{0.34} & \multicolumn{1}{c}{0.37} &       & \multicolumn{1}{c}{0.01} & \multicolumn{1}{c}{0.01} & \multicolumn{1}{c}{0.05} & \multicolumn{1}{c}{0.02} \\          &       & \multicolumn{3}{c}{\textbf{LD}} &       & \multicolumn{3}{c}{\textbf{MHB}} &       & \multicolumn{3}{c}{\textbf{CS}} &       & \multicolumn{3}{c}{\textbf{GWD}} & \multicolumn{1}{c}{\textbf{Bias}} \\\cmidrule{3-5}\cmidrule{7-9}\cmidrule{11-13}\cmidrule{15-17}    \multicolumn{2}{l}{Unweighted} & \multicolumn{1}{c}{0.12} & \multicolumn{1}{c}{0.01} & \multicolumn{1}{c}{0.11} &       & \multicolumn{1}{c}{1.09} & \multicolumn{1}{c}{0.00} & \multicolumn{1}{c}{1.01} &       & \multicolumn{1}{c}{0.54} & \multicolumn{1}{c}{0.53} & \multicolumn{1}{c}{0.53} &       & \multicolumn{1}{c}{0.17} & \multicolumn{1}{c}{0.01} & \multicolumn{1}{c}{0.15} & \multicolumn{1}{c}{0.29} \\    \multicolumn{2}{l}{$W_0\times W_1$} & \multicolumn{1}{c}{0.01} & \multicolumn{1}{c}{0.01} & \multicolumn{1}{c}{0.01} &       & \multicolumn{1}{c}{0.00} & \multicolumn{1}{c}{0.00} & \multicolumn{1}{c}{0.00} &       & \multicolumn{1}{c}{0.05} & \multicolumn{1}{c}{0.04} & \multicolumn{1}{c}{0.04} &       & \multicolumn{1}{c}{0.01} & \multicolumn{1}{c}{0.01} & \multicolumn{1}{c}{0.01} & \multicolumn{1}{c}{0.01} \\    \multicolumn{2}{l}{$W_1$} & \multicolumn{1}{c}{0.12} & \multicolumn{1}{c}{0.01} & \multicolumn{1}{c}{0.01} &       & \multicolumn{1}{c}{1.09} & \multicolumn{1}{c}{0.00} & \multicolumn{1}{c}{0.00} &       & \multicolumn{1}{c}{0.54} & \multicolumn{1}{c}{0.05} & \multicolumn{1}{c}{0.05} &       & \multicolumn{1}{c}{0.17} & \multicolumn{1}{c}{0.01} & \multicolumn{1}{c}{0.01} & \multicolumn{1}{c}{0.11} \\    \multicolumn{2}{l}{$W_0$} & \multicolumn{1}{c}{0.01} & \multicolumn{1}{c}{0.01} & \multicolumn{1}{c}{0.11} &       & \multicolumn{1}{c}{0.00} & \multicolumn{1}{c}{0.00} & \multicolumn{1}{c}{1.01} &       & \multicolumn{1}{c}{0.05} & \multicolumn{1}{c}{0.52} & \multicolumn{1}{c}{0.52} &       & \multicolumn{1}{c}{0.01} & \multicolumn{1}{c}{0.01} & \multicolumn{1}{c}{0.15} & \multicolumn{1}{c}{0.09} \\    \multicolumn{2}{l}{$W_0$tr90$\times W_1$} & \multicolumn{1}{c}{0.04} & \multicolumn{1}{c}{0.01} & \multicolumn{1}{c}{0.01} &       & \multicolumn{1}{c}{0.13} & \multicolumn{1}{c}{0.00} & \multicolumn{1}{c}{0.00} &       & \multicolumn{1}{c}{0.54} & \multicolumn{1}{c}{0.04} & \multicolumn{1}{c}{0.04} &       & \multicolumn{1}{c}{0.06} & \multicolumn{1}{c}{0.01} & \multicolumn{1}{c}{0.01} & \multicolumn{1}{c}{0.02} \\    \multicolumn{2}{l}{$W_0\times W_1$tr90} & \multicolumn{1}{c}{0.01} & \multicolumn{1}{c}{0.01} & \multicolumn{1}{c}{0.04} &       & \multicolumn{1}{c}{0.00} & \multicolumn{1}{c}{0.00} & \multicolumn{1}{c}{0.13} &       & \multicolumn{1}{c}{0.04} & \multicolumn{1}{c}{0.52} & \multicolumn{1}{c}{0.52} &       & \multicolumn{1}{c}{0.01} & \multicolumn{1}{c}{0.01} & \multicolumn{1}{c}{0.06} & \multicolumn{1}{c}{0.02} \\    \midrule    \multicolumn{18}{c}{Censored data} \\    \midrule          &       & \multicolumn{3}{c}{\textbf{D}} &       & \multicolumn{3}{c}{\textbf{SMD}} &       & \multicolumn{3}{c}{\textbf{OVL}} &       & \multicolumn{3}{c}{\textbf{KS}} & \multicolumn{1}{c}{\textbf{Bias}} \\\cmidrule{3-5}\cmidrule{7-9}\cmidrule{11-13}\cmidrule{15-17}    \multicolumn{2}{l}{Unweighted} & 1.38  & 0.08  & 1.42  &       & 0.29  & 0.02  & 0.29  &       & 0.42  & 0.33  & 0.41  &       & 0.12  & 0.02  & 0.12  & 0.28 \\    \multicolumn{2}{l}{$W_{0}^{A,C}\times W_{1}^{A,C}$} & 0.06  & 0.05  & 0.05  &       & 0.02  & 0.02  & 0.02  &       & 0.37  & 0.34  & 0.36  &       & 0.02  & 0.02  & 0.02  & 0.01 \\    \multicolumn{2}{l}{$W_{1}^{A,C}$} & 1.44  & 0.02  & 0.03  &       & 0.32  & 0.01  & 0.01  &       & 0.43  & 0.34  & 0.35  &       & 0.14  & 0.01  & 0.01  & 0.12 \\    \multicolumn{2}{l}{$W_{0}^{A,C}$} & 0.04  & 0.06  & 1.42  &       & 0.01  & 0.02  & 0.29  &       & 0.37  & 0.34  & 0.41  &       & 0.01  & 0.02  & 0.12  & 0.10 \\    \multicolumn{2}{l}{$W_{0}^{A,C}$tr90$\times W_{1}^{A,C}$} & 0.51  & 0.02  & 0.04  &       & 0.11  & 0.01  & 0.01  &       & 0.38  & 0.34  & 0.35  &       & 0.05  & 0.01  & 0.01  & 0.03 \\    \multicolumn{2}{l}{$W_{0}^{A,C}\times W_{1}^{A,C}$tr90} & 0.04  & 0.05  & 0.50  &       & 0.02  & 0.02  & 0.10  &       & 0.37  & 0.34  & 0.37  &       & 0.02  & 0.02  & 0.05  & 0.03 \\          &       & \multicolumn{3}{c}{\textbf{LD}} &       & \multicolumn{3}{c}{\textbf{MHB}} &       & \multicolumn{3}{c}{\textbf{CS}} &       & \multicolumn{3}{c}{\textbf{GWD}} & \multicolumn{1}{c}{\textbf{Bias}} \\\cmidrule{3-5}\cmidrule{7-9}\cmidrule{11-13}\cmidrule{15-17}    \multicolumn{2}{l}{Unweighted} & 0.11  & 0.01  & 0.11  &       & 1.00  & 0.00  & 1.01  &       & 0.52  & 0.53  & 0.53  &       & 0.15  & 0.01  & 0.15  & 0.28 \\    \multicolumn{2}{l}{$W_{0}^{A,C}\times W_{1}^{A,C}$} & 0.01  & 0.01  & 0.01  &       & 0.00  & 0.00  & 0.00  &       & 0.03  & 0.03  & 0.03  &       & 0.01  & 0.01  & 0.01  & 0.01 \\    \multicolumn{2}{l}{$W_{1}^{A,C}$} & 0.13  & 0.01  & 0.01  &       & 1.09  & 0.00  & 0.00  &       & 0.52  & 0.04  & 0.04  &       & 0.17  & 0.01  & 0.01  & 0.12 \\    \multicolumn{2}{l}{$W_{0}^{A,C}$} & 0.01  & 0.01  & 0.11  &       & 0.00  & 0.01  & 1.01  &       & 0.02  & 0.52  & 0.52  &       & 0.01  & 0.01  & 0.15  & 0.10 \\    \multicolumn{2}{l}{$W_{0}^{A,C}$tr90$\times W_{1}^{A,C}$} & 0.04  & 0.01  & 0.01  &       & 0.13  & 0.00  & 0.00  &       & 0.52  & 0.03  & 0.03  &       & 0.06  & 0.01  & 0.01  & 0.03 \\    \multicolumn{2}{l}{$W_{0}^{A,C}\times W_{1}^{A,C}$tr90} & 0.01  & 0.01  & 0.04  &       & 0.00  & 0.00  & 0.13  &       & 0.02  & 0.52  & 0.52  &       & 0.01  & 0.01  & 0.06  & 0.03  \\ \hline   \end{tabular} 
		\begin{tablenotes}
			\item For each imbalance metric, the three columns represent the mean imbalance between groups $A_0$ by covariates $X_0$, mean imbalance between groups $A_1$ by covariates $X_0$ and mean imbalance between groups $A_1$ by covariates $X_1$, respectively. $W_0\mbox{trunc}90$ is $W_0$ truncated at the 90th percentile and   $W_1\mbox{trunc}90$ is $W_1$ truncated at the 90th percentile. 
		\end{tablenotes}
\end{threeparttable} \end{table}%

	 \begin{table}[h]  \centering  \begin{threeparttable} \caption{Estimated intercept ($\beta_0$) for each balance metrics in ten simulation scenarios and the explained variation (R$^2$) for each balance metric.\label{tab2}}    \begin{tabular}{rrrrrrrrrrrrrrrrr}          \cmidrule{2-16}          & \multicolumn{15}{c}{\textbf{Simple propensity score}}                                                                 &  \\\cmidrule{2-16}          &       & \multicolumn{2}{c}{\textbf{Scenario 1}} &       & \multicolumn{2}{c}{\textbf{Scenario 2}} &       & \multicolumn{2}{c}{\textbf{Scenario 3}} &       & \multicolumn{2}{c}{\textbf{Scenario 4}} &       & \multicolumn{2}{c}{\textbf{Scenario 5}} &  \\\cmidrule{3-4}\cmidrule{6-7}\cmidrule{9-10}\cmidrule{12-13}\cmidrule{15-16}          &       & \multicolumn{1}{c}{R$^2$} & \multicolumn{1}{c}{$\beta_0$} &       & \multicolumn{1}{c}{R$^2$} & \multicolumn{1}{c}{$\beta_0$} &       & \multicolumn{1}{c}{R$^2$} & \multicolumn{1}{c}{$\beta_0$} &       & \multicolumn{1}{c}{R$^2$} & \multicolumn{1}{c}{$\beta_0$} &       & \multicolumn{1}{c}{R$^2$} & \multicolumn{1}{c}{$\beta_0$} &  \\          & \multicolumn{1}{l}{\textbf{D}} & \multicolumn{1}{c}{0.94} & \multicolumn{1}{c}{-0.05} &       & \multicolumn{1}{c}{0.89} & \multicolumn{1}{c}{-0.04} &       & \multicolumn{1}{c}{0.66} & \multicolumn{1}{c}{-0.05} &       & \multicolumn{1}{c}{0.08} & \multicolumn{1}{c}{0.01} &       & \multicolumn{1}{c}{0.60} & \multicolumn{1}{c}{0.00} &  \\          & \multicolumn{1}{l}{\textbf{SMD}} & \multicolumn{1}{c}{0.94} & \multicolumn{1}{c}{-0.07} &       & \multicolumn{1}{c}{0.90} & \multicolumn{1}{c}{-0.06} &       & \multicolumn{1}{c}{0.67} & \multicolumn{1}{c}{-0.07} &       & \multicolumn{1}{c}{0.09} & \multicolumn{1}{c}{0.03} &       & \multicolumn{1}{c}{0.61} & \multicolumn{1}{c}{0.00} &  \\          & \multicolumn{1}{l}{\textbf{OVL}} & \multicolumn{1}{c}{0.91} & \multicolumn{1}{c}{-11.24} &       & \multicolumn{1}{c}{0.89} & \multicolumn{1}{c}{-0.84} &       & \multicolumn{1}{c}{0.57} & \multicolumn{1}{c}{-1.62} &       & \multicolumn{1}{c}{0.11} & \multicolumn{1}{c}{1.10} &       & \multicolumn{1}{c}{0.54} & \multicolumn{1}{c}{80.01} &  \\          & \multicolumn{1}{l}{\textbf{KS}} & \multicolumn{1}{c}{0.94} & \multicolumn{1}{c}{-0.14} &       & \multicolumn{1}{c}{0.89} & \multicolumn{1}{c}{-0.11} &       & \multicolumn{1}{c}{0.65} & \multicolumn{1}{c}{-0.15} &       & \multicolumn{1}{c}{0.11} & \multicolumn{1}{c}{0.02} &       & \multicolumn{1}{c}{0.59} & \multicolumn{1}{c}{-0.09} &  \\          & \multicolumn{1}{l}{\textbf{LD}} & \multicolumn{1}{c}{0.94} & \multicolumn{1}{c}{-0.10} &       & \multicolumn{1}{c}{0.89} & \multicolumn{1}{c}{-0.09} &       & \multicolumn{1}{c}{0.66} & \multicolumn{1}{c}{-0.12} &       & \multicolumn{1}{c}{0.11} & \multicolumn{1}{c}{0.02} &       & \multicolumn{1}{c}{0.59} & \multicolumn{1}{c}{-0.03} &  \\          & \multicolumn{1}{l}{\textbf{MHB}} & \multicolumn{1}{c}{0.93} & \multicolumn{1}{c}{-0.03} &       & \multicolumn{1}{c}{0.89} & \multicolumn{1}{c}{-0.03} &       & \multicolumn{1}{c}{0.66} & \multicolumn{1}{c}{-0.02} &       & \multicolumn{1}{c}{0.11} & \multicolumn{1}{c}{0.01} &       & \multicolumn{1}{c}{0.59} & \multicolumn{1}{c}{0.01} &  \\          & \multicolumn{1}{l}{\textbf{CS}} & \multicolumn{1}{c}{0.58} & \multicolumn{1}{c}{-0.09} &       & \multicolumn{1}{c}{0.58} & \multicolumn{1}{c}{-0.10} &       & \multicolumn{1}{c}{0.46} & \multicolumn{1}{c}{-0.09} &       & \multicolumn{1}{c}{0.01} & \multicolumn{1}{c}{0.02} &       & \multicolumn{1}{c}{0.38} & \multicolumn{1}{c}{-0.04} &  \\          & \multicolumn{1}{l}{\textbf{GWD}} & \multicolumn{1}{c}{0.94} & \multicolumn{1}{c}{-0.08} &       & \multicolumn{1}{c}{0.90} & \multicolumn{1}{c}{-0.07} &       & \multicolumn{1}{c}{0.67} & \multicolumn{1}{c}{-0.10} &       & \multicolumn{1}{c}{0.10} & \multicolumn{1}{c}{0.02} &       & \multicolumn{1}{c}{0.61} & \multicolumn{1}{c}{-0.02} &  \\          &       & \multicolumn{2}{c}{\textbf{Scenario 6}} &       & \multicolumn{2}{c}{\textbf{Scenario 7}} &       & \multicolumn{2}{c}{\textbf{Scenario 8}} &       & \multicolumn{2}{c}{\textbf{Scenario 9}} &       & \multicolumn{2}{c}{\textbf{Scenario 10}} &  \\\cmidrule{3-4}\cmidrule{6-7}\cmidrule{9-10}\cmidrule{12-13}\cmidrule{15-16}          &       & \multicolumn{1}{c}{R$^2$} & \multicolumn{1}{c}{$\beta_0$} &       & \multicolumn{1}{c}{R$^2$} & \multicolumn{1}{c}{$\beta_0$} &       & \multicolumn{1}{c}{R$^2$} & \multicolumn{1}{c}{$\beta_0$} &       & \multicolumn{1}{c}{R$^2$} & \multicolumn{1}{c}{$\beta_0$} &       & \multicolumn{1}{c}{R$^2$} & \multicolumn{1}{c}{$\beta_0$} &  \\          & \multicolumn{1}{l}{\textbf{D}} & \multicolumn{1}{c}{0.87} & \multicolumn{1}{c}{-0.55} &       & \multicolumn{1}{c}{0.94} & \multicolumn{1}{c}{-0.07} &       & \multicolumn{1}{c}{0.93} & \multicolumn{1}{c}{-0.06} &       & \multicolumn{1}{c}{0.87} & \multicolumn{1}{c}{-0.02} &       & \multicolumn{1}{c}{0.93} & \multicolumn{1}{c}{-0.03} &  \\          & \multicolumn{1}{l}{\textbf{SMD}} & \multicolumn{1}{c}{0.89} & \multicolumn{1}{c}{-0.52} &       & \multicolumn{1}{c}{0.94} & \multicolumn{1}{c}{-0.09} &       & \multicolumn{1}{c}{0.94} & \multicolumn{1}{c}{-0.08} &       & \multicolumn{1}{c}{0.88} & \multicolumn{1}{c}{-0.02} &       & \multicolumn{1}{c}{0.93} & \multicolumn{1}{c}{-0.02} &  \\          & \multicolumn{1}{l}{\textbf{OVL}} & \multicolumn{1}{c}{0.84} & \multicolumn{1}{c}{-13.98} &       & \multicolumn{1}{c}{0.90} & \multicolumn{1}{c}{-0.57} &       & \multicolumn{1}{c}{0.89} & \multicolumn{1}{c}{-0.33} &       & \multicolumn{1}{c}{0.81} & \multicolumn{1}{c}{3.63} &       & \multicolumn{1}{c}{0.89} & \multicolumn{1}{c}{5.03} &  \\          & \multicolumn{1}{l}{\textbf{KS}} & \multicolumn{1}{c}{0.88} & \multicolumn{1}{c}{-0.79} &       & \multicolumn{1}{c}{0.94} & \multicolumn{1}{c}{-0.20} &       & \multicolumn{1}{c}{0.94} & \multicolumn{1}{c}{-0.18} &       & \multicolumn{1}{c}{0.87} & \multicolumn{1}{c}{-0.05} &       & \multicolumn{1}{c}{0.93} & \multicolumn{1}{c}{-0.03} &  \\          & \multicolumn{1}{l}{\textbf{LD}} & \multicolumn{1}{c}{0.87} & \multicolumn{1}{c}{-0.69} &       & \multicolumn{1}{c}{0.94} & \multicolumn{1}{c}{-0.13} &       & \multicolumn{1}{c}{0.94} & \multicolumn{1}{c}{-0.11} &       & \multicolumn{1}{c}{0.88} & \multicolumn{1}{c}{-0.03} &       & \multicolumn{1}{c}{0.93} & \multicolumn{1}{c}{-0.02} &  \\          & \multicolumn{1}{l}{\textbf{MHB}} & \multicolumn{1}{c}{0.90} & \multicolumn{1}{c}{-0.32} &       & \multicolumn{1}{c}{0.93} & \multicolumn{1}{c}{-0.04} &       & \multicolumn{1}{c}{0.93} & \multicolumn{1}{c}{-0.04} &       & \multicolumn{1}{c}{0.87} & \multicolumn{1}{c}{-0.01} &       & \multicolumn{1}{c}{0.93} & \multicolumn{1}{c}{-0.01} &  \\          & \multicolumn{1}{l}{\textbf{CS}} & \multicolumn{1}{c}{0.57} & \multicolumn{1}{c}{-0.65} &       & \multicolumn{1}{c}{0.58} & \multicolumn{1}{c}{-0.11} &       & \multicolumn{1}{c}{0.59} & \multicolumn{1}{c}{-0.09} &       & \multicolumn{1}{c}{0.57} & \multicolumn{1}{c}{-0.04} &       & \multicolumn{1}{c}{0.63} & \multicolumn{1}{c}{-0.06} &  \\          & \multicolumn{1}{l}{\textbf{GWD}} & \multicolumn{1}{c}{0.89} & \multicolumn{1}{c}{-0.57} &       & \multicolumn{1}{c}{0.94} & \multicolumn{1}{c}{-0.11} &       & \multicolumn{1}{c}{0.94} & \multicolumn{1}{c}{-0.10} &       & \multicolumn{1}{c}{0.88} & \multicolumn{1}{c}{-0.03} &       & \multicolumn{1}{c}{0.93} & \multicolumn{1}{c}{-0.03} &  \\\cmidrule{2-16}      &     
	 			\multicolumn{15}{c}{\textbf{Complex propensity score														}}                                                  &  \\\cmidrule{2-16}          &       & \multicolumn{2}{c}{\textbf{Scenario 1}} &       & \multicolumn{2}{c}{\textbf{Scenario 2}} &       & \multicolumn{2}{c}{\textbf{Scenario 3}} &       & \multicolumn{2}{c}{\textbf{Scenario 4}} &       & \multicolumn{2}{c}{\textbf{Scenario 5}} &  \\\cmidrule{3-4}\cmidrule{6-7}\cmidrule{9-10}\cmidrule{12-13}\cmidrule{15-16}          &       & \multicolumn{1}{c}{R$^2$} & \multicolumn{1}{c}{$\beta_0$} &       & \multicolumn{1}{c}{R$^2$} & \multicolumn{1}{c}{$\beta_0$} &       & \multicolumn{1}{c}{R$^2$} & \multicolumn{1}{c}{$\beta_0$} &       & \multicolumn{1}{c}{R$^2$} & \multicolumn{1}{c}{$\beta_0$} &       & \multicolumn{1}{c}{R$^2$} & \multicolumn{1}{c}{$\beta_0$} &  \\          & \multicolumn{1}{l}{\textbf{D}} & \multicolumn{1}{c}{0.94} & \multicolumn{1}{c}{-0.05} &       & \multicolumn{1}{c}{0.89} & \multicolumn{1}{c}{-0.04} &       & \multicolumn{1}{c}{0.66} & \multicolumn{1}{c}{-0.05} &       & \multicolumn{1}{c}{0.08} & \multicolumn{1}{c}{0.00} &       & \multicolumn{1}{c}{0.60} & \multicolumn{1}{c}{0.00} &  \\          & \multicolumn{1}{l}{\textbf{SMD}} & \multicolumn{1}{c}{0.94} & \multicolumn{1}{c}{-0.06} &       & \multicolumn{1}{c}{0.89} & \multicolumn{1}{c}{-0.06} &       & \multicolumn{1}{c}{0.66} & \multicolumn{1}{c}{-0.07} &       & \multicolumn{1}{c}{0.09} & \multicolumn{1}{c}{0.02} &       & \multicolumn{1}{c}{0.61} & \multicolumn{1}{c}{0.00} &  \\          & \multicolumn{1}{l}{\textbf{OVL}} & \multicolumn{1}{c}{0.91} & \multicolumn{1}{c}{-9.55} &       & \multicolumn{1}{c}{0.88} & \multicolumn{1}{c}{-1.57} &       & \multicolumn{1}{c}{0.56} & \multicolumn{1}{c}{-1.88} &       & \multicolumn{1}{c}{0.10} & \multicolumn{1}{c}{1.08} &       & \multicolumn{1}{c}{0.54} & \multicolumn{1}{c}{68.73} &  \\          & \multicolumn{1}{l}{\textbf{KS}} & \multicolumn{1}{c}{0.94} & \multicolumn{1}{c}{-0.13} &       & \multicolumn{1}{c}{0.89} & \multicolumn{1}{c}{-0.11} &       & \multicolumn{1}{c}{0.65} & \multicolumn{1}{c}{-0.13} &       & \multicolumn{1}{c}{0.10} & \multicolumn{1}{c}{0.01} &       & \multicolumn{1}{c}{0.59} & \multicolumn{1}{c}{-0.11} &  \\          & \multicolumn{1}{l}{\textbf{LD}} & \multicolumn{1}{c}{0.94} & \multicolumn{1}{c}{-0.09} &       & \multicolumn{1}{c}{0.89} & \multicolumn{1}{c}{-0.09} &       & \multicolumn{1}{c}{0.66} & \multicolumn{1}{c}{-0.11} &       & \multicolumn{1}{c}{0.10} & \multicolumn{1}{c}{0.01} &       & \multicolumn{1}{c}{0.59} & \multicolumn{1}{c}{-0.03} &  \\          & \multicolumn{1}{l}{\textbf{MHB}} & \multicolumn{1}{c}{0.93} & \multicolumn{1}{c}{-0.03} &       & \multicolumn{1}{c}{0.89} & \multicolumn{1}{c}{-0.03} &       & \multicolumn{1}{c}{0.66} & \multicolumn{1}{c}{-0.02} &       & \multicolumn{1}{c}{0.10} & \multicolumn{1}{c}{0.01} &       & \multicolumn{1}{c}{0.59} & \multicolumn{1}{c}{0.01} &  \\          & \multicolumn{1}{l}{\textbf{CS}} & \multicolumn{1}{c}{0.65} & \multicolumn{1}{c}{-0.08} &       & \multicolumn{1}{c}{0.66} & \multicolumn{1}{c}{-0.09} &       & \multicolumn{1}{c}{0.52} & \multicolumn{1}{c}{-0.08} &       & \multicolumn{1}{c}{0.01} & \multicolumn{1}{c}{0.01} &       & \multicolumn{1}{c}{0.41} & \multicolumn{1}{c}{-0.04} &  \\          & \multicolumn{1}{l}{\textbf{GWD}} & \multicolumn{1}{c}{0.94} & \multicolumn{1}{c}{-0.07} &       & \multicolumn{1}{c}{0.90} & \multicolumn{1}{c}{-0.07} &       & \multicolumn{1}{c}{0.66} & \multicolumn{1}{c}{-0.09} &       & \multicolumn{1}{c}{0.10} & \multicolumn{1}{c}{0.02} &       & \multicolumn{1}{c}{0.61} & \multicolumn{1}{c}{-0.01} &  \\          &       & \multicolumn{2}{c}{\textbf{Scenario 6}} &       & \multicolumn{2}{c}{\textbf{Scenario 7}} &       & \multicolumn{2}{c}{\textbf{Scenario 8}} &       & \multicolumn{2}{c}{\textbf{Scenario 9}} &       & \multicolumn{2}{c}{\textbf{Scenario 10}} &  \\\cmidrule{3-4}\cmidrule{6-7}\cmidrule{9-10}\cmidrule{12-13}\cmidrule{15-16}          &       & \multicolumn{1}{c}{R$^2$} & \multicolumn{1}{c}{$\beta_0$} &       & \multicolumn{1}{c}{R$^2$} & \multicolumn{1}{c}{$\beta_0$} &       & \multicolumn{1}{c}{R$^2$} & \multicolumn{1}{c}{$\beta_0$} &       & \multicolumn{1}{c}{R$^2$} & \multicolumn{1}{c}{$\beta_0$} &       & \multicolumn{1}{c}{R$^2$} & \multicolumn{1}{c}{$\beta_0$} &  \\          & \multicolumn{1}{l}{\textbf{D}} & \multicolumn{1}{c}{0.87} & \multicolumn{1}{c}{-0.55} &       & \multicolumn{1}{c}{0.94} & \multicolumn{1}{c}{-0.06} &       & \multicolumn{1}{c}{0.93} & \multicolumn{1}{c}{-0.06} &       & \multicolumn{1}{c}{0.88} & \multicolumn{1}{c}{-0.03} &       & \multicolumn{1}{c}{0.93} & \multicolumn{1}{c}{-0.03} &  \\          & \multicolumn{1}{l}{\textbf{SMD}} & \multicolumn{1}{c}{0.90} & \multicolumn{1}{c}{-0.53} &       & \multicolumn{1}{c}{0.94} & \multicolumn{1}{c}{-0.08} &       & \multicolumn{1}{c}{0.94} & \multicolumn{1}{c}{-0.07} &       & \multicolumn{1}{c}{0.88} & \multicolumn{1}{c}{-0.03} &       & \multicolumn{1}{c}{0.94} & \multicolumn{1}{c}{-0.03} &  \\          & \multicolumn{1}{l}{\textbf{OVL}} & \multicolumn{1}{c}{0.84} & \multicolumn{1}{c}{-14.59} &       & \multicolumn{1}{c}{0.90} & \multicolumn{1}{c}{-0.96} &       & \multicolumn{1}{c}{0.89} & \multicolumn{1}{c}{-1.56} &       & \multicolumn{1}{c}{0.81} & \multicolumn{1}{c}{2.53} &       & \multicolumn{1}{c}{0.90} & \multicolumn{1}{c}{5.03} &  \\          & \multicolumn{1}{l}{\textbf{KS}} & \multicolumn{1}{c}{0.88} & \multicolumn{1}{c}{-0.80} &       & \multicolumn{1}{c}{0.94} & \multicolumn{1}{c}{-0.18} &       & \multicolumn{1}{c}{0.94} & \multicolumn{1}{c}{-0.16} &       & \multicolumn{1}{c}{0.88} & \multicolumn{1}{c}{-0.06} &       & \multicolumn{1}{c}{0.94} & \multicolumn{1}{c}{-0.06} &  \\          & \multicolumn{1}{l}{\textbf{LD}} & \multicolumn{1}{c}{0.87} & \multicolumn{1}{c}{-0.70} &       & \multicolumn{1}{c}{0.94} & \multicolumn{1}{c}{-0.12} &       & \multicolumn{1}{c}{0.94} & \multicolumn{1}{c}{-0.11} &       & \multicolumn{1}{c}{0.88} & \multicolumn{1}{c}{-0.03} &       & \multicolumn{1}{c}{0.94} & \multicolumn{1}{c}{-0.03} &  \\          & \multicolumn{1}{l}{\textbf{MHB}} & \multicolumn{1}{c}{0.91} & \multicolumn{1}{c}{-0.32} &       & \multicolumn{1}{c}{0.93} & \multicolumn{1}{c}{-0.04} &       & \multicolumn{1}{c}{0.93} & \multicolumn{1}{c}{-0.04} &       & \multicolumn{1}{c}{0.87} & \multicolumn{1}{c}{-0.02} &       & \multicolumn{1}{c}{0.93} & \multicolumn{1}{c}{-0.01} &  \\          & \multicolumn{1}{l}{\textbf{CS}} & \multicolumn{1}{c}{0.66} & \multicolumn{1}{c}{-0.63} &       & \multicolumn{1}{c}{0.65} & \multicolumn{1}{c}{-0.11} &       & \multicolumn{1}{c}{0.05} & \multicolumn{1}{c}{0.05} &       & \multicolumn{1}{c}{0.65} & \multicolumn{1}{c}{-0.04} &       & \multicolumn{1}{c}{0.70} & \multicolumn{1}{c}{-0.06} &  \\          & \multicolumn{1}{l}{\textbf{GWD}} & \multicolumn{1}{c}{0.89} & \multicolumn{1}{c}{-0.58} &       & \multicolumn{1}{c}{0.94} & \multicolumn{1}{c}{-0.10} &       & \multicolumn{1}{c}{0.93} & \multicolumn{1}{c}{-0.09} &       & \multicolumn{1}{c}{0.88} & \multicolumn{1}{c}{-0.04} &       & \multicolumn{1}{c}{0.94} & \multicolumn{1}{c}{-0.04} &  \\  \hline  \end{tabular}  
		\begin{tablenotes}
			\item Scenario 1 is the base case, Scenario 2 is the low prevalence of exposure case, Scenario 3 is the small sample case, Scenario 4 is the high imbalance, no confounding  case, Scenario 5 is the low imbalance sample case, Scenario 6 is the high imbalance-high confounding case, Scenario 7 is the nonlinear outcome case, Scenario 8 is the nonlinear outcome and exposure case, Scenario 9 is the redundant covariates case, Scenario 10 is the instrumental variables case.
		\end{tablenotes}
	\end{threeparttable}
	\label{tab:addlabel}\end{table}

 \begin{table}[h]  \centering  \caption{Estimated intercept ($\beta_0$) for each balance metrics in the base case scenario with censoring and the explained variation (R$^2$) for each balance metric.\label{tab3}}    \begin{tabular}{rrr}          & \multicolumn{2}{c}{} \\    \midrule          & \multicolumn{1}{c}{R$^{2}$} & \multicolumn{1}{c}{$\beta_0$} \\\cmidrule{2-3}    \multicolumn{1}{l}{\textbf{D}} & \multicolumn{1}{c}{0.92} & \multicolumn{1}{c}{-0.03} \\    \multicolumn{1}{l}{\textbf{SMD}} & \multicolumn{1}{c}{0.92} & \multicolumn{1}{c}{-0.06} \\    \multicolumn{1}{l}{\textbf{OVL}} & \multicolumn{1}{c}{0.88} & \multicolumn{1}{c}{10.90} \\    \multicolumn{1}{l}{\textbf{KS}} & \multicolumn{1}{c}{0.92} & \multicolumn{1}{c}{-0.13} \\    \multicolumn{1}{l}{\textbf{LD}} & \multicolumn{1}{c}{0.92} & \multicolumn{1}{c}{-0.09} \\    \multicolumn{1}{l}{\textbf{MHB}} & \multicolumn{1}{c}{0.92} & \multicolumn{1}{c}{-0.02} \\    \multicolumn{1}{l}{\textbf{CS}} & \multicolumn{1}{c}{0.59} & \multicolumn{1}{c}{-0.07} \\    \multicolumn{1}{l}{\textbf{GWD}} & \multicolumn{1}{c}{0.92} & \multicolumn{1}{c}{-0.08}   \\   \hline  \end{tabular} \label{tab:addlabel} \end{table}

				\section{Illustration} \label{sec4}
				We now illustrate the use of our methods in real data concerning the effectiveness of statins for the prevention of a first cardiovascular event among older adults, that is, for primary prevention. Several randomized studies indicate that statins have major benefits for primary prevention in certain populations \cite{10.1001/jama.282.24.2340,law2003quantifying,10.1053/euhj.2001.2775,key2}. However, few studies inform about the real-world benefits in people aged 65 years and older for primary prevention \cite{al}. Population-wide administrative data could provide crucial evidence because their large sample size and population representativeness allow studying populations that are typically excluded or under-represented in randomized studies. We thus conducted a retrospective cohort study using medico-administrative data from Quebec, Canada. This project was approved by the research ethics board of the research center of the CHU of Quebec (decision $\#2020-4892$). The purpose of the illustrative study was to estimate the hazard reduction of a first cardiovascular event or death associated to patients’ treatment compliance within the first year of treatment. 
				
				\subsection{Data}
				We formed a retrospective cohort using data available at the \textit{Institut de la statistique du Qu\'ebec} by merging data from five medical administrative databases in Qu\'ebec, Canada using a unique anonymized personal identifier: the health insurance registry, the pharmaceutical services database, the physician claims database, the hospitalization database and the death registry. 
				The cohort included Quebecers aged 66 or older as of April 1, 2013, who were beneficiaries of Quebec's public drug insurance plan without interruption for the previous year. Individuals with a statin dispensation in the previous year (between April 1, 2012, and March 31, 2013) and those with a history of cardiovascular disease (myocardial infarction, heart failure and other ischemic heart disease, cerebrovascular disease, and atherosclerosis) in the past 5 years (April 1, 2008, to March 31, 2013) were excluded to ensure that only primary prevention statin users were included. To control the risk of confounding by indication, only individuals who received at least one statin dispensation between April 2012 and March 2017 were included in this analysis resulting in a sample of 32,690 statin initiators. Each individual's exposure status was updated monthly 
				up to 12 months after their first statin dispensation up to 12 months later. An individual was considered as exposed in a given month if they had a filled statin prescription covering at least one day of that month, after accounting for a 50\% grace period.   Follow-up for the outcome began immediately after the 12-month 
				exposure follow-up and until the earliest of the following: occurrence of a cardiovascular event, death, termination of public drug plan membership, admission to a long-term care facility, or March 31, 2018, for a total of 386,422 person-months of follow-up. To adjust for selection bias when estimating the effect of statins, we used inverse probability censoring weights. We considered that all individuals having one of the following events: cardiovascular disease during exposure follow-up, death during exposure follow-up or termination of public drug plan membership during exposure follow-up were censored.
				
				Using the notation introduced in this paper, the data set contained the following information about each subject:
				\begin{enumerate}
					\item $C_t = 1$, if an individual is censored at the beginning of month $t$, ($C_t = 0$ otherwise). 
					\item $X_0,...,X_{11}$ the set of covariates (potential confounders) measured at months $0,...,11$. These include both time-fixed covariates measured at baseline only (sex, age and region of residence) and time-varying covariates whose values were updated monthly (prevalent diabetes, hypertension, chronic renal failure, use of aspirin and other antiplatelet agents and pharmacological treatment of blood pressure). These factors were identified as potential confounders on the basis of a literature review and experts' opinions.   
					\item Statin exposure ($A_t=1$) or non-exposure ($A_t=0$) in month $t$, $t=0,..,11$. 
					\item A binary outcome ($Y$) measured after exposure follow-up ($Y=1$ if the individual experiences a cardiovascular event or all-cause death during the outcome follow-up and $0$ otherwise).
				\end{enumerate}
				\subsection{Analysis}
				Because of the time-to-event nature of the outcome, our goal was to estimate the parameters of the marginal structural Cox model $\lambda_{\bar{a}_T}(\tau) = \lambda_0(\tau)\exp(\beta \sum_{t=0}^{11} a_t+\beta_1V)$, where $V$ is set of baseline or time-invariant confounders using an inverse probability weighting estimator \cite{ba,bb}. We first estimated the probability of statin exposure at each time-point (month $t = 0$ to month $t = 11$) as a function of pretreatment characteristics (baseline and time-varying covariates) using logistic regressions. The treatment weights were then estimated as the inverse of the probabilities of the observed exposure status at each time-point. Censoring weights were calculated analogously. Marginal stabilized weights were used to check for covariate balance, and stabilized weights were used to estimate the effect of statin history. For each individual, we multiplied their treatment weight and censoring weight from month 0 to month 11 to obtain a total weight. Because of the presence of extreme weights, we truncated the total weights at the 99th percentile as suggested in previous studies \cite{bb,bc}. 
				
				Based on our simulation results, we decided to first check balance between treatment groups at each time-point using a global metric, the MHB. As we proposed in previous sections, we consider all the covariates measured at a given time-point jointly, but covariates measured at different time-points separately. If a global imbalance was found at a particular time-point, we would then use SMDs to more finely assess which covariates were imbalanced. Using this procedure, we assessed the balance both in the unweighted data and in the weighted data. As we did previously \cite{bc},  we only assessed covariate balance between treatment groups after period 3, because very few participants were non-users in first three periods ($0\%$, $0.3\%$ and $6.3\%$ respectively). 
				
				\subsection{Results}
				Table \ref{tab4} describes the baseline characteristics of the population. As required by \textit{Institut de la statistique du Qu\'ebec} confidentiality rules, note that all frequencies were rounded in base five. The mean age of the participants was 71.96 years, 57.1\% were women, 16\% had a diabetes diagnosis and 61.2\% received blood pressure treatment.    
				
				Table \ref{tab5} reports the balance results. Given that $19$ variables were examined (after considering dummy variables as separate variables) for time-fixed covariates and that  $7$ variables were examined for time-varying covariates
				the threshold for assessing that distribution of the covariates differed meaningfully between treatment group was $0.19$ and $0.07$ respectively. All time-fixed covariates were well balanced in both unweighted and weighted data.
				For time-varying covariatess exposure groups were relatively imbalanced before weighting at each time-point, with the largest MHB being $0.17$. 
				Weighting improved overall covariate balance at all time-points except balance of covariates at time $t = 1, 2, 3, 7$ between the $A_7$ treatment groups. To detect problematic covariates, we used the SMD. We found that the number of medications in the year of statin initiation, number of days of hospitalization in the year of statin initiation and Use of aspirin and other antiplatelet agents were imbalanced. We recalculated the weights by including in the treatment model quadratic terms for the number of medications in the year of statin initiation and number of days of hospitalization in the year of statin initiation, and an interaction term between Use of aspirin and other antiplatelet agents and number of medications in the year of statin initiation on one hand and between Use of aspirin and other antiplatelet agents and number of days of hospitalization in the year of statin initiation on the other hand. The overall balance improved at times $t = 1,2$, from $0.08$ to $0.07$, but remained unchanged at times $t = 3,7$. Despite several attempts to modify the treatment model, these imbalances remained.
				
				Before adjustment, each additional month of statin exposure during the first year of treatment was associated, within the following 5 years, with $3\%$ reduction of the hazard of a first cardiovascular event or death (HR = 0.97, 95$\%$ CI: 0.96 -- 0.98). Very similar results were obtained after adjustment for potential confounding and censoring bias (HR~=~0.97, 95$\%$ CI: 0.96 -- 0.98). It should be noted that the treatment compliance within the first year is most likely correlated with compliance in the following years. Because we were unable to fully balance treatment groups, some relatively low residual confounding due to measured confounders is expected.
				
				{\renewcommand{\arraystretch}{1.2} 
					{\setlength{\tabcolsep}{0.2cm}  
				\begin{table}[h]  
					\centering  \begin{threeparttable} 
					\caption{Participant characteristics at baseline \label{tab4}}    
					\begin{tabular}{rrrrr}    \cmidrule{2-4}          & \multicolumn{2}{l}{\textbf{Characteristic}} & \multicolumn{1}{c}{\textbf{N = 32,690}} &  \\\cmidrule{2-4}          & \multicolumn{2}{l}{Age (mean (SD))    } & \multicolumn{1}{c}{71.96 (5.69) } &  \\          & \multicolumn{2}{l}{Women n  (\%)} & \multicolumn{1}{c}{18,670 (57.1)} &  \\          & \multicolumn{2}{l}{Socio-sanitary area n (\%)} &       &  \\    \textcolor[rgb]{ .02,  .388,  .757}{} &       & \multicolumn{1}{l}{Bas Saint Laurent} & \multicolumn{1}{c}{1,080 (3.3) } &  \\          &       & \multicolumn{1}{l}{Saguenay-Lac Saint Jean	} & \multicolumn{1}{c}{1,060 (3.2) } &  \\          &       & \multicolumn{1}{l}{Capitale Nationale 	} & \multicolumn{1}{c}{2,345 (7.2)} &  \\          &       & \multicolumn{1}{l}{Mauricie et Centre du Qu\'{e}bec 	} & \multicolumn{1}{c}{2,440 (7.5)} &  \\          &       & \multicolumn{1}{l}{Estrie } & \multicolumn{1}{c}{2,090 (6.4)} &  \\          &       & \multicolumn{1}{l}{Montr\'{e}al} & \multicolumn{1}{c}{7,585 (23.2)} &  \\          &       & \multicolumn{1}{l}{Outaouais } & \multicolumn{1}{c}{1,040 (3.2)} &  \\          &       & \multicolumn{1}{l}{Abitibi-T\'{e}miscamingue 	} & \multicolumn{1}{c}{580 (1.8)} &  \\          &       & \multicolumn{1}{l}{C\^{o}te-Nord } & \multicolumn{1}{c}{355 (1.1)} &  \\          &       & \multicolumn{1}{l}{Gasp\'{e}sie-les de la Madeleine} & \multicolumn{1}{c}{555 (1.7)} &  \\          &       & \multicolumn{1}{l}{Chaudi\`{e}re Appalaches 	} & \multicolumn{1}{c}{2,065 (6.3)} &  \\          &       & \multicolumn{1}{l}{Laval } & \multicolumn{1}{c}{1,655 (5.1)} &  \\          &       & \multicolumn{1}{l}{Lanaudi\`{e}re } & \multicolumn{1}{c}{2,185 (6.7)} &  \\          &       & \multicolumn{1}{l}{Laurentides } & \multicolumn{1}{c}{2,370 (7.2)} &  \\          &       & \multicolumn{1}{l}{Mont\'{e}r\'{e}gie } & \multicolumn{1}{c}{5,130 (15.7)} &  \\          &       & \multicolumn{1}{p{13.585em}}{Terres Cries de la Baie James and Nord du Qu\'{e}bec} & \multicolumn{1}{c}{55 (0.2)} &  \\ & \multicolumn{2}{l}{Prevalent diabetes, (yes, n(\%)) } & \multicolumn{1}{c}{5,240 (16)} &  \\          & \multicolumn{2}{l}{Hypertension, (yes, n(\%)) } & \multicolumn{1}{c}{8,120 (24.8)} &  \\          & \multicolumn{2}{p{18.585em}}{Number of filled prescriptions in the year of statin initiation, (mean (SD))  } & \multicolumn{1}{c}{4.21 (3.80)} &  \\          & \multicolumn{2}{p{18.585em}}{Number of days of hospitalization in the year of statin initiation, (mean (SD))  } & \multicolumn{1}{c}{0.21 (0.59)} &  \\          & \multicolumn{2}{l}{Use of aspirin and other antiplatelet agents (yes, n(\%)) } & \multicolumn{1}{c}{1,005 (3.1)} &  \\          & \multicolumn{2}{l}{Chronic kidney disease, (yes, n(\%)) } & \multicolumn{1}{c}{135 (0.4)} &  \\          & \multicolumn{2}{l}{Blood pressure treatment, (yes, n(\%)) } & \multicolumn{1}{c}{19,995 (61.2)} &  \\
						\cmidrule{2-4}          &       &       &       &  \\
					\end{tabular}  \end{threeparttable}  \end{table}
				
			\begin{table}[h]  \centering \begin{threeparttable}  \caption{Covariates balance at each time-point between statin users and non-users in unweighted data and in data weighted by IPTW$\times$IPCW \label{tab5}}    \begin{tabular}{rrrrrrrrrrrr}  \cmidrule{2-11}          &       &       & \multicolumn{2}{c}{$A_3 = 1$ vs $A_3 = 0$} &       & \multicolumn{2}{c}{$A_4 = 1$ vs $A_4 = 0$} &       & \multicolumn{2}{c}{$A_5 = 1$ vs $A_5 = 0$} &  \\\cmidrule{4-5}\cmidrule{7-8}\cmidrule{10-11}          &       &       & \multicolumn{1}{l}{\textbf{Unweighted}} & \multicolumn{1}{l}{\textbf{Weighted}} &       & \multicolumn{1}{l}{\textbf{Unweighted}} & \multicolumn{1}{l}{\textbf{Weighted}} &       & \multicolumn{1}{l}{\textbf{Unweighted}} & \multicolumn{1}{l}{\textbf{Weighted}} &  \\\cmidrule{2-11}          & \multicolumn{2}{l}{Time-fixed covariates} & \multicolumn{1}{c}{0.06} & \multicolumn{1}{c}{0.07} &       & \multicolumn{1}{c}{0.06} & \multicolumn{1}{c}{0.10} &       & \multicolumn{1}{c}{0.05} & \multicolumn{1}{c}{0.11} &  \\          & \multicolumn{2}{l}{Time-varying covariates} &       &       &       &       &       &       &       &       &  \\          & \multicolumn{1}{c}{\multirow{6}[1]{*}{\begin{turn}{1}Time-point\end{turn}}} & \multicolumn{1}{c}{0} & \multicolumn{1}{c}{0.04} & \multicolumn{1}{c}{0.07} &       & \multicolumn{1}{c}{0.04} & \multicolumn{1}{c}{0.04} &       & \multicolumn{1}{c}{0.04} & \multicolumn{1}{c}{0.01} &  \\          &       & \multicolumn{1}{c}{1} & \multicolumn{1}{c}{0.07} & \multicolumn{1}{c}{0.06} &       & \multicolumn{1}{c}{0.06} & \multicolumn{1}{c}{0.03} &       & \multicolumn{1}{c}{0.06} & \multicolumn{1}{c}{0.02} &  \\          &       & \multicolumn{1}{c}{2} & \multicolumn{1}{c}{0.12} & \multicolumn{1}{c}{0.05} &       & \multicolumn{1}{c}{0.09} & \multicolumn{1}{c}{0.03} &       & \multicolumn{1}{c}{0.08} & \multicolumn{1}{c}{0.02} &  \\          &       & \multicolumn{1}{c}{3} & \multicolumn{1}{c}{0.17} & \multicolumn{1}{c}{0.03} &       & \multicolumn{1}{c}{0.13} & \multicolumn{1}{c}{0.02} &       & \multicolumn{1}{c}{0.11} & \multicolumn{1}{c}{0.02} &  \\          &       & \multicolumn{1}{c}{4} &       &       &       & \multicolumn{1}{c}{0.15} & \multicolumn{1}{c}{0.02} &       & \multicolumn{1}{c}{0.13} & \multicolumn{1}{c}{0.02} &  \\          &       & \multicolumn{1}{c}{5} &       &       &       &       &       &       & \multicolumn{1}{c}{0.15} & \multicolumn{1}{c}{0.02} &  \\\cmidrule{2-11}          &       &       & \multicolumn{2}{c}{$A_6 = 1$ vs $A_6 = 0$} &       & \multicolumn{2}{c}{$A_7 = 1$ vs $A_7 = 0$} &       & \multicolumn{2}{c}{$A_8 = 1$ vs $A_8 = 0$} &  \\\cmidrule{4-5}\cmidrule{7-8}\cmidrule{10-11}          &       &       & \multicolumn{1}{l}{\textbf{Unweighted}} & \multicolumn{1}{l}{\textbf{Weighted}} &       & \multicolumn{1}{l}{\textbf{Unweighted}} & \multicolumn{1}{l}{\textbf{Weighted}} &       & \multicolumn{1}{l}{\textbf{Unweighted}} & \multicolumn{1}{l}{\textbf{Weighted}} &  \\\cmidrule{2-11}          & \multicolumn{2}{l}{Time-fixed covariates} & \multicolumn{1}{c}{0.05} & \multicolumn{1}{c}{0.05} &       & \multicolumn{1}{c}{0.05} & \multicolumn{1}{c}{0.03} &       & \multicolumn{1}{c}{0.05} & \multicolumn{1}{c}{0.05} &  \\          & \multicolumn{2}{l}{Time-varying covariates} &       &       &       &       &       &       &       &       &  \\          & \multicolumn{1}{c}{\multirow{9}[1]{*}{\begin{turn}{1}Time-point\end{turn}}} & \multicolumn{1}{c}{0} & \multicolumn{1}{c}{0.04} & \multicolumn{1}{c}{0.01} &       & \multicolumn{1}{c}{0.04} & \multicolumn{1}{c}{0.06} &       & \multicolumn{1}{c}{0.04} & \multicolumn{1}{c}{0.01} &  \\          &       & \multicolumn{1}{c}{1} & \multicolumn{1}{c}{0.06} & \multicolumn{1}{c}{0.01} &       & \multicolumn{1}{c}{0.06} & \multicolumn{1}{c}{0.07} &       & \multicolumn{1}{c}{0.05} & \multicolumn{1}{c}{0.02} &  \\          &       & \multicolumn{1}{c}{2} & \multicolumn{1}{c}{0.08} & \multicolumn{1}{c}{0.01} &       & \multicolumn{1}{c}{0.08} & \multicolumn{1}{c}{0.07} &       & \multicolumn{1}{c}{0.07} & \multicolumn{1}{c}{0.02} &  \\          &       & \multicolumn{1}{c}{3} & \multicolumn{1}{c}{0.10} & \multicolumn{1}{c}{0.01} &       & \multicolumn{1}{c}{0.10} & \multicolumn{1}{c}{0.08} &       & \multicolumn{1}{c}{0.09} & \multicolumn{1}{c}{0.01} &  \\          &       & \multicolumn{1}{c}{4} & \multicolumn{1}{c}{0.11} & \multicolumn{1}{c}{0.02} &       & \multicolumn{1}{c}{0.11} & \multicolumn{1}{c}{0.07} &       & \multicolumn{1}{c}{0.10} & \multicolumn{1}{c}{0.01} &  \\          &       & \multicolumn{1}{c}{5} & \multicolumn{1}{c}{0.13} & \multicolumn{1}{c}{0.02} &       & \multicolumn{1}{c}{0.12} & \multicolumn{1}{c}{0.07} &       & \multicolumn{1}{c}{0.12} & \multicolumn{1}{c}{0.02} &  \\          &       & \multicolumn{1}{c}{6} & \multicolumn{1}{c}{0.15} & \multicolumn{1}{c}{0.03} &       & \multicolumn{1}{c}{0.14} & \multicolumn{1}{c}{0.05} &       & \multicolumn{1}{c}{0.13} & \multicolumn{1}{c}{0.02} &  \\          &       & \multicolumn{1}{c}{7} &       &       &       & \multicolumn{1}{c}{0.15} & \multicolumn{1}{c}{0.08} &       & \multicolumn{1}{c}{0.14} & \multicolumn{1}{c}{0.01} &  \\          &       & \multicolumn{1}{c}{8} &       &       &       &       &       &       & \multicolumn{1}{c}{0.15} & \multicolumn{1}{c}{0.01} &  \\\cmidrule{2-11}          &       &       & \multicolumn{2}{c}{$A_9 = 1$ vs $A_9 = 0$} &       & \multicolumn{2}{c}{$A_{10} = 1$ vs $A_{10} = 0$} &       & \multicolumn{2}{c}{$A_{11} = 1$ vs $A_{11} = 0$} &  \\\cmidrule{4-5}\cmidrule{7-8}\cmidrule{10-11}          &       &       & \multicolumn{1}{l}{\textbf{Unweighted}} & \multicolumn{1}{l}{\textbf{Weighted}} &       & \multicolumn{1}{l}{\textbf{Unweighted}} & \multicolumn{1}{l}{\textbf{Weighted}} &       & \multicolumn{1}{l}{\textbf{Unweighted}} & \multicolumn{1}{l}{\textbf{Weighted}} &  \\\cmidrule{2-11}          & \multicolumn{2}{l}{Time-fixed covariates} & \multicolumn{1}{c}{0.05} & \multicolumn{1}{c}{0.05} &       & \multicolumn{1}{c}{0.05} & \multicolumn{1}{c}{0.03} &       & \multicolumn{1}{c}{0.05} & \multicolumn{1}{c}{0.05} &  \\          & \multicolumn{2}{l}{Time-varying covariates} &       &       &       &       &       &       &       &       &  \\          & \multicolumn{1}{c}{\multirow{12}[1]{*}{\begin{turn}{1}Time-point\end{turn}}} & \multicolumn{1}{c}{0} & \multicolumn{1}{c}{0.04} & \multicolumn{1}{c}{0.04} &       & \multicolumn{1}{c}{0.04} & \multicolumn{1}{c}{0.02} &       & \multicolumn{1}{c}{0.04} & \multicolumn{1}{c}{0.02} &  \\          &       & \multicolumn{1}{c}{1} & \multicolumn{1}{c}{0.05} & \multicolumn{1}{c}{0.04} &       & \multicolumn{1}{c}{0.05} & \multicolumn{1}{c}{0.02} &       & \multicolumn{1}{c}{0.05} & \multicolumn{1}{c}{0.02} &  \\          &       & \multicolumn{1}{c}{2} & \multicolumn{1}{c}{0.07} & \multicolumn{1}{c}{0.04} &       & \multicolumn{1}{c}{0.07} & \multicolumn{1}{c}{0.02} &       & \multicolumn{1}{c}{0.07} & \multicolumn{1}{c}{0.02} &  \\          &       & \multicolumn{1}{c}{3} & \multicolumn{1}{c}{0.09} & \multicolumn{1}{c}{0.04} &       & \multicolumn{1}{c}{0.09} & \multicolumn{1}{c}{0.02} &       & \multicolumn{1}{c}{0.08} & \multicolumn{1}{c}{0.02} &  \\          &       & \multicolumn{1}{c}{4} & \multicolumn{1}{c}{0.10} & \multicolumn{1}{c}{0.04} &       & \multicolumn{1}{c}{0.10} & \multicolumn{1}{c}{0.03} &       & \multicolumn{1}{c}{0.09} & \multicolumn{1}{c}{0.02} &  \\          &       & \multicolumn{1}{c}{5} & \multicolumn{1}{c}{0.11} & \multicolumn{1}{c}{0.05} &       & \multicolumn{1}{c}{0.11} & \multicolumn{1}{c}{0.03} &       & \multicolumn{1}{c}{0.10} & \multicolumn{1}{c}{0.02} &  \\          &       & \multicolumn{1}{c}{6} & \multicolumn{1}{c}{0.12} & \multicolumn{1}{c}{0.05} &       & \multicolumn{1}{c}{0.11} & \multicolumn{1}{c}{0.03} &       & \multicolumn{1}{c}{0.11} & \multicolumn{1}{c}{0.02} &  \\          &       & \multicolumn{1}{c}{7} & \multicolumn{1}{c}{0.13} & \multicolumn{1}{c}{0.04} &       & \multicolumn{1}{c}{0.12} & \multicolumn{1}{c}{0.02} &       & \multicolumn{1}{c}{0.11} & \multicolumn{1}{c}{0.02} &  \\          &       & \multicolumn{1}{c}{8} & \multicolumn{1}{c}{0.14} & \multicolumn{1}{c}{0.04} &       & \multicolumn{1}{c}{0.13} & \multicolumn{1}{c}{0.02} &       & \multicolumn{1}{c}{0.12} & \multicolumn{1}{c}{0.02} &  \\          &       & \multicolumn{1}{c}{9} & \multicolumn{1}{c}{0.15} & \multicolumn{1}{c}{0.03} &       & \multicolumn{1}{c}{0.14} & \multicolumn{1}{c}{0.02} &       & \multicolumn{1}{c}{0.13} & \multicolumn{1}{c}{0.03} &  \\          &       & \multicolumn{1}{c}{10} &       &       &       & \multicolumn{1}{c}{0.16} & \multicolumn{1}{c}{0.02} &       & \multicolumn{1}{c}{0.14} & \multicolumn{1}{c}{0.04} &  \\          &       & \multicolumn{1}{c}{11} &       &       &       &       &       &       & \multicolumn{1}{c}{0.16} & \multicolumn{1}{c}{0.02} &  \\\cmidrule{2-11}          &       &       &       &       &       &       &       &       &       &       &  \\    \end{tabular}
					\end{threeparttable} 
			\end{table}
\newpage
\section{Discussion}\label{sec5}

In this article, we have first extended the definition of several balance metrics from the point-exposure setting to the case of a time-varying treatment in the presence of censoring, including a new variant of the Mahalanobis balance. We have compared the performance of these balance metrics at signaling potential bias in a simulation study with two time-points.  We found that, apart from the overlap coefficient and the post-weighted C-statistic, all other imbalance metrics were strongly associated with bias, but the Mahalanobis balance offered the best performance. In the light of these results, we recommend using the Mahalanobis balance first as a global measure of imbalance, and then using individual-level metrics such the standardised mean difference to determine which variables are imbalanced if required. To support the practical implementation of this recommendation, we provide R code on GitHub (\url{https://github.com/detal9/LongitudinalBalanceMetrics}). We have also illustrated the implementation of our recommended approach in estimating the effect of cumulative statin exposure on cardiovascular disease risk in the older population using an inverse probability weighted marginal structural model. In this application, weighting substantially improved the overall balance of covariates as measured with Mahalanobis balance.  After adjustment for confounding and censoring, each additional month of statin exposure was associated with a $3\%$ reduction of the hazard of a first cardiovascular event or death.

To the best of our knowledge, only \textit{Jackson et al} \cite{d} and Greifer N \cite{NoahGreifer} had previously studied balance metrics with a time-varying treatment. Unlike them, we propose checking the balance between treatment groups using marginal stabilized weights instead of the usual stabilized weights. This choice significantly reduces the burden of checking balance between groups. As mentioned in Section \ref{sec3}, stabilized weights only balance treatment groups conditional on past treatment. As such, one must check balance between treatment groups conditional on each possible treatment history when considering stabilized weights. Marginal stabilized weights do not share this limitation and allows checking balance between groups unconditionally of previous treatment history. Our simulation study supports the validity of our proposed approach. Our application to population-wide administrative data with 12 time-points further supports the applicability of our methods in practice in large databases with multiple time-points. In this application with 12 time-points, we have to check 77 balances using marginal stabilized weight, compared with 22,528 if we had used stabilized weights. This choice has thus considerably reduced the computation burden and improved feasibility.

Despite the strengths of the proposed method, certain limitations must be taken into account. A first important limitation is that it is not known what limit of Mahalanobis balance would be acceptable from the point of view of bias. Our proposed threshold is based on connections with the standardized mean difference, for which the a commonly accepted threshold is 0.1 \cite{l,m}, under the assumptions that covariates are uncorrelated. Extensive simulations would be required to assess the validity of this threshold under various scenarios, or to propose an alternative threshold. However, as illustrated in our simulations, a second limitation is that there are scenarios where bias is poorly associated with imbalance. For example, if the imbalance concerns a variable with little association with the response, then even a high level of imbalance can generate a low or moderate bias (scenario 4). However, in the case of a variable highly associated with the response, even a small imbalance can generate a considerable bias (scenario 5). This second limitation highlights the challenge of establishing an acceptable threshold of imbalance for any balance metric. A potential solution would be to use metrics that take into account the covariate-outcome relationship. Such metrics have been proposed for the cross-sectional case \cite{a2}, but have not yet been extended to the longitudinal case. However, this approach would contradict the principle accepted by many of separating the study design stage (i.e., balancing the treatment groups) from the response analysis step \cite{aj}.

Our current work could be extended in several directions. For example, apart from the general weighted difference, the balance metrics described in this article, in particular the Mahalanobis balance, do not take into account the overall imbalance on all covariate squares and pairwise interactions. As suggested by\cite{l}, higher-order moments and interactions between variables should be similar between treatment groups in the weighted sample. Future work would be to extend the Mahalanobis balance to account for global imbalance on all covariate squares and all pairwise interactions using a matrix version of the general weighted difference metric. In conclusion, we believe that better checking of covariates balance following our recommendations will improve the current practice of using IPTW to estimate time-varying  treatment effects.

\bmsection*{FUNDING INFORMATION}
This work was supported by a grant from the Canadian Institutes of Health Research (\# 420060). JRG, CS and DT are supported by research career awards from the \textit{Fonds de recherche du Qu\'ebec -- Sant\'e}.

\bmsection*{DATA AVAILABILITY STATEMENT}
All R codes used in the simulation study are available on \url{https://github.com/detal9/LongitudinalBalanceMetrics}. The data from the \textit{Institut de la statistique du Qu\'ebec} are not publicly available. Access to these data is managed by the Data Access Service of \textit{Institut de la statistique du Qu\'ebec} under Qu\'ebec provincial laws.

\bmsection*{ORCID}

\bibliography{wileyNJD-AMA2}

\begin{center}
	{\huge  Supplementary material for ``Evaluation and comparison of covariate balance metrics in studies with time-dependent confounding'' by David Adenyo, Jason~R~Guertin, Bernard Candas, Caroline Sirois and Denis Talbot.}
\end{center}

\section*{Appendix 1 -- detailed parameters for the simulation studies.\label{app1}}
\vspace*{12pt}

In this section, we present the parameters used to simulate the data in the different scenarios. 

\begin{center}
	\begin{tabular}{ | l | p{8.7cm} |}
		\hline
		\textbf{Simulation scenario}
		& \textbf{Parameter values}  \\ \hline
		Scenario 1 (Base case) & Covariates and exposure at time-point 0: \par $\delta_0=-0.05, \alpha_0=-1.3,
		\phi_{L_0}=0.05, 	\phi_{M_0}=0.05, 	\phi_{N_0}=0.1, 	\phi_{O_0}=0.75,	\phi_{P_0}=0.5$ , $\phi_{Q_0}=0.4$, $\phi_{T_0}=\phi_{R_0}=	\phi_{V_0}=	\phi_{Z_0}=0$ 		
		\par
		Covariates and exposure at time-point 1: \par
		$\beta=0, \gamma_0=-1, \gamma_1=-0.5, \gamma_2=-0.25,  \delta_1=1.23, \gamma_3=-0.5$, \par $\mu_0=-1.1, \gamma_4=-0.75$
		$\alpha_1=-1.7, \phi_{L_1}=0.05, 	\phi_{M_1}=0.05, 	\phi_{N_1}=0.1, 	\phi_{O_1}=0.75,	\phi_{P_1}=0.5$ , $\phi_{Q_1}=0.4$, $\phi_{T_1}=\phi_{R_1}=	\phi_{V_1}=	\phi_{Z_1}=0,  \theta=0.69$ 	
		\par
		Outcome: \par
		$\alpha_Y=-4.3, \beta_{L_0}=\beta_{L_1}=0.5, 	\beta_{M_0}=\beta_{M_1}=0.5,	\beta_{N_0}=\beta_{N_1}=0.05, \beta_{O_0}=\beta_{O_1}=1$, 
		$
		\beta_{P_0}=\beta_{P_1}=1,
		\beta_{Q_0}=\beta_{Q_1}=0.2,
		\beta_{T_0}=\beta_{T_1}=0,
		\beta_{R_0}=\beta_{R_1}=0$, 
		$
		\beta_{V_0}=\beta_{V_1}=0,
		\beta_{Z_0}=\beta_{Z_1}=0,
		\beta_{A_0}=\beta_{A_1}=-0.69$ 	
		\\ \hline
		Scenario 4 (High imbalance, no confounding) &  Covariates and exposure at time-point 0: \par $\delta_0=-0.05, \alpha_0=-3.25,
		\phi_{L_0}=1, 	\phi_{M_0}=1, 	\phi_{N_0}=0.1, 	\phi_{O_0}=2,	\phi_{P_0}=2$ , $\phi_{Q_0}=0.4$, $\phi_{T_0}=\phi_{R_0}=	\phi_{V_0}=	\phi_{Z_0}=0$ 		
		\par
		Covariates and exposure at time-point 1: \par
		$\beta=0, \gamma_0=-1, \gamma_1=-0.5, \gamma_2=-0.25,  \delta_1=1.23, \gamma_3=-0.5$, \par $\mu_0=-1.1, \gamma_4=-0.75$
		$\alpha_1=-2.93, \phi_{L_1}=1, 	\phi_{M_1}=1, 	\phi_{N_1}=0.1, 	\phi_{O_1}=2,	\phi_{P_1}=2$ , $\phi_{Q_1}=0.4$, $\phi_{T_1}=\phi_{R_1}=	\phi_{V_1}=	\phi_{Z_1}=0,  \theta=0.69$ 	
		\par
		Outcome: \par
		$\alpha_Y=-0.75, \beta_{L_0}=\beta_{L_1}=0, 	\beta_{M_0}=\beta_{M_1}=0,	\beta_{N_0}=\beta_{N_1}=0, \beta_{O_0}=\beta_{O_1}=0$, 
		$
		\beta_{P_0}=\beta_{P_1}=0,
		\beta_{Q_0}=\beta_{Q_1}=0,
		\beta_{T_0}=\beta_{T_1}=0,
		\beta_{R_0}=\beta_{R_1}=0$, 
		$
		\beta_{V_0}=\beta_{V_1}=0,
		\beta_{Z_0}=\beta_{Z_1}=0,
		\beta_{A_0}=\beta_{A_1}=-0.69$ 	
		\\ \hline
		Scenario 5 (Low imbalance, moderate confounding) & Covariates and exposure at time-point 0: \par $\delta_0=-0.05, \alpha_0=-0.05,
		\phi_{L_0}=0.01, 	\phi_{M_0}=0.01, 	\phi_{N_0}=0.02, 	\phi_{O_0}=0.02,	\phi_{P_0}=0.01$ , $\phi_{Q_0}=0.01$, $\phi_{T_0}=\phi_{R_0}=	\phi_{V_0}=	\phi_{Z_0}=0$ 		
		\par
		Covariates and exposure at time-point 1: \par
		$\beta=0, \gamma_0=0, \gamma_1=0, \gamma_2=0,  \delta_1=-4.5, \gamma_3=0$, \par $\mu_0=0, \gamma_4=0$
		$\alpha_1=-0.2, \phi_{L_1}=0.01, 	\phi_{M_1}=0.01, 	\phi_{N_1}=0.02, 	\phi_{O_1}=0.02,	\phi_{P_1}=0.01$ , $\phi_{Q_1}=0.01$, $\phi_{T_1}=\phi_{R_1}=	\phi_{V_1}=	\phi_{Z_1}=0,  \theta=0$ 	
		\par
		Outcome: \par
		$\alpha_Y=-20.5, \beta_{L_0}=\beta_{L_1}=1, 	\beta_{M_0}=\beta_{M_1}=1,	\beta_{N_0}=\beta_{N_1}=0.1, \beta_{O_0}=\beta_{O_1}=2$, 
		$
		\beta_{P_0}=\beta_{P_1}=2,
		\beta_{Q_0}=\beta_{Q_1}=0.4,
		\beta_{T_0}=\beta_{T_1}=0,
		\beta_{R_0}=\beta_{R_1}=0$, 
		$
		\beta_{V_0}=\beta_{V_1}=0,
		\beta_{Z_0}=\beta_{Z_1}=0,
		\beta_{A_0}=\beta_{A_1}=0$ 	
		\\ \hline
	\end{tabular}
\end{center}
\newpage
\begin{center}
	\begin{tabular}{ |l| p{8.7cm} |}
		\hline
		\textbf{Simulation scenario}
		& \textbf{Parameter values}  \\ \hline
		Scenario 6 (High imbalance-high confounding) &  Covariates and exposure at time-point 0: \par $\delta_0=-0.05, \alpha_0=-3.25,
		\phi_{L_0}=1, 	\phi_{M_0}=1, 	\phi_{N_0}=0.1, 	\phi_{O_0}=2,	\phi_{P_0}=2$ , $\phi_{Q_0}=0.4$, $\phi_{T_0}=\phi_{R_0}=	\phi_{V_0}=	\phi_{Z_0}=0$ 		
		\par
		Covariates and exposure at time-point 1: \par
		$\beta=0, \gamma_0=-1, \gamma_1=-0.5, \gamma_2=-0.25,  \delta_1=1.23, \gamma_3=-0.5$, \par $\mu_0=-1.1, \gamma_4=-0.75$
		$\alpha_1=-2.95, \phi_{L_1}=1, 	\phi_{M_1}=1, 	\phi_{N_1}=0.1, 	\phi_{O_1}=2,	\phi_{P_1}=2$ , $\phi_{Q_1}=0.4$, $\phi_{T_1}=\phi_{R_1}=	\phi_{V_1}=	\phi_{Z_1}=0,  \theta=0$ 	
		\par
		Outcome: \par
		$\alpha_Y=-4.07, \beta_{L_0}=\beta_{L_1}=0.5, 	\beta_{M_0}=\beta_{M_1}=0.5,	\beta_{N_0}=\beta_{N_1}=0.05, \beta_{O_0}=\beta_{O_1}=1$, 
		$
		\beta_{P_0}=\beta_{P_1}=1,
		\beta_{Q_0}=\beta_{Q_1}=0.2,
		\beta_{T_0}=\beta_{T_1}=0,
		\beta_{R_0}=\beta_{R_1}=0$, 
		$
		\beta_{V_0}=\beta_{V_1}=0,
		\beta_{Z_0}=\beta_{Z_1}=0,
		\beta_{A_0}=\beta_{A_1}=-0.69$ 	
		\\ \hline
		Scenario 7 (Nonlinear outcome) &  Covariates and exposure at time-point 0: \par $\delta_0=-0.05, \alpha_0=-1.3,
		\phi_{L_0}=0.05, 	\phi_{M_0}=0.05, 	\phi_{N_0}=0.1, 	\phi_{O_0}=0.75,	\phi_{P_0}=0.5$ , $\phi_{Q_0}=0.4$, $\phi_{T_0}=\phi_{R_0}=	\phi_{V_0}=	\phi_{Z_0}=0$ 		
		\par
		Covariates and exposure at time-point 1: \par
		$\beta=0, \gamma_0=-1, \gamma_1=-0.5, \gamma_2=-0.25,  \delta_1=1.23, \gamma_3=-0.5$, \par $\mu_0=-1.1, \gamma_4=-0.75$
		$\alpha_1=-1.7, \phi_{L_1}=0.05, 	\phi_{M_1}=0.05, 	\phi_{N_1}=0.1, 	\phi_{O_1}=0.75,	\phi_{P_1}=0.5$ , $\phi_{Q_1}=0.4$, $\phi_{T_1}=\phi_{R_1}=	\phi_{V_1}=	\phi_{Z_1}=0,  \theta=0$ 	
		\par
		Outcome: \par
		$\alpha_Y=-3.1, \beta_{L_0}=\beta_{L_1}=0.4, 	\beta_{M_0}=\beta_{M_1}=0.03,	\beta_{N_0}=\beta_{N_1}=0.03, \beta_{O_0}=\beta_{O_1}=0.75$, 
		$
		\beta_{P_0}=\beta_{P_1}=0.75,
		\beta_{Q_0}=\beta_{Q_1}=0.2,
		\beta_{T_0}=\beta_{T_1}=0.4,
		\beta_{R_0}=\beta_{R_1}=0.02$, 
		$
		\beta_{V_0}=\beta_{V_1}=0.04,
		\beta_{Z_0}=\beta_{Z_1}=0.5,
		\beta_{A_0}=\beta_{A_1}=-0.69$ 	
		\\ \hline
		Scenario 8 (Nonlinear outcome and exposure) &  Covariates and exposure at time-point 0: \par $\delta_0=-0.05, \alpha_0=-1.14,
		\phi_{L_0}=0.05, 	\phi_{M_0}=0.05, 	\phi_{N_0}=0.1, 	\phi_{O_0}=0.5,	\phi_{P_0}=0.25$ , $\phi_{Q_0}=0.4$, $\phi_{T_0}=0.01, \phi_{R_0}=	0.02, 
		\phi_{V_0}=0.01, \phi_{Z_0}=0.1$ 	\par
		Covariates and exposure at time-point 1: \par
		$\beta=0, \gamma_0=-1, \gamma_1=-0.5, \gamma_2=-0.25,  \delta_1=1.23, \gamma_3=-0.5$, \par $\mu_0=-1.08, \gamma_4=-0.75$
		$\alpha_1=-1.5, \phi_{L_1}=0.05, 	\phi_{M_1}=0.05, 	\phi_{N_1}=0.1, 	\phi_{O_1}=0.5,	\phi_{P_1}=0.25$ , $\phi_{Q_1}=0.4$, $\phi_{T_1}=0.01,
		\phi_{R_1}=0.02, \phi_{V_1}=0.01, \phi_{Z_1}=0.1,  \theta=0.69$ 	
		\par
		Outcome: \par
		$\alpha_Y=-3.1, \beta_{L_0}=\beta_{L_1}=0.4, 	\beta_{M_0}=\beta_{M_1}=0.03,	\beta_{N_0}=\beta_{N_1}=0.03, \beta_{O_0}=\beta_{O_1}=0.75$, 
		$
		\beta_{P_0}=\beta_{P_1}=0.75,
		\beta_{Q_0}=\beta_{Q_1}=0.2,
		\beta_{T_0}=\beta_{T_1}=0.4,
		\beta_{R_0}=\beta_{R_1}=0.02$, 
		$
		\beta_{V_0}=\beta_{V_1}=0.04,
		\beta_{Z_0}=\beta_{Z_1}=0.5,
		\beta_{A_0}=\beta_{A_1}=-0.69$ 	
		\\ \hline
		Scenario 9 (Redundant Covariates) & Covariates and exposure at time-point 0: \par  $\delta_0=-0.05, \alpha_0=-0.37,
		\phi_{L_0}=0.2, 	\phi_{M_0}=0.03, 	\phi_{N_0}=0.02, 	\phi_{O_0}=0,	\phi_{P_0}=1.5$ , $\phi_{Q_0}=0.01$, $\phi_{T_0}=0.01, \phi_{R_0}=	0.02, 
		\phi_{V_0}=0, \phi_{Z_0}=0$ 	\par
		Covariates and exposure at time-point 1: \par
		$\beta=0, \gamma_0=-1, \gamma_1=-0.5, \gamma_2=-0.25,  \delta_1=1.23, \gamma_3=-0.5$, \par $\mu_0=-1.08, \gamma_4=-0.75$
		$\alpha_1=-0.59, \phi_{L_1}=0.2, 	\phi_{M_1}=0.03, 	\phi_{N_1}=0.02, 	\phi_{O_1}=0,	\phi_{P_1}=1.5$ , $\phi_{Q_1}=0.01$, $\phi_{T_1}=0.01,
		\phi_{R_1}=0.02, \phi_{V_1}=0, \phi_{Z_1}=0,  \theta=0.69$ 	
		\par
		Outcome: \par
		$\alpha_Y=-2, \beta_{L_0}=\beta_{L_1}=0.4, 	\beta_{M_0}=\beta_{M_1}=0.03,	\beta_{N_0}=\beta_{N_1}=0.03, \beta_{O_0}=\beta_{O_1}=0$, 
		$
		\beta_{P_0}=\beta_{P_1}=0.75,
		\beta_{Q_0}=\beta_{Q_1}=0.2,
		\beta_{T_0}=\beta_{T_1}=0.4,
		\beta_{R_0}=\beta_{R_1}=0.02$, 
		$
		\beta_{V_0}=\beta_{V_1}=0,
		\beta_{Z_0}=\beta_{Z_1}=0,
		\beta_{A_0}=\beta_{A_1}=-0.69$ 	
		\\ \hline
	\end{tabular}
\end{center}
\newpage
\begin{center}
	\begin{tabular}{ |l| p{8.7cm} |}
		\hline
		\textbf{Simulation scenario}
		& \textbf{Parameter values}  \\ \hline
		Scenario 10 (Instrumental variables) &  Covariates and exposure at time-point 0: \par $\delta_0=-0.05, \alpha_0=-0.4,
		\phi_{L_0}=0.2, 	\phi_{M_0}=0.03, 	\phi_{N_0}=0.02, 	\phi_{O_0}=0.5,	\phi_{P_0}=0.25$ , $\phi_{Q_0}=0.01$, $\phi_{T_0}=0.01, \phi_{R_0}=	0.02, 
		\phi_{V_0}=0.01, \phi_{Z_0}=0.1$ 	\par
		Covariates and exposure at time-point 1: \par
		$\beta=0, \gamma_0=-1, \gamma_1=-0.5, \gamma_2=-0.25,  \delta_1=1.21, \gamma_3=-0.5$, \par $\mu_0=-1.08, \gamma_4=-0.75$
		$\alpha_1=-0.6, \phi_{L_1}=0.2, 	\phi_{M_1}=0.03, 	\phi_{N_1}=0.02, 	\phi_{O_1}=0.5,	\phi_{P_1}=0.25$ , $\phi_{Q_1}=0.01$, $\phi_{T_1}=0.01,
		\phi_{R_1}=0.02, \phi_{V_1}=0.01, \phi_{Z_1}=0.1,  \theta=0.69$ 	
		\par
		Outcome: \par
		$\alpha_Y=-2, \beta_{L_0}=\beta_{L_1}=0.4, 	\beta_{M_0}=\beta_{M_1}=0,	\beta_{N_0}=\beta_{N_1}=0.03, \beta_{O_0}=\beta_{O_1}=0.75$, 
		$
		\beta_{P_0}=\beta_{P_1}=0.75,
		\beta_{Q_0}=\beta_{Q_1}=0.2,
		\beta_{T_0}=\beta_{T_1}=0.4,
		\beta_{R_0}=\beta_{R_1}=0$, 
		$
		\beta_{V_0}=\beta_{V_1}=0.04,
		\beta_{Z_0}=\beta_{Z_1}=0.5,
		\beta_{A_0}=\beta_{A_1}=-0.69$ 	
		\\ \hline
		Base case Scenario with censoring & Censoring indicator at time-point 1: \par $\mu_1=-2.7, \mu_{L_0}=0.04,  \mu_{M_0}=0.05, \mu_{N_0}=0.02, 	 \mu_{O_0}=1, \mu_{P_0}=0.02$ , $ \mu_{Q_0}=0.01, \lambda=1$ 		
		\\ \hline
	\end{tabular}
\end{center}

\newpage 
\section*{Appendix 2 -- detailed simulation results for scenarios $2$ to $10$ \label{app1.1a}}

In this section, we present the average covariate imbalances across covariates at each time-point and average bias for Scenarios $2$ to $10$. Averages are taken over $1,\!000$ simulated datasets in the unweighted data and in data weighted according to the five weights.

\begin{table}[h!]  \centering  \caption{Average covariate imbalances in Scenario 2 (Low prevalence of exposure) }    \begin{tabular}{rrrrrrrrrrrrrrrrrr}   \multicolumn{18}{c}{} \\    \midrule          &       & \multicolumn{3}{c}{\textbf{D}} &       & \multicolumn{3}{c}{\textbf{SMD}} &       & \multicolumn{3}{c}{\textbf{OVL}} &       & \multicolumn{3}{c}{\textbf{KS}} & \multicolumn{1}{c}{\textbf{Bias}} \\\cmidrule{3-5}\cmidrule{7-9}\cmidrule{11-13}\cmidrule{15-17}    \multicolumn{2}{l}{Unweighted} & 1.53  & 0.07  & 1.51  &       & 0.34  & 0.02  & 0.32  &       & 0.46  & 0.37  & 0.43  &       & 0.14  & 0.02  & 0.13  & 0.33 \\    \multicolumn{2}{l}{$W_0\times W_1$} & 0.08  & 0.07  & 0.08  &       & 0.03  & 0.03  & 0.03  &       & 0.40  & 0.37  & 0.38  &       & 0.02  & 0.02  & 0.02  & 0.02 \\    \multicolumn{2}{l}{$W_1$} & 1.53  & 0.05  & 0.06  &       & 0.34  & 0.02  & 0.02  &       & 0.46  & 0.37  & 0.38  &       & 0.15  & 0.02  & 0.02  & 0.12 \\    \multicolumn{2}{l}{$W_0$} & 0.06  & 0.06  & 1.52  &       & 0.02  & 0.02  & 0.32  &       & 0.40  & 0.37  & 0.43  &       & 0.02  & 0.02  & 0.14  & 0.10 \\    \multicolumn{2}{l}{$W_0$tr90$\times W_1$} & 0.69  & 0.05  & 0.07  &       & 0.15  & 0.02  & 0.02  &       & 0.42  & 0.37  & 0.38  &       & 0.07  & 0.02  & 0.02  & 0.04 \\    \multicolumn{2}{l}{$W_0\times W_1$tr90} & 0.07  & 0.05  & 0.68  &       & 0.02  & 0.02  & 0.14  &       & 0.40  & 0.37  & 0.40  &       & 0.02  & 0.02  & 0.07  & 0.03 \\          &       & \multicolumn{3}{c}{\textbf{LD}} &       & \multicolumn{3}{c}{\textbf{MHB}} &       & \multicolumn{3}{c}{\textbf{CS}} &       & \multicolumn{3}{c}{\textbf{GWD}} & \multicolumn{1}{c}{\textbf{Bias}} \\\cmidrule{3-5}\cmidrule{7-9}\cmidrule{11-13}\cmidrule{15-17}    \multicolumn{2}{l}{Unweighted} & 0.13  & 0.01  & 0.12  &       & 1.18  & 0.01  & 1.11  &       & 0.56  & 0.55  & 0.55  &       & 0.19  & 0.01  & 0.18  & 0.33 \\    \multicolumn{2}{l}{$W_0\times W_1$} & 0.02  & 0.02  & 0.02  &       & 0.01  & 0.01  & 0.01  &       & 0.04  & 0.03  & 0.03  &       & 0.02  & 0.02  & 0.02  & 0.02 \\    \multicolumn{2}{l}{$W_1$} & 0.13  & 0.01  & 0.01  &       & 1.18  & 0.00  & 0.00  &       & 0.56  & 0.03  & 0.03  &       & 0.19  & 0.01  & 0.01  & 0.12 \\    \multicolumn{2}{l}{$W_0$} & 0.01  & 0.01  & 0.12  &       & 0.00  & 0.01  & 1.11  &       & 0.04  & 0.54  & 0.54  &       & 0.01  & 0.02  & 0.18  & 0.10 \\    \multicolumn{2}{l}{$W_0$tr90$\times W_1$} & 0.06  & 0.02  & 0.02  &       & 0.22  & 0.00  & 0.01  &       & 0.56  & 0.03  & 0.03  &       & 0.08  & 0.02  & 0.02  & 0.04 \\    \multicolumn{2}{l}{$W_0\times W_1$tr90} & 0.01  & 0.01  & 0.06  &       & 0.00  & 0.00  & 0.21  &       & 0.04  & 0.54  & 0.54  &       & 0.01  & 0.01  & 0.08  & 0.03 \\    \midrule          &       &       &       &       &       &       &       &       &       &       &       &       &       &       &       &       &  \\    \end{tabular}  \label{tab:addlabel} \end{table}

\begin{table}[h!]  \centering  \caption{Average covariate imbalances in Scenario 3 (Small sample)}    \begin{tabular}{rrrrrrrrrrrrrrrrrr} \multicolumn{18}{c}{} \\    \midrule          &       & \multicolumn{3}{c}{\textbf{D}} &       & \multicolumn{3}{c}{\textbf{SMD}} &       & \multicolumn{3}{c}{\textbf{OVL}} &       & \multicolumn{3}{c}{\textbf{KS}} & \multicolumn{1}{c}{\textbf{Bias}} \\\cmidrule{3-5}\cmidrule{7-9}\cmidrule{11-13}\cmidrule{15-17}    \multicolumn{2}{l}{Unweighted} & 1.45  & 0.13  & 1.42  &       & 0.33  & 0.05  & 0.29  &       & 0.42  & 0.35  & 0.40  &       & 0.15  & 0.04  & 0.13  & 0.29 \\    \multicolumn{2}{l}{$W_0\times W_1$} & 0.14  & 0.15  & 0.14  &       & 0.06  & 0.06  & 0.05  &       & 0.37  & 0.36  & 0.36  &       & 0.05  & 0.05  & 0.05  & 0.04 \\    \multicolumn{2}{l}{$W_1$} & 1.45  & 0.06  & 0.09  &       & 0.33  & 0.02  & 0.03  &       & 0.42  & 0.35  & 0.36  &       & 0.15  & 0.04  & 0.04  & 0.11 \\    \multicolumn{2}{l}{$W_0$} & 0.08  & 0.17  & 1.42  &       & 0.03  & 0.07  & 0.30  &       & 0.36  & 0.35  & 0.41  &       & 0.04  & 0.05  & 0.14  & 0.10 \\    \multicolumn{2}{l}{$W_0$tr90$\times W_1$} & 0.52  & 0.07  & 0.11  &       & 0.12  & 0.03  & 0.04  &       & 0.38  & 0.35  & 0.36  &       & 0.07  & 0.04  & 0.04  & 0.04 \\    \multicolumn{2}{l}{$W_0\times W_1$tr90} & 0.10  & 0.14  & 0.52  &       & 0.04  & 0.05  & 0.11  &       & 0.37  & 0.35  & 0.37  &       & 0.04  & 0.04  & 0.07  & 0.04 \\          &       & \multicolumn{3}{c}{\textbf{LD}} &       & \multicolumn{3}{c}{\textbf{MHB}} &       & \multicolumn{3}{c}{\textbf{CS}} &       & \multicolumn{3}{c}{\textbf{GWD}} & \multicolumn{1}{c}{\textbf{Bias}} \\\cmidrule{3-5}\cmidrule{7-9}\cmidrule{11-13}\cmidrule{15-17}    \multicolumn{2}{l}{Unweighted} & 0.13  & 0.03  & 0.12  &       & 1.13  & 0.03  & 1.03  &       & 0.55  & 0.54  & 0.54  &       & 0.17  & 0.03  & 0.15  & 0.29 \\    \multicolumn{2}{l}{$W_0\times W_1$} & 0.04  & 0.04  & 0.04  &       & 0.03  & 0.03  & 0.03  &       & 0.08  & 0.08  & 0.08  &       & 0.04  & 0.04  & 0.04  & 0.04 \\    \multicolumn{2}{l}{$W_1$} & 0.14  & 0.03  & 0.03  &       & 1.15  & 0.01  & 0.01  &       & 0.54  & 0.09  & 0.09  &       & 0.17  & 0.02  & 0.03  & 0.11 \\    \multicolumn{2}{l}{$W_0$} & 0.03  & 0.04  & 0.13  &       & 0.01  & 0.04  & 1.05  &       & 0.09  & 0.53  & 0.53  &       & 0.02  & 0.04  & 0.15  & 0.10 \\    \multicolumn{2}{l}{$W_0$tr90$\times W_1$} & 0.06  & 0.03  & 0.03  &       & 0.16  & 0.01  & 0.01  &       & 0.52  & 0.08  & 0.08  &       & 0.07  & 0.03  & 0.03  & 0.04 \\    \multicolumn{2}{l}{$W_0\times W_1$tr90} & 0.03  & 0.03  & 0.06  &       & 0.01  & 0.02  & 0.15  &       & 0.08  & 0.49  & 0.49  &       & 0.03  & 0.03  & 0.06  & 0.04 \\    \midrule          &       &       &       &       &       &       &       &       &       &       &       &       &       &       &       &       &  \\    \end{tabular}  \label{tab:addlabel} \end{table}

\begin{table}[h!]  \centering  \caption{Average covariate imbalances in Scenario 4 (High imbalance, no confounding)}    \begin{tabular}{llrrrrrrrrrrrrrrrr} \multicolumn{18}{c}{} \\    \midrule          &       & \multicolumn{3}{c}{\textbf{D}} &       & \multicolumn{3}{c}{\textbf{SMD}} &       & \multicolumn{3}{c}{\textbf{OVL}} &       & \multicolumn{3}{c}{\textbf{KS}} & \multicolumn{1}{c}{\textbf{Bias}} \\\cmidrule{3-5}\cmidrule{7-9}\cmidrule{11-13}\cmidrule{15-17}    \multicolumn{2}{l}{Unweighted} & 1.22  & 0.31  & 1.25  &       & 0.62  & 0.14  & 0.68  &       & 0.47  & 0.36  & 0.48  &       & 0.25  & 0.07  & 0.28  & 0.02 \\    \multicolumn{2}{l}{$W_0\times W_1$} & 0.17  & 0.16  & 0.18  &       & 0.08  & 0.07  & 0.08  &       & 0.38  & 0.37  & 0.38  &       & 0.05  & 0.05  & 0.05  & 0.08 \\    \multicolumn{2}{l}{$W_1$} & 1.22  & 0.07  & 0.10  &       & 0.62  & 0.03  & 0.04  &       & 0.47  & 0.35  & 0.36  &       & 0.26  & 0.03  & 0.03  & 0.05 \\    \multicolumn{2}{l}{$W_0$} & 0.09  & 0.10  & 1.25  &       & 0.04  & 0.04  & 0.68  &       & 0.36  & 0.35  & 0.48  &       & 0.03  & 0.03  & 0.28  & 0.04 \\    \multicolumn{2}{l}{$W_0$tr90$\times W_1$} & 0.74  & 0.08  & 0.11  &       & 0.36  & 0.03  & 0.05  &       & 0.42  & 0.35  & 0.36  &       & 0.15  & 0.03  & 0.03  & 0.05 \\    \multicolumn{2}{l}{$W_0\times W_1$tr90} & 0.10  & 0.11  & 0.77  &       & 0.04  & 0.05  & 0.41  &       & 0.36  & 0.35  & 0.42  &       & 0.03  & 0.03  & 0.17  & 0.04 \\          &       & \multicolumn{3}{c}{\textbf{LD}} &       & \multicolumn{3}{c}{\textbf{MHB}} &       & \multicolumn{3}{c}{\textbf{CS}} &       & \multicolumn{3}{c}{\textbf{GWD}} & \multicolumn{1}{c}{\textbf{Bias}} \\\cmidrule{3-5}\cmidrule{7-9}\cmidrule{11-13}\cmidrule{15-17}    \multicolumn{2}{l}{Unweighted} & 0.22  & 0.06  & 0.24  &       & 3.16  & 0.12  & 3.31  &       & 0.79  & 0.80  & 0.80  &       & 0.31  & 0.08  & 0.33  & 0.02 \\    \multicolumn{2}{l}{$W_0\times W_1$} & 0.04  & 0.04  & 0.04  &       & 0.05  & 0.04  & 0.05  &       & 0.21  & 0.19  & 0.19  &       & 0.04  & 0.04  & 0.05  & 0.08 \\    \multicolumn{2}{l}{$W_1$} & 0.22  & 0.02  & 0.02  &       & 3.19  & 0.01  & 0.02  &       & 0.79  & 0.15  & 0.15  &       & 0.31  & 0.02  & 0.03  & 0.05 \\    \multicolumn{2}{l}{$W_0$} & 0.02  & 0.02  & 0.24  &       & 0.01  & 0.02  & 3.33  &       & 0.15  & 0.79  & 0.79  &       & 0.03  & 0.03  & 0.33  & 0.04 \\    \multicolumn{2}{l}{$W_0$tr90$\times W_1$} & 0.13  & 0.02  & 0.02  &       & 0.98  & 0.01  & 0.02  &       & 0.78  & 0.15  & 0.15  &       & 0.19  & 0.02  & 0.03  & 0.05 \\    \multicolumn{2}{l}{$W_0\times W_1$tr90} & 0.02  & 0.02  & 0.15  &       & 0.02  & 0.02  & 1.05  &       & 0.15  & 0.78  & 0.78  &       & 0.03  & 0.03  & 0.21  & 0.04 \\    \bottomrule    \end{tabular}  \label{tab:addlabel} \end{table}

\begin{table}[h!]  \centering  \caption{Average covariate imbalances in Scenario 5 (Low imbalance, moderate confounding)}    \begin{tabular}{llrrrrrrrrrrrrrrrr} \multicolumn{18}{c}{} \\    \midrule          &       & \multicolumn{3}{c}{\textbf{D}} &       & \multicolumn{3}{c}{\textbf{SMD}} &       & \multicolumn{3}{c}{\textbf{OVL}} &       & \multicolumn{3}{c}{\textbf{KS}} & \multicolumn{1}{c}{\textbf{Bias}} \\\cmidrule{3-5}\cmidrule{7-9}\cmidrule{11-13}\cmidrule{15-17}    \multicolumn{2}{l}{Unweighted} & 0.34  & 0.04  & 0.37  &       & 0.05  & 0.02  & 0.05  &       & 0.34  & 0.33  & 0.34  &       & 0.02  & 0.01  & 0.03  & 0.23 \\    \multicolumn{2}{l}{$W_0\times W_1$} & 0.00  & 0.00  & 0.01  &       & 0.00  & 0.00  & 0.00  &       & 0.33  & 0.33  & 0.33  &       & 0.01  & 0.01  & 0.01  & 0.05 \\    \multicolumn{2}{l}{$W_1$} & 0.34  & 0.00  & 0.00  &       & 0.05  & 0.00  & 0.00  &       & 0.34  & 0.33  & 0.33  &       & 0.02  & 0.01  & 0.01  & 0.08 \\    \multicolumn{2}{l}{$W_0$} & 0.00  & 0.04  & 0.37  &       & 0.00  & 0.02  & 0.05  &       & 0.33  & 0.33  & 0.34  &       & 0.01  & 0.01  & 0.03  & 0.15 \\    \multicolumn{2}{l}{$W_0$tr90$\times W_1$} & 0.05  & 0.00  & 0.00  &       & 0.01  & 0.00  & 0.00  &       & 0.33  & 0.33  & 0.33  &       & 0.01  & 0.01  & 0.01  & 0.05 \\    \multicolumn{2}{l}{$W_0\times W_1$tr90} & 0.00  & 0.01  & 0.05  &       & 0.00  & 0.00  & 0.01  &       & 0.33  & 0.33  & 0.33  &       & 0.01  & 0.01  & 0.01  & 0.05 \\          &       & \multicolumn{3}{c}{\textbf{LD}} &       & \multicolumn{3}{c}{\textbf{MHB}} &       & \multicolumn{3}{c}{\textbf{CS}} &       & \multicolumn{3}{c}{\textbf{GWD}} & \multicolumn{1}{c}{\textbf{Bias}} \\\cmidrule{3-5}\cmidrule{7-9}\cmidrule{11-13}\cmidrule{15-17}    \multicolumn{2}{l}{Unweighted} & 0.02  & 0.01  & 0.02  &       & 0.04  & 0.00  & 0.04  &       & 0.11  & 0.12  & 0.12  &       & 0.03  & 0.01  & 0.03  & 0.23 \\    \multicolumn{2}{l}{$W_0\times W_1$} & 0.00  & 0.00  & 0.00  &       & 0.00  & 0.00  & 0.00  &       & 0.00  & 0.00  & 0.00  &       & 0.00  & 0.00  & 0.00  & 0.05 \\    \multicolumn{2}{l}{$W_1$} & 0.02  & 0.00  & 0.00  &       & 0.04  & 0.00  & 0.00  &       & 0.11  & 0.00  & 0.00  &       & 0.03  & 0.00  & 0.00  & 0.08 \\    \multicolumn{2}{l}{$W_0$} & 0.00  & 0.01  & 0.02  &       & 0.00  & 0.00  & 0.04  &       & 0.00  & 0.12  & 0.12  &       & 0.00  & 0.01  & 0.03  & 0.15 \\    \multicolumn{2}{l}{$W_0$tr90$\times W_1$} & 0.00  & 0.00  & 0.00  &       & 0.00  & 0.00  & 0.00  &       & 0.11  & 0.00  & 0.00  &       & 0.01  & 0.00  & 0.00  & 0.05 \\    \multicolumn{2}{l}{$W_0\times W_1$tr90} & 0.00  & 0.00  & 0.00  &       & 0.00  & 0.00  & 0.00  &       & 0.00  & 0.12  & 0.12  &       & 0.00  & 0.00  & 0.01  & 0.05 \\    \bottomrule    \end{tabular}  \label{tab:addlabel} \end{table}

\begin{table}[h!]  \centering  \caption{Average covariate imbalances in Scenario 6 (High imbalance-high confounding)}    \begin{tabular}{rrrrrrrrrrrrrrrrrr} \multicolumn{18}{c}{} \\    \midrule          &       & \multicolumn{3}{c}{\textbf{D}} &       & \multicolumn{3}{c}{\textbf{SMD}} &       & \multicolumn{3}{c}{\textbf{OVL}} &       & \multicolumn{3}{c}{\textbf{KS}} & \multicolumn{1}{c}{\textbf{Bias}} \\\cmidrule{3-5}\cmidrule{7-9}\cmidrule{11-13}\cmidrule{15-17}    \multicolumn{2}{l}{Unweighted} & 1.22  & 0.31  & 1.25  &       & 0.62  & 0.14  & 0.68  &       & 0.47  & 0.36  & 0.48  &       & 0.25  & 0.07  & 0.28  & 1.81 \\    \multicolumn{2}{l}{$W_0\times W_1$} & 0.17  & 0.17  & 0.18  &       & 0.08  & 0.07  & 0.08  &       & 0.38  & 0.37  & 0.38  &       & 0.05  & 0.05  & 0.05  & 0.04 \\    \multicolumn{2}{l}{$W_1$} & 1.22  & 0.07  & 0.10  &       & 0.62  & 0.03  & 0.04  &       & 0.47  & 0.35  & 0.36  &       & 0.26  & 0.03  & 0.03  & 0.30 \\    \multicolumn{2}{l}{$W_0$} & 0.09  & 0.10  & 1.25  &       & 0.04  & 0.04  & 0.68  &       & 0.36  & 0.35  & 0.48  &       & 0.03  & 0.03  & 0.28  & 0.44 \\    \multicolumn{2}{l}{$W_0$tr90$\times W_1$} & 0.74  & 0.08  & 0.11  &       & 0.36  & 0.03  & 0.05  &       & 0.42  & 0.35  & 0.36  &       & 0.15  & 0.03  & 0.03  & 0.12 \\    \multicolumn{2}{l}{$W_0\times W_1$tr90} & 0.10  & 0.11  & 0.77  &       & 0.04  & 0.05  & 0.41  &       & 0.36  & 0.35  & 0.42  &       & 0.03  & 0.03  & 0.17  & 0.20 \\          &       & \multicolumn{3}{c}{\textbf{LD}} &       & \multicolumn{3}{c}{\textbf{MHB}} &       & \multicolumn{3}{c}{\textbf{CS}} &       & \multicolumn{3}{c}{\textbf{GWD}} & \multicolumn{1}{c}{\textbf{Bias}} \\\cmidrule{3-5}\cmidrule{7-9}\cmidrule{11-13}\cmidrule{15-17}    \multicolumn{2}{l}{Unweighted} & 0.22  & 0.06  & 0.24  &       & 3.16  & 0.12  & 3.31  &       & 0.79  & 0.80  & 0.80  &       & 0.31  & 0.08  & 0.33  & 1.81 \\    \multicolumn{2}{l}{$W_0\times W_1$} & 0.04  & 0.04  & 0.04  &       & 0.05  & 0.05  & 0.05  &       & 0.21  & 0.19  & 0.19  &       & 0.04  & 0.04  & 0.05  & 0.04 \\    \multicolumn{2}{l}{$W_1$} & 0.22  & 0.02  & 0.02  &       & 3.19  & 0.01  & 0.02  &       & 0.79  & 0.15  & 0.15  &       & 0.31  & 0.02  & 0.03  & 0.30 \\    \multicolumn{2}{l}{$W_0$} & 0.02  & 0.02  & 0.24  &       & 0.01  & 0.02  & 3.33  &       & 0.15  & 0.79  & 0.79  &       & 0.03  & 0.03  & 0.33  & 0.44 \\    \multicolumn{2}{l}{$W_0$tr90$\times W_1$} & 0.13  & 0.02  & 0.02  &       & 0.98  & 0.01  & 0.02  &       & 0.78  & 0.15  & 0.15  &       & 0.19  & 0.02  & 0.03  & 0.12 \\    \multicolumn{2}{l}{$W_0\times W_1$tr90} & 0.02  & 0.02  & 0.15  &       & 0.02  & 0.02  & 1.05  &       & 0.15  & 0.78  & 0.78  &       & 0.03  & 0.03  & 0.21  & 0.20 \\    \midrule          &       &       &       &       &       &       &       &       &       &       &       &       &       &       &       &       &  \\    \end{tabular}  \label{tab:addlabel} \end{table}

\begin{table}[h!]  \centering  \caption{Average covariate imbalances in Scenario 7 (Nonlinear outcome)}    \begin{tabular}{rrrrrrrrrrrrrrrrrr} \multicolumn{18}{c}{} \\    \midrule          &       & \multicolumn{3}{c}{\textbf{D}} &       & \multicolumn{3}{c}{\textbf{SMD}} &       & \multicolumn{3}{c}{\textbf{OVL}} &       & \multicolumn{3}{c}{\textbf{KS}} & \multicolumn{1}{c}{\textbf{Bias}} \\\cmidrule{3-5}\cmidrule{7-9}\cmidrule{11-13}\cmidrule{15-17}    \multicolumn{2}{l}{Unweighted} & 1.28  & 0.07  & 1.26  &       & 0.30  & 0.02  & 0.26  &       & 0.39  & 0.32  & 0.37  &       & 0.12  & 0.02  & 0.11  & 0.37 \\    \multicolumn{2}{l}{$W_0\times W_1$} & 0.05  & 0.05  & 0.05  &       & 0.02  & 0.02  & 0.02  &       & 0.34  & 0.32  & 0.33  &       & 0.02  & 0.02  & 0.02  & 0.02 \\    \multicolumn{2}{l}{$W_1$} & 1.28  & 0.02  & 0.03  &       & 0.30  & 0.01  & 0.01  &       & 0.39  & 0.32  & 0.33  &       & 0.13  & 0.01  & 0.01  & 0.13 \\    \multicolumn{2}{l}{$W_0$} & 0.03  & 0.06  & 1.26  &       & 0.01  & 0.02  & 0.26  &       & 0.34  & 0.32  & 0.38  &       & 0.01  & 0.02  & 0.11  & 0.12 \\    \multicolumn{2}{l}{$W_0$tr90$\times W_1$} & 0.45  & 0.03  & 0.04  &       & 0.10  & 0.01  & 0.01  &       & 0.35  & 0.32  & 0.33  &       & 0.05  & 0.01  & 0.01  & 0.03 \\    \multicolumn{2}{l}{$W_0\times W_1$tr90} & 0.04  & 0.05  & 0.44  &       & 0.01  & 0.02  & 0.09  &       & 0.34  & 0.32  & 0.34  &       & 0.01  & 0.01  & 0.04  & 0.02 \\          &       & \multicolumn{3}{c}{\textbf{LD}} &       & \multicolumn{3}{c}{\textbf{MHB}} &       & \multicolumn{3}{c}{\textbf{CS}} &       & \multicolumn{3}{c}{\textbf{GWD}} & \multicolumn{1}{c}{\textbf{Bias}} \\\cmidrule{3-5}\cmidrule{7-9}\cmidrule{11-13}\cmidrule{15-17}    \multicolumn{2}{l}{Unweighted} & 0.11  & 0.01  & 0.10  &       & 1.09  & 0.01  & 1.02  &       & 0.54  & 0.53  & 0.53  &       & 0.14  & 0.01  & 0.12  & 0.37 \\    \multicolumn{2}{l}{$W_0\times W_1$} & 0.01  & 0.01  & 0.01  &       & 0.01  & 0.01  & 0.01  &       & 0.04  & 0.04  & 0.04  &       & 0.01  & 0.01  & 0.01  & 0.02 \\    \multicolumn{2}{l}{$W_1$} & 0.11  & 0.01  & 0.01  &       & 1.10  & 0.00  & 0.00  &       & 0.54  & 0.04  & 0.04  &       & 0.14  & 0.01  & 0.01  & 0.13 \\    \multicolumn{2}{l}{$W_0$} & 0.01  & 0.01  & 0.10  &       & 0.00  & 0.01  & 1.02  &       & 0.05  & 0.52  & 0.52  &       & 0.01  & 0.01  & 0.12  & 0.12 \\    \multicolumn{2}{l}{$W_0$tr90$\times W_1$} & 0.04  & 0.01  & 0.01  &       & 0.14  & 0.00  & 0.00  &       & 0.54  & 0.04  & 0.04  &       & 0.05  & 0.01  & 0.01  & 0.03 \\    \multicolumn{2}{l}{$W_0\times W_1$tr90} & 0.01  & 0.01  & 0.03  &       & 0.00  & 0.01  & 0.13  &       & 0.04  & 0.52  & 0.52  &       & 0.01  & 0.01  & 0.04  & 0.02 \\    \midrule          &       &       &       &       &       &       &       &       &       &       &       &       &       &       &       &       &  \\    \end{tabular}  \label{tab:addlabel} \end{table}

\begin{table}[h!]  \centering  \caption{Average covariate imbalances in Scenario 8 (Nonlinear outcome and exposure)}    \begin{tabular}{rrrrrrrrrrrrrrrrrr} \multicolumn{18}{c}{} \\    \midrule          &       & \multicolumn{3}{c}{\textbf{D}} &       & \multicolumn{3}{c}{\textbf{SMD}} &       & \multicolumn{3}{c}{\textbf{OVL}} &       & \multicolumn{3}{c}{\textbf{KS}} & \multicolumn{1}{c}{\textbf{Bias}} \\\cmidrule{3-5}\cmidrule{7-9}\cmidrule{11-13}\cmidrule{15-17}    \multicolumn{2}{l}{Unweighted} & 1.34  & 0.09  & 1.30  &       & 0.28  & 0.02  & 0.24  &       & 0.38  & 0.32  & 0.36  &       & 0.11  & 0.02  & 0.10  & 0.32 \\    \multicolumn{2}{l}{$W_0\times W_1$} & 0.06  & 0.05  & 0.06  &       & 0.02  & 0.02  & 0.02  &       & 0.33  & 0.32  & 0.32  &       & 0.02  & 0.02  & 0.02  & 0.01 \\    \multicolumn{2}{l}{$W_1$} & 1.34  & 0.02  & 0.05  &       & 0.28  & 0.01  & 0.01  &       & 0.38  & 0.32  & 0.32  &       & 0.11  & 0.01  & 0.01  & 0.12 \\    \multicolumn{2}{l}{$W_0$} & 0.05  & 0.06  & 1.30  &       & 0.01  & 0.02  & 0.24  &       & 0.33  & 0.32  & 0.36  &       & 0.01  & 0.02  & 0.10  & 0.11 \\    \multicolumn{2}{l}{$W_0$tr90$\times W_1$} & 0.48  & 0.03  & 0.05  &       & 0.10  & 0.01  & 0.02  &       & 0.34  & 0.32  & 0.32  &       & 0.04  & 0.01  & 0.01  & 0.03 \\    \multicolumn{2}{l}{$W_0\times W_1$tr90} & 0.05  & 0.05  & 0.47  &       & 0.02  & 0.02  & 0.09  &       & 0.33  & 0.32  & 0.33  &       & 0.01  & 0.01  & 0.04  & 0.02 \\          &       & \multicolumn{3}{c}{\textbf{LD}} &       & \multicolumn{3}{c}{\textbf{MHB}} &       & \multicolumn{3}{c}{\textbf{CS}} &       & \multicolumn{3}{c}{\textbf{GWD}} & \multicolumn{1}{c}{\textbf{Bias}} \\\cmidrule{3-5}\cmidrule{7-9}\cmidrule{11-13}\cmidrule{15-17}    \multicolumn{2}{l}{Unweighted} & 0.10  & 0.01  & 0.09  &       & 1.09  & 0.01  & 1.03  &       & 0.54  & 0.53  & 0.53  &       & 0.13  & 0.01  & 0.11  & 0.32 \\    \multicolumn{2}{l}{$W_0\times W_1$} & 0.01  & 0.01  & 0.01  &       & 0.01  & 0.01  & 0.01  &       & 0.04  & 0.04  & 0.04  &       & 0.01  & 0.01  & 0.01  & 0.01 \\    \multicolumn{2}{l}{$W_1$} & 0.10  & 0.01  & 0.01  &       & 1.09  & 0.00  & 0.01  &       & 0.54  & 0.04  & 0.04  &       & 0.13  & 0.01  & 0.01  & 0.12 \\    \multicolumn{2}{l}{$W_0$} & 0.01  & 0.01  & 0.09  &       & 0.01  & 0.01  & 1.03  &       & 0.03  & 0.53  & 0.53  &       & 0.01  & 0.01  & 0.11  & 0.11 \\    \multicolumn{2}{l}{$W_0$tr90$\times W_1$} & 0.04  & 0.01  & 0.01  &       & 0.14  & 0.00  & 0.01  &       & 0.53  & 0.03  & 0.03  &       & 0.05  & 0.01  & 0.01  & 0.03 \\    \multicolumn{2}{l}{$W_0\times W_1$tr90} & 0.01  & 0.01  & 0.03  &       & 0.01  & 0.01  & 0.13  &       & 0.03  & 0.52  & 0.52  &       & 0.01  & 0.01  & 0.04  & 0.02 \\    \midrule          &       &       &       &       &       &       &       &       &       &       &       &       &       &       &       &       &  \\    \end{tabular}  \label{tab:addlabel} \end{table}

\begin{table}[h!]  \centering  \caption{Average covariate imbalances in Scenario 9 (Redundant Covariates)}    \begin{tabular}{rrrrrrrrrrrrrrrrrr} \multicolumn{18}{c}{} \\    \midrule          &       & \multicolumn{3}{c}{\textbf{D}} &       & \multicolumn{3}{c}{\textbf{SMD}} &       & \multicolumn{3}{c}{\textbf{OVL}} &       & \multicolumn{3}{c}{\textbf{KS}} & \multicolumn{1}{c}{\textbf{Bias}} \\\cmidrule{3-5}\cmidrule{7-9}\cmidrule{11-13}\cmidrule{15-17}    \multicolumn{2}{l}{Unweighted} & 0.36  & 0.04  & 0.32  &       & 0.18  & 0.02  & 0.14  &       & 0.26  & 0.23  & 0.25  &       & 0.07  & 0.02  & 0.06  & 0.12 \\    \multicolumn{2}{l}{$W_0\times W_1$} & 0.02  & 0.02  & 0.02  &       & 0.01  & 0.01  & 0.01  &       & 0.24  & 0.23  & 0.23  &       & 0.01  & 0.01  & 0.01  & 0.01 \\    \multicolumn{2}{l}{$W_1$} & 0.36  & 0.01  & 0.01  &       & 0.18  & 0.00  & 0.00  &       & 0.26  & 0.23  & 0.23  &       & 0.07  & 0.01  & 0.01  & 0.05 \\    \multicolumn{2}{l}{$W_0$} & 0.01  & 0.04  & 0.32  &       & 0.00  & 0.02  & 0.14  &       & 0.24  & 0.23  & 0.25  &       & 0.01  & 0.02  & 0.06  & 0.05 \\    \multicolumn{2}{l}{$W_0$tr90$\times W_1$} & 0.08  & 0.01  & 0.01  &       & 0.05  & 0.01  & 0.01  &       & 0.24  & 0.22  & 0.23  &       & 0.03  & 0.01  & 0.01  & 0.01 \\    \multicolumn{2}{l}{$W_0\times W_1$tr90} & 0.01  & 0.02  & 0.08  &       & 0.01  & 0.01  & 0.04  &       & 0.24  & 0.23  & 0.23  &       & 0.01  & 0.01  & 0.02  & 0.01 \\          &       & \multicolumn{3}{c}{\textbf{LD}} &       & \multicolumn{3}{c}{\textbf{MHB}} &       & \multicolumn{3}{c}{\textbf{CS}} &       & \multicolumn{3}{c}{\textbf{GWD}} & \multicolumn{1}{c}{\textbf{Bias}} \\\cmidrule{3-5}\cmidrule{7-9}\cmidrule{11-13}\cmidrule{15-17}    \multicolumn{2}{l}{Unweighted} & 0.06  & 0.01  & 0.05  &       & 0.40  & 0.00  & 0.33  &       & 0.32  & 0.29  & 0.29  &       & 0.09  & 0.01  & 0.07  & 0.12 \\    \multicolumn{2}{l}{$W_0\times W_1$} & 0.01  & 0.01  & 0.01  &       & 0.00  & 0.00  & 0.00  &       & 0.00  & 0.00  & 0.00  &       & 0.01  & 0.01  & 0.01  & 0.01 \\    \multicolumn{2}{l}{$W_1$} & 0.06  & 0.00  & 0.00  &       & 0.40  & 0.00  & 0.00  &       & 0.32  & 0.01  & 0.01  &       & 0.09  & 0.01  & 0.01  & 0.05 \\    \multicolumn{2}{l}{$W_0$} & 0.00  & 0.01  & 0.05  &       & 0.00  & 0.00  & 0.33  &       & 0.01  & 0.28  & 0.28  &       & 0.01  & 0.01  & 0.07  & 0.05 \\    \multicolumn{2}{l}{$W_0$tr90$\times W_1$} & 0.02  & 0.01  & 0.01  &       & 0.04  & 0.00  & 0.00  &       & 0.32  & 0.00  & 0.00  &       & 0.03  & 0.01  & 0.01  & 0.01 \\    \multicolumn{2}{l}{$W_0\times W_1$tr90} & 0.01  & 0.01  & 0.02  &       & 0.00  & 0.00  & 0.03  &       & 0.00  & 0.28  & 0.28  &       & 0.01  & 0.01  & 0.02  & 0.01 \\    \midrule          &       &       &       &       &       &       &       &       &       &       &       &       &       &       &       &       &  \\    \end{tabular}  \label{tab:addlabel} \end{table}

\begin{table}[h!]  \centering  \caption{Average covariate imbalances in Scenario 10 (Instrumental variables)}    \begin{tabular}{rrrrrrrrrrrrrrrrrr}   \multicolumn{18}{c}{} \\    \midrule          &       & \multicolumn{3}{c}{\textbf{D}} &       & \multicolumn{3}{c}{\textbf{SMD}} &       & \multicolumn{3}{c}{\textbf{OVL}} &       & \multicolumn{3}{c}{\textbf{KS}} & \multicolumn{1}{c}{\textbf{Bias}} \\\cmidrule{3-5}\cmidrule{7-9}\cmidrule{11-13}\cmidrule{15-17}    \multicolumn{2}{l}{Unweighted} & 0.58  & 0.05  & 0.55  &       & 0.23  & 0.02  & 0.18  &       & 0.41  & 0.36  & 0.40  &       & 0.11  & 0.01  & 0.09  & 0.21 \\    \multicolumn{2}{l}{$W_0\times W_1$} & 0.04  & 0.02  & 0.04  &       & 0.01  & 0.01  & 0.01  &       & 0.37  & 0.36  & 0.37  &       & 0.01  & 0.01  & 0.01  & 0.01 \\    \multicolumn{2}{l}{$W_1$} & 0.58  & 0.01  & 0.04  &       & 0.23  & 0.00  & 0.01  &       & 0.41  & 0.36  & 0.37  &       & 0.11  & 0.01  & 0.01  & 0.09 \\    \multicolumn{2}{l}{$W_0$} & 0.03  & 0.04  & 0.55  &       & 0.01  & 0.02  & 0.18  &       & 0.37  & 0.36  & 0.40  &       & 0.01  & 0.01  & 0.09  & 0.08 \\    \multicolumn{2}{l}{$W_0$tr90$\times W_1$} & 0.13  & 0.02  & 0.04  &       & 0.04  & 0.01  & 0.01  &       & 0.37  & 0.36  & 0.37  &       & 0.02  & 0.01  & 0.01  & 0.01 \\    \multicolumn{2}{l}{$W_0\times W_1$tr90} & 0.04  & 0.02  & 0.13  &       & 0.01  & 0.01  & 0.04  &       & 0.37  & 0.36  & 0.37  &       & 0.01  & 0.01  & 0.02  & 0.01 \\          &       & \multicolumn{3}{c}{\textbf{LD}} &       & \multicolumn{3}{c}{\textbf{MHB}} &       & \multicolumn{3}{c}{\textbf{CS}} &       & \multicolumn{3}{c}{\textbf{GWD}} & \multicolumn{1}{c}{\textbf{Bias}} \\\cmidrule{3-5}\cmidrule{7-9}\cmidrule{11-13}\cmidrule{15-17}    \multicolumn{2}{l}{Unweighted} & 0.09  & 0.01  & 0.07  &       & 0.25  & 0.00  & 0.16  &       & 0.28  & 0.23  & 0.23  &       & 0.10  & 0.01  & 0.08  & 0.21 \\    \multicolumn{2}{l}{$W_0\times W_1$} & 0.00  & 0.00  & 0.00  &       & 0.00  & 0.00  & 0.00  &       & 0.00  & 0.01  & 0.01  &       & 0.01  & 0.01  & 0.01  & 0.01 \\    \multicolumn{2}{l}{$W_1$} & 0.09  & 0.00  & 0.00  &       & 0.25  & 0.00  & 0.00  &       & 0.28  & 0.02  & 0.02  &       & 0.10  & 0.01  & 0.01  & 0.09 \\    \multicolumn{2}{l}{$W_0$} & 0.00  & 0.01  & 0.07  &       & 0.00  & 0.00  & 0.16  &       & 0.01  & 0.22  & 0.22  &       & 0.01  & 0.01  & 0.08  & 0.08 \\    \multicolumn{2}{l}{$W_0$tr90$\times W_1$} & 0.02  & 0.00  & 0.00  &       & 0.01  & 0.00  & 0.00  &       & 0.26  & 0.01  & 0.01  &       & 0.02  & 0.01  & 0.01  & 0.01 \\    \multicolumn{2}{l}{$W_0\times W_1$tr90} & 0.00  & 0.00  & 0.02  &       & 0.00  & 0.00  & 0.01  &       & 0.00  & 0.22  & 0.22  &       & 0.01  & 0.01  & 0.02  & 0.01 \\    \midrule          &       &       &       &       &       &       &       &       &       &       &       &       &       &       &       &       &  \\    \end{tabular}  \label{tab:addlabel} \end{table}

\clearpage

\section*{Appendix 3 -- Connection between the standardized mean difference and the Mahalanobis balance\label{app2}}%
\vspace*{12pt}

In this section we present a mathematical connection between the standardized mean difference and the Mahalanobis balance on which is based our proposed threshold for the Mahalanobis balance. To simplify the presentation, we consider two uncorrelated covariates.

Let $X_T = (X_{1T},X_{2T})^\top$ be the vector of covariates in treated subjects, with $X_{1T}=(x_{11T},x_{12T},...,x_{1n_1T})$, $X_{2T}=(x_{21T},x_{22T},...,x_{2n_1T})$ and $n_1$ is the number of treated subjects. We also note $X_C = (X_{1C},X_{2C})^\top$ the vector of covariates in the untreated subjects, with $X_{1C}=(x_{11C},x_{12C},...,x_{1n_2C})$, $X_{2C}=(x_{21C},x_{22C},...,x_{2n_2C})$ and $n_2$ the number of untreated subjects. The Mahalanobis balance is given by:
\begin{equation*}
	MHB = (\bar{X}_T-\bar{X}_C)^\top \Sigma^{-1}(\bar{X}_T-\bar{X}_C), \ \ \mbox{where}
\end{equation*}
$\bar{X}_T = (\bar{X}_{1T},\bar{X}_{2T})^\top$, $\bar{X}_C = (\bar{X}_{1C},\bar{X}_{2C})^\top$ are the sample averages and 
\begin{equation*}
	\Sigma = \dfrac{var(X_T)+var(X_C)}{2}.
\end{equation*}
\noindent We have
\begin{equation*}
	\Sigma = 
	\begin{pmatrix}
		\dfrac{var(X_{1T})+var(X_{1C})}{2} & \dfrac{cov(X_{1T},X_{2T})+cov(X_{1C},X_{2C})}{2}  \\
		\dfrac{cov(X_{1T},X_{2T})+cov(X_{1C},X_{2C})}{2}  &  \dfrac{var(X_{2T})+var(X_{2C})}{2} 
	\end{pmatrix},
\end{equation*}
\begin{equation*}
	\Sigma^{-1} = \dfrac{1}{A}
	\begin{pmatrix}
		\dfrac{var(X_{2T})+var(X_{2C})}{2} & -\dfrac{cov(X_{1T},X_{2T})+cov(X_{1C},X_{2C})}{2}  \\
		-\dfrac{cov(X_{1T},X_{2T})+cov(X_{1C},X_{2C})}{2}  &  
		\dfrac{var(X_{1T})+var(X_{1C})}{2}
	\end{pmatrix}, \ \ \mbox{where}
\end{equation*}
\begin{equation*}
	A = \dfrac{[var(X_{1T})+var(X_{1C})][var(X_{2T})+var(X_{2C})]}{2^{2}}-\left(\frac{cov(X_{1T},X_{2T})+cov(X_{1C},X_{2C})}{2}\right)^{2}.
\end{equation*}
As such,
\begin{align*}
	MHB &= \dfrac{1}{A}
	(\bar{X}_{1T}-\bar{X}_{1C}, \bar{X}_{2T}-\bar{X}_{2C}) \\
	& \qquad \times 
	\begin{pmatrix}
		\dfrac{var(X_{2T})+var(X_{2C})}{2} & -\dfrac{cov(X_{1T},X_{2T})+cov(X_{1C},X_{2C})}{2}  \\
		-\dfrac{cov(X_{1T},X_{2T})+cov(X_{1C},X_{2C})}{2}  &  
		\dfrac{var(X_{1T})+var(X_{1C})}{2}
	\end{pmatrix}
	\begin{pmatrix}
		\bar{X}_{1T}-\bar{X}_{1C} \\ \bar{X}_{2T}-\bar{X}_{2C}
	\end{pmatrix}  \\   
	&= \dfrac{1}{A}\Biggl[  \frac{(\bar{X}_{1T}-\bar{X}_{1C})^{2}var(X_{2T})+(\bar{X}_{1T}-\bar{X}_{1C})^{2}var(X_{2C})}{2}- \frac{(\bar{X}_{2T}-\bar{X}_{2C})(	\bar{X}_{1T}-\bar{X}_{1C})cov(X_{1T},X_{2T})}{2}  \\
	& \quad - \frac{(\bar{X}_{2T}-\bar{X}_{2C})(	\bar{X}_{1T}-\bar{X}_{1C})cov(X_{1C},X_{2C})}{2}- \frac{(\bar{X}_{1T}-\bar{X}_{1C})(\bar{X}_{2T}-\bar{X}_{2C})cov(X_{1T},X_{2T})}{2}\\
	& \quad-  \frac{(\bar{X}_{1T}-\bar{X}_{1C})(\bar{X}_{2T}-\bar{X}_{2C})cov(X_{1C},X_{2C})}{2}+
	\frac{(\bar{X}_{2T}-\bar{X}_{2C})^{2}var(X_{1T})}{2}+ \frac{(\bar{X}_{2T}-\bar{X}_{2C})^{2}var(X_{1C})}{2}\\
	& =  \frac{1}{A}\Biggl[  \dfrac{(\bar{X}_{1T}-\bar{X}_{1C})^{2}[var(X_{2T})+var(X_{2C})]}{2}- 2\times \frac{(\bar{X}_{2T}-\bar{X}_{2C})(	\bar{X}_{1T}-\bar{X}_{1C})[cov(X_{1T},X_{2T})+cov(X_{1C},X_{2C})]}{2}  \\
	& \quad + 
	\frac{(\bar{X}_{2T}-\bar{X}_{2C})^{2}[var(X_{1T})+var(X_{1C})]}{2}
	\Biggr] 
\end{align*}
Since $X_{1} \ \mbox{and} \ X_{2}$ are uncorrelated, $cov(X_{1T},X_{2T})=cov(X_{1C},X_{2C})=0$ and we have
\begin{eqnarray*}
	MHB  & = & \frac{1}{B}\Biggl[  \dfrac{(\bar{X}_{1T}-\bar{X}_{1C})^{2}[var(X_{2T})+var(X_{2C})]}{2} + 
	\dfrac{(\bar{X}_{2T}-\bar{X}_{2C})^{2}[var(X_{1T})+var(X_{1C})]}{2}
	\Biggr], \ \ \mbox{where}  \ \ \\
\end{eqnarray*}
$$B = \dfrac{[var(X_{1T})+var(X_{1C})][var(X_{2T})+var(X_{2C})]}{2^{2}}.$$ 

\noindent As a result,
\begin{eqnarray*}
	MHB  & = &  \dfrac{2(\bar{X}_{1T}-\bar{X}_{1C})^{2}[var(X_{2T})+var(X_{2C})]}{[var(X_{1T})+var(X_{1C})][var(X_{2T})+var(X_{2C})]} + 
	\dfrac{2(\bar{X}_{2T}-\bar{X}_{2C})^{2}[var(X_{1T})+var(X_{1C})]}{[var(X_{1T})+var(X_{1C})][var(X_{2T})+var(X_{2C})]}\\
	& = &
	\dfrac{2(\bar{X}_{1T}-\bar{X}_{1C})^{2}}{var(X_{1T})+var(X_{1C})}+	\dfrac{2(\bar{X}_{2T}-\bar{X}_{2C})^{2}}{var(X_{2T})+var(X_{2C})} \\
	& = &  \dfrac{(\bar{X}_{1T}-\bar{X}_{1C})^{2}}{\dfrac{var(X_{1T})+var(X_{1C})}{2}}+\dfrac{(\bar{X}_{2T}-\bar{X}_{2C})^{2}}{\dfrac{var(X_{2T})+var(X_{2C})}{2}}.
\end{eqnarray*}
This correspond to the sum of the squared standardized mean differences of the covariates. Consequently, the commonly used threshold of 0.1 for each covariate for the standardized mean difference corresponds to
\begin{eqnarray*}
	\dfrac{(\bar{X}_{1T}-\bar{X}_{1C})^{2}}{\dfrac{var(X_{1T})+var(X_{1C})}{2}}+\dfrac{(\bar{X}_{2T}-\bar{X}_{2C})^{2}}{\dfrac{var(X_{2T})+var(X_{2C})}{2}}
	& \leq &  (0.1)^{2}+(0.1)^{2} = 0.02 \\
\end{eqnarray*}
according to the MHB. Similarly, when $p$ uncorrelated covariates are considered, the corresponding threshold for the MHB is $p \times 0.01$.

\clearpage

\section*{Appendix 4 -- Distributions of the propensity scores in the different simulation scenarios \label{app2}}%
\vspace*{12pt}

In this section, we provide figures that depict the distributions of the propensity score in each treatment group in a single simulated dataset for each of the simulation scenarios. The red curve corresponds to the treated group and the blue curve to the untreated. Odd panels correspond to the distribution at $t = 0$ and even panels to the distribution at $t = 1$. The first row (panels 1-2) correspond to unweighted data, the second (panels 3-4) to data weighted by $W_0\times W_1$, the third (panels 5-6) to data weighted by $W_1$ only, the fourth (panels 7-8) to data weighted according to $W_0$ only, the fifth (panels 9-10) to the product of $W_0$ truncated at the 90th percentile with $W_1$ and the last row (panels 11-12) to the product of $W_0$ with $W_1$ truncated at the 90th percentile

\begin{figure}[h!]
	\centerline{\includegraphics[width=10in,width=\textwidth]{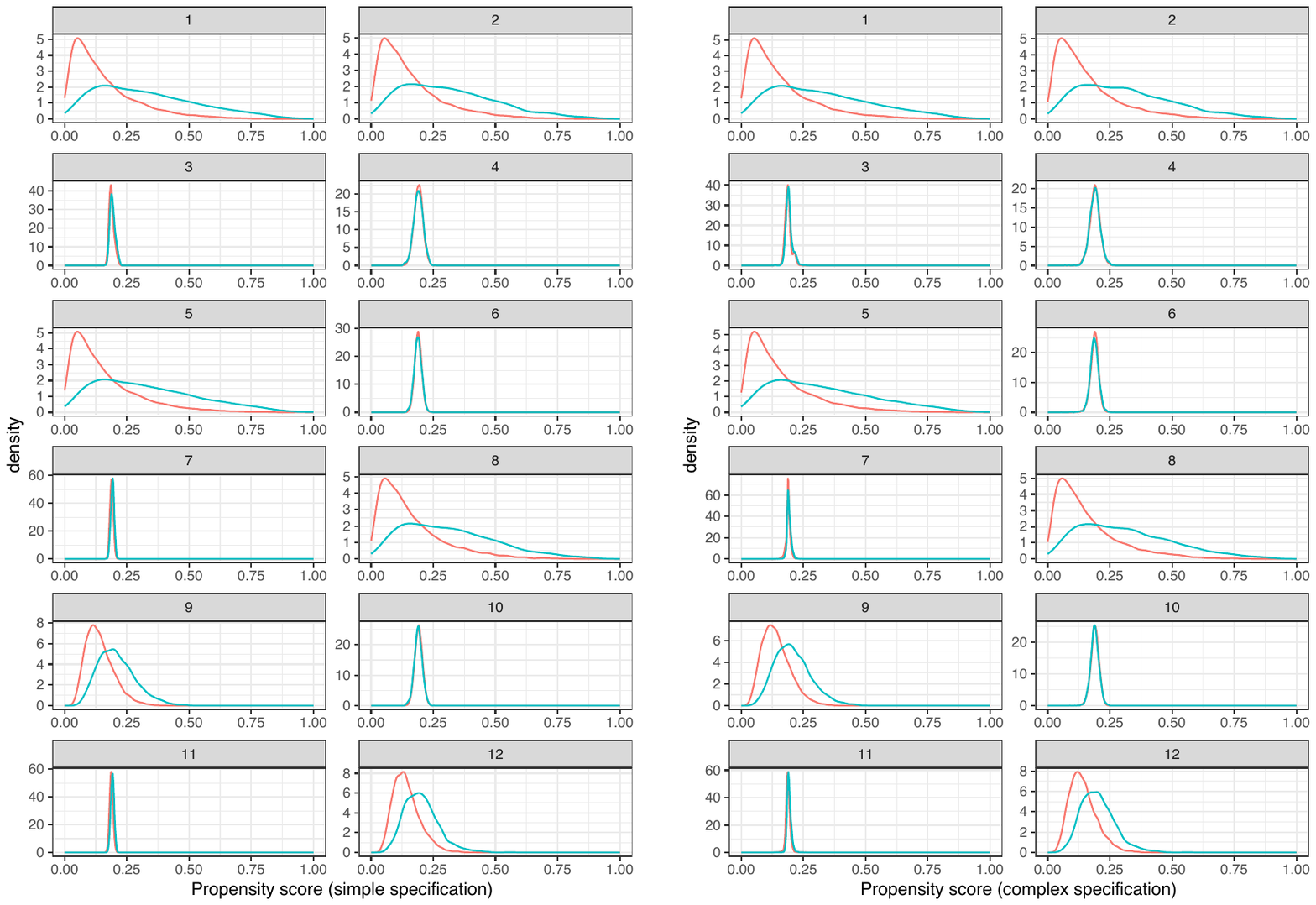}}
	\caption{Distribution of the propensity score in Scenario 2 (Low prevalence of exposure).\label{fig5}}
\end{figure}

\begin{figure}[h!]
	\centerline{\includegraphics[width=10in,width=\textwidth]{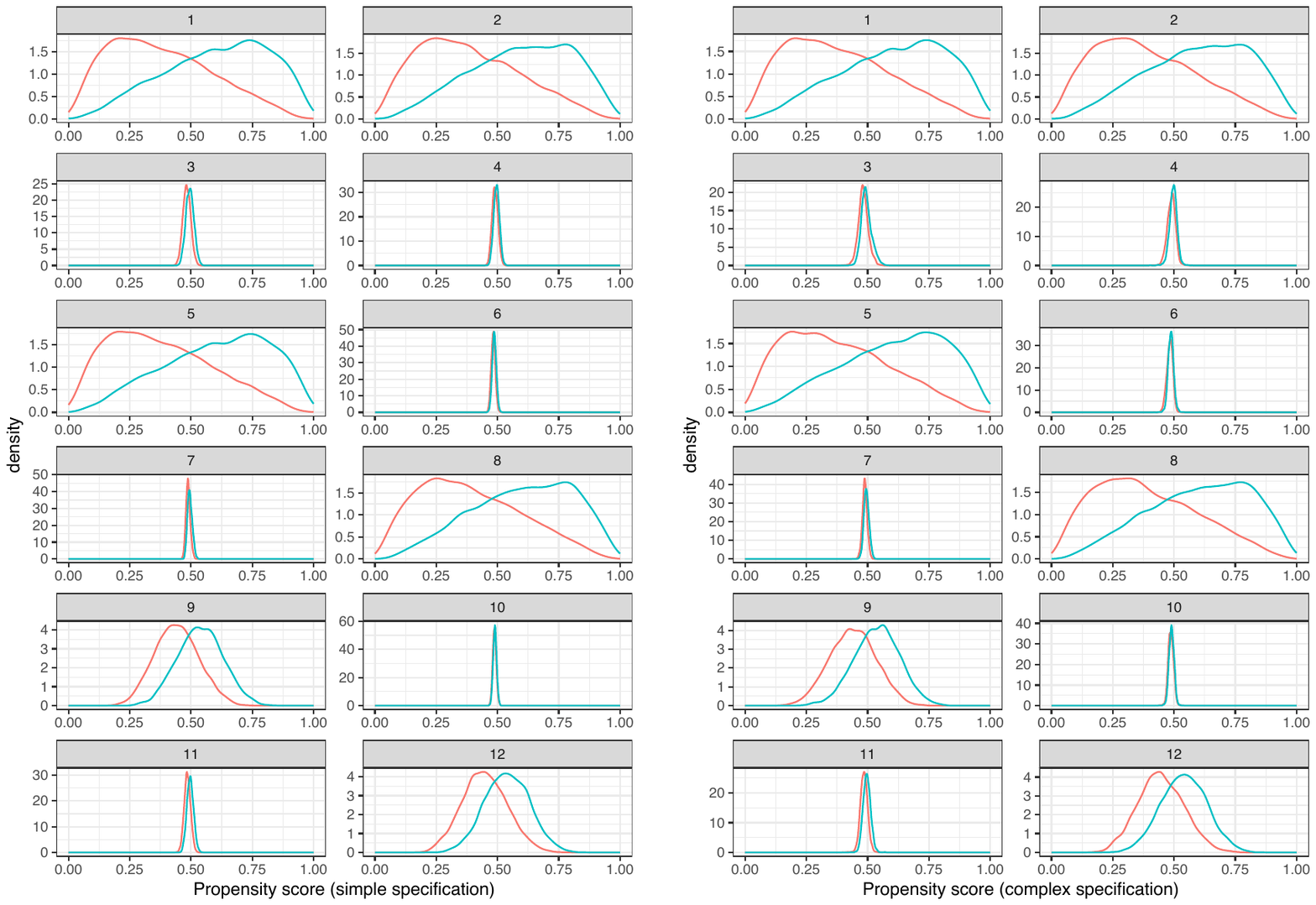}}
	\caption{Distribution of the propensity score in Scenario 3 (Small sample).\label{fig5}}
\end{figure}

\begin{figure}[h!]
	\centerline{\includegraphics[width=10in,width=\textwidth]{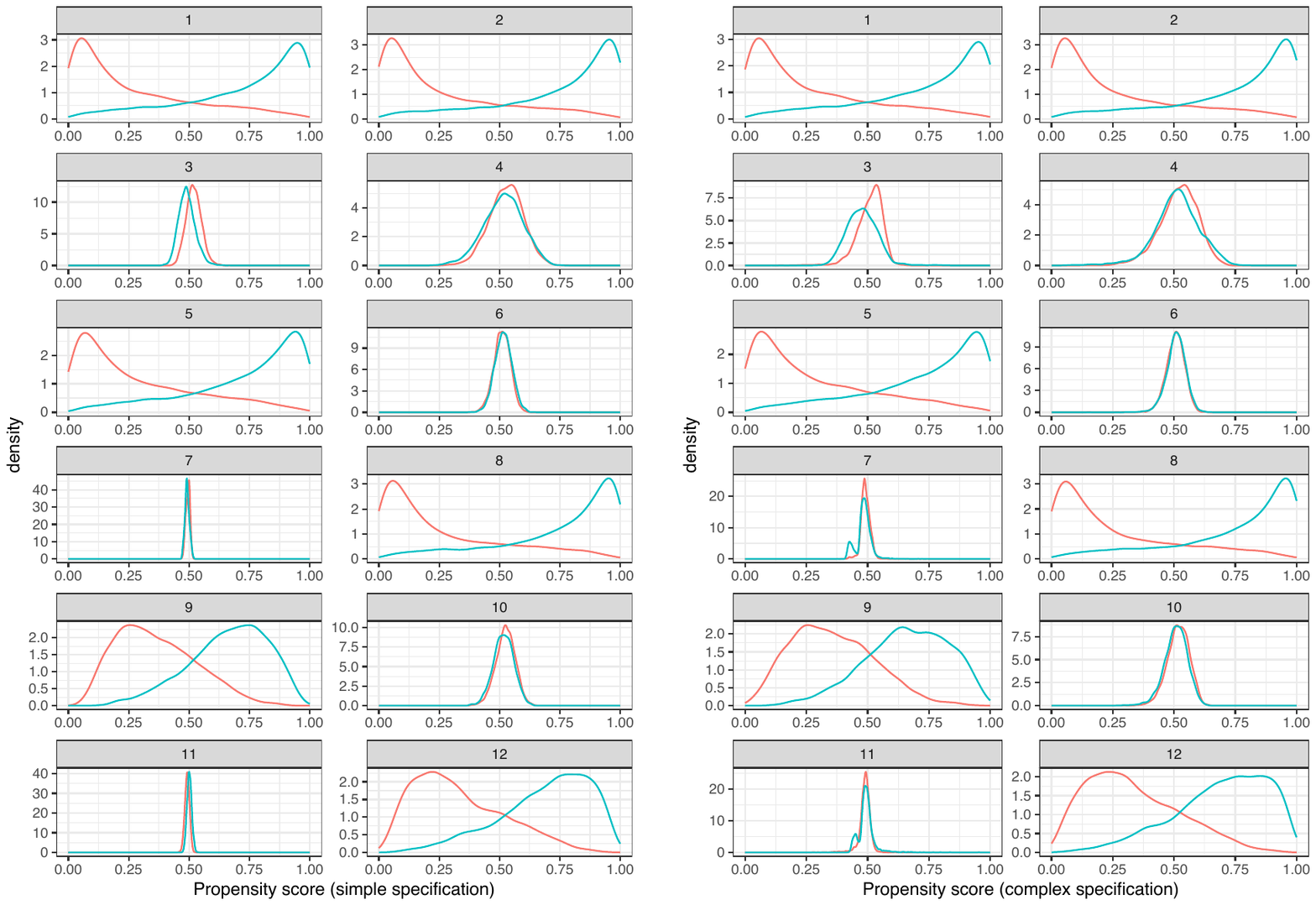}}
	\caption{Distribution of the propensity score in Scenario 4 (High imbalance, no confounding).\label{fig5}}
\end{figure}

\begin{figure}[h!]
	\centerline{\includegraphics[width=10in,width=\textwidth]{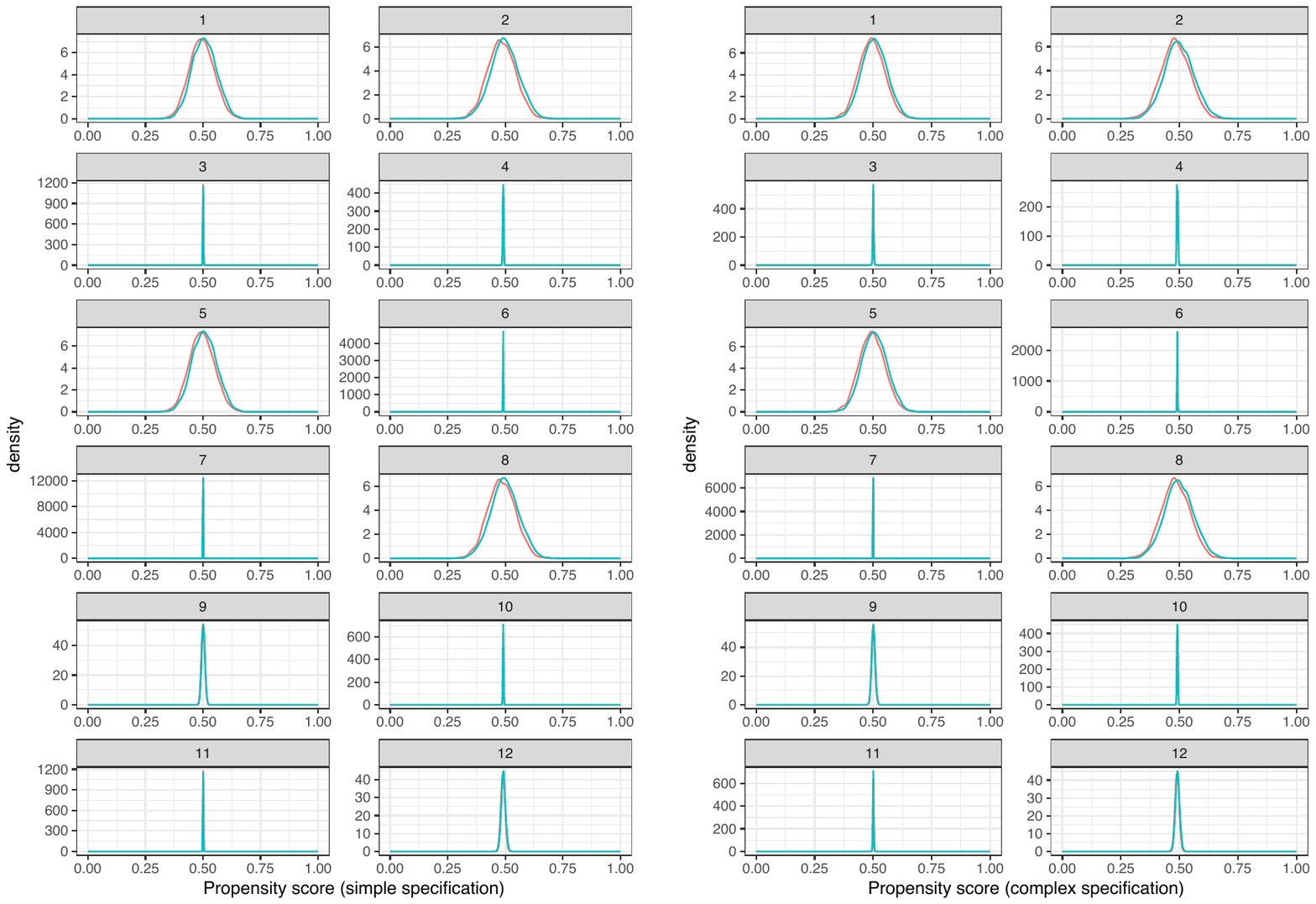}}
	\caption{Distribution of the propensity score in Scenario 5 (Low imbalance, moderate confounding).\label{fig5}}
\end{figure}

\begin{figure}[h!]
	\centerline{\includegraphics[width=10in,width=\textwidth]{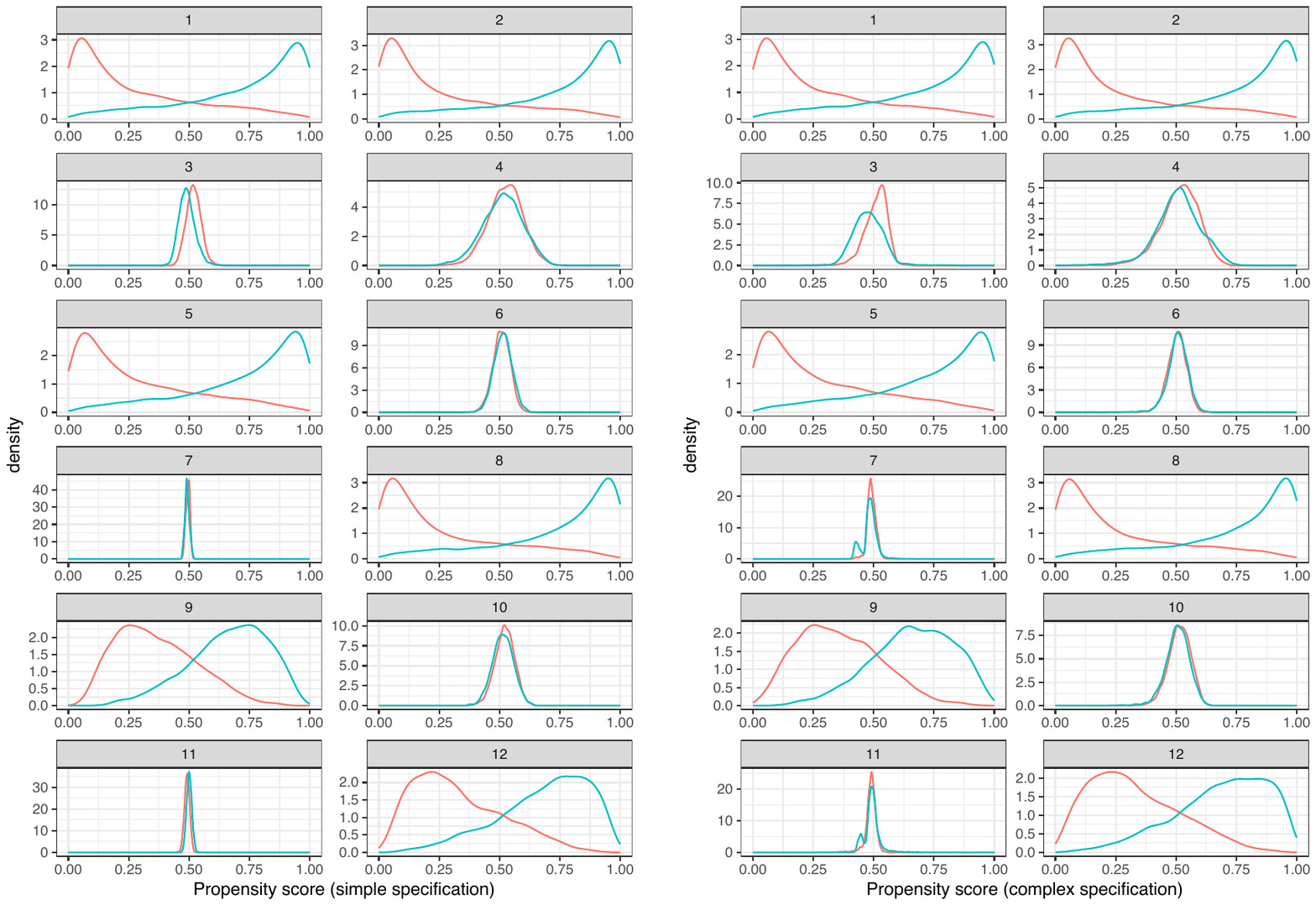}}
	\caption{Distribution of the propensity score in Scenario 6 (High imbalance-high confounding).\label{fig5}}
\end{figure}

\begin{figure}[h!]
	\centerline{\includegraphics[width=10in,width=\textwidth]{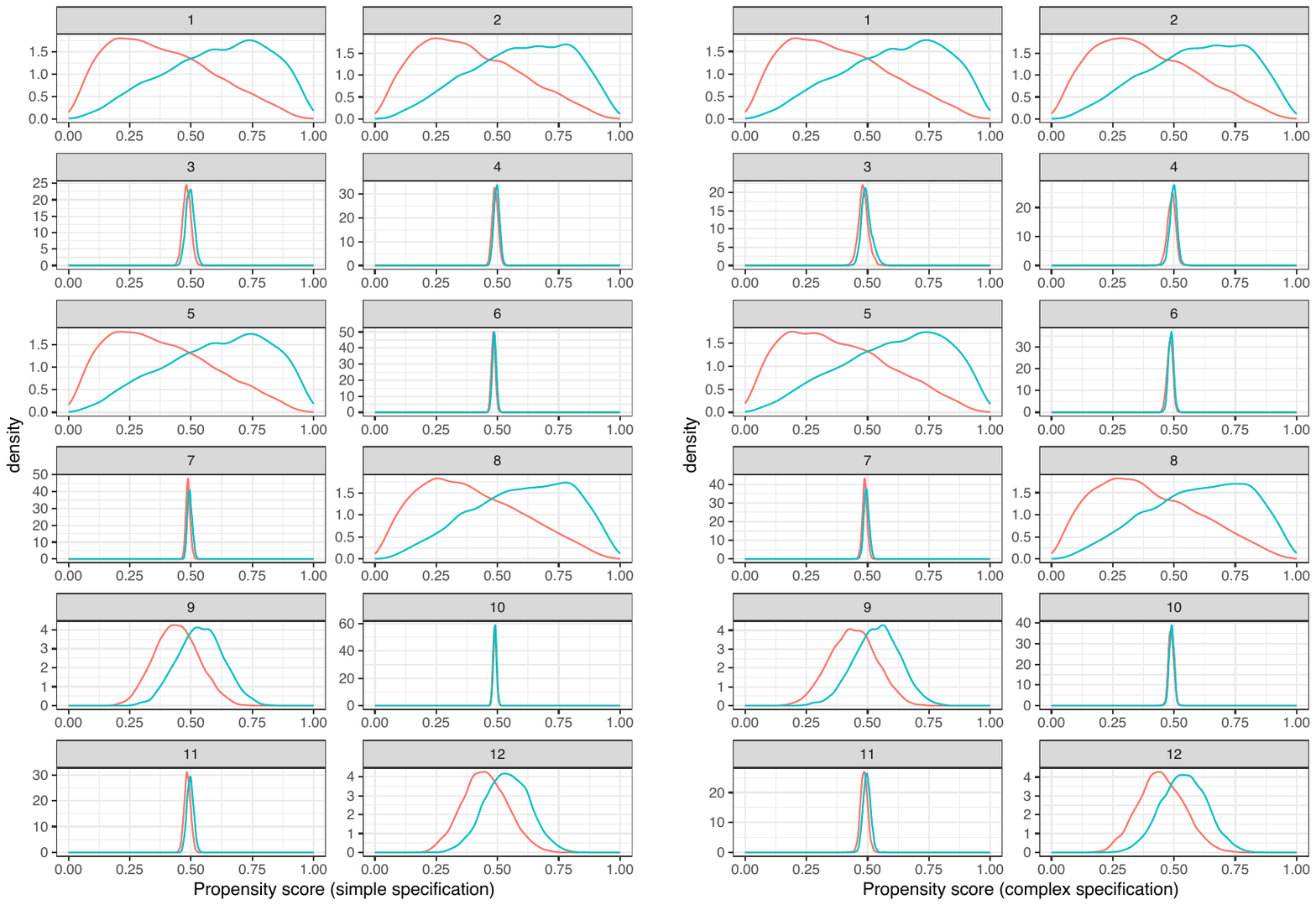}}
	\caption{Distribution of the propensity score in Scenario 7 (Nonlinear outcome).\label{fig5}}
\end{figure}

\begin{figure}[h!]
	\centerline{\includegraphics[width=10in,width=\textwidth]{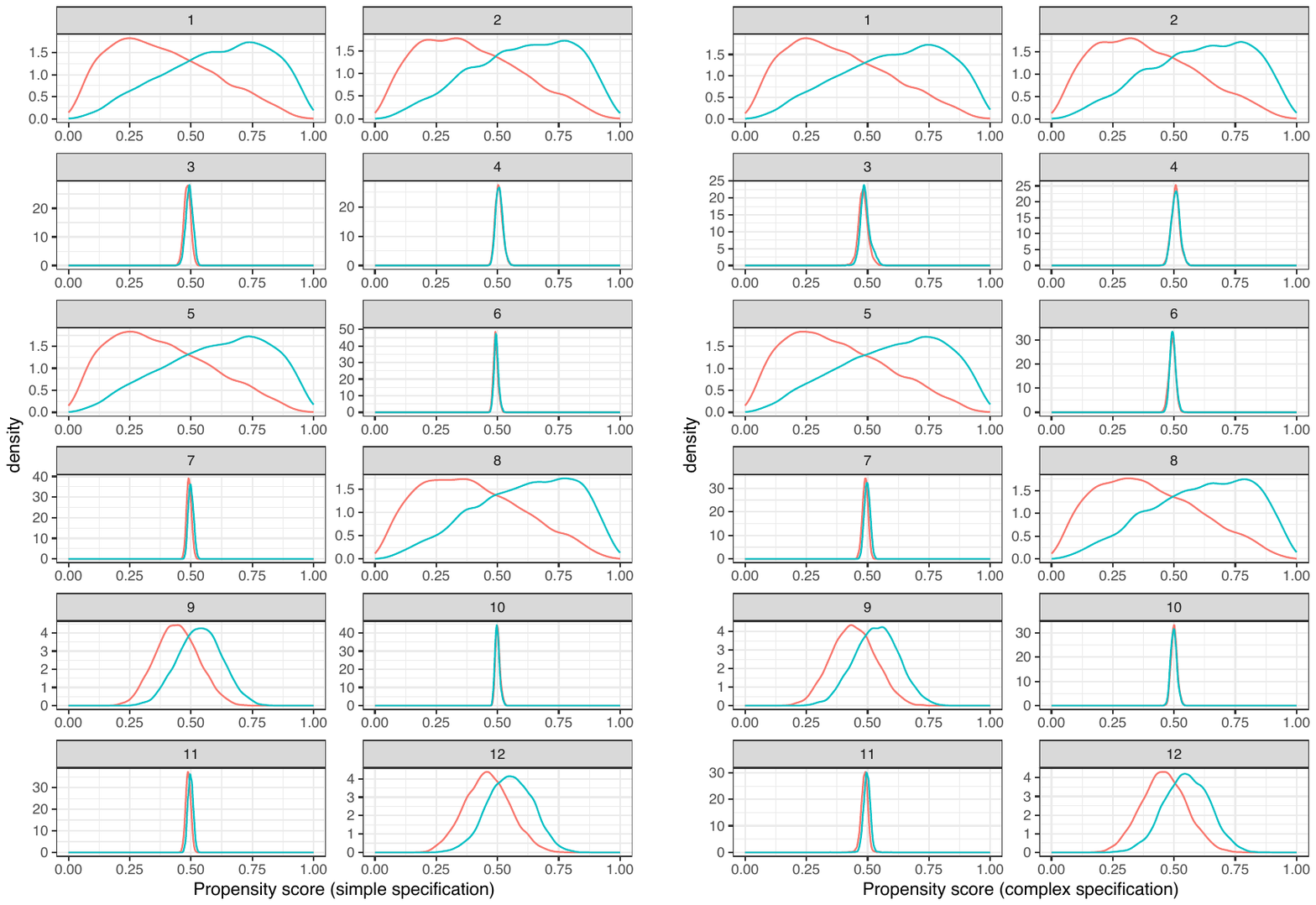}}
	\caption{Distribution of the propensity score in Scenario 8 (Nonlinear outcome and exposure).\label{fig5}}
\end{figure}

\begin{figure}[h!]
	\centerline{\includegraphics[width=10in,width=\textwidth]{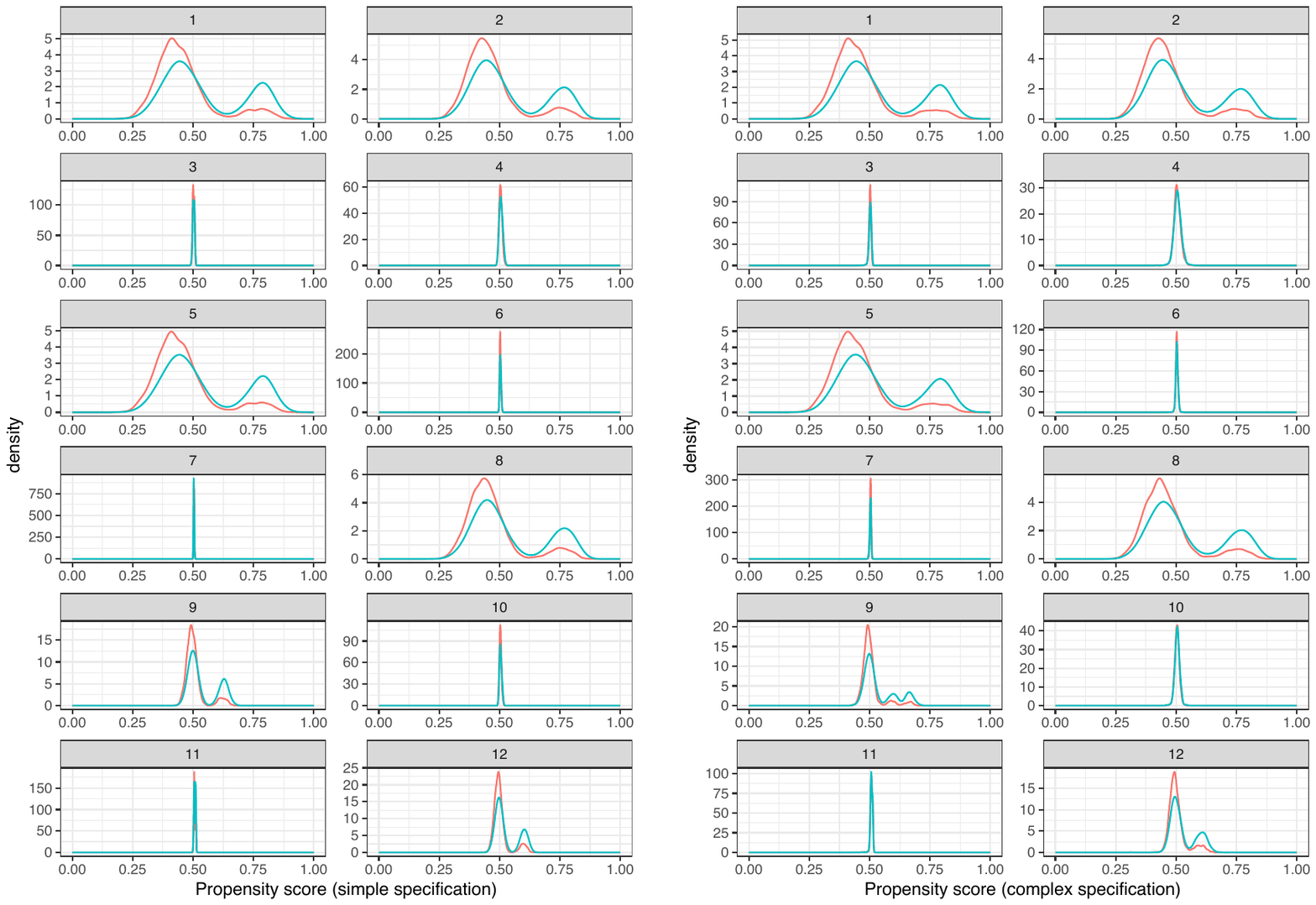}}
	\caption{Distribution of the propensity score in Scenario 9 (Redundant Covariates).\label{fig5}}
\end{figure}

\begin{figure}[h!]
	\centerline{\includegraphics[width=10in,width=\textwidth]{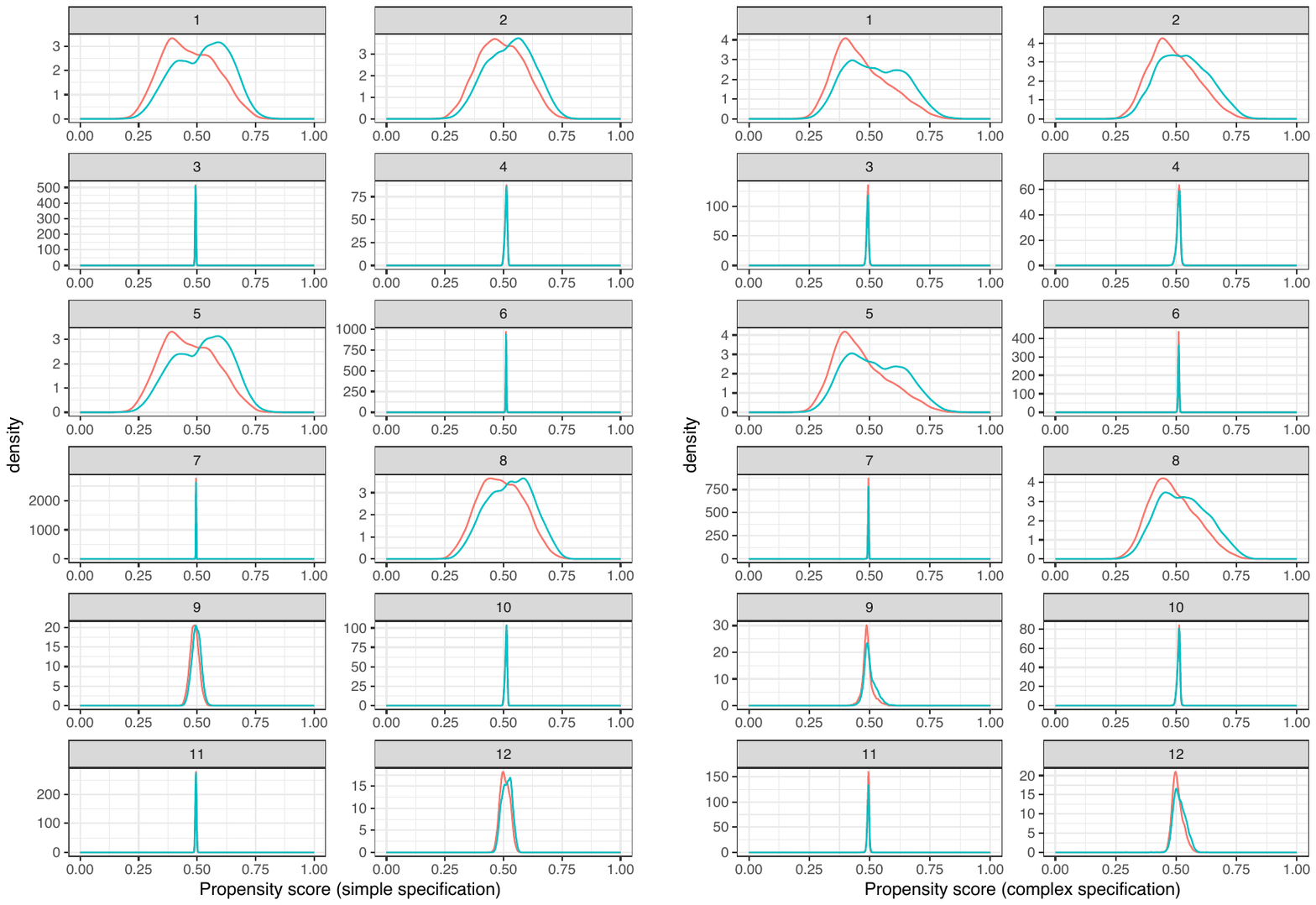}}
	\caption{Distribution of the propensity score in Scenario 10 (Instrumental variables).\label{fig5}}
\end{figure}

\begin{figure}[h!]
	\centerline{\includegraphics[width=10in,width=\textwidth]{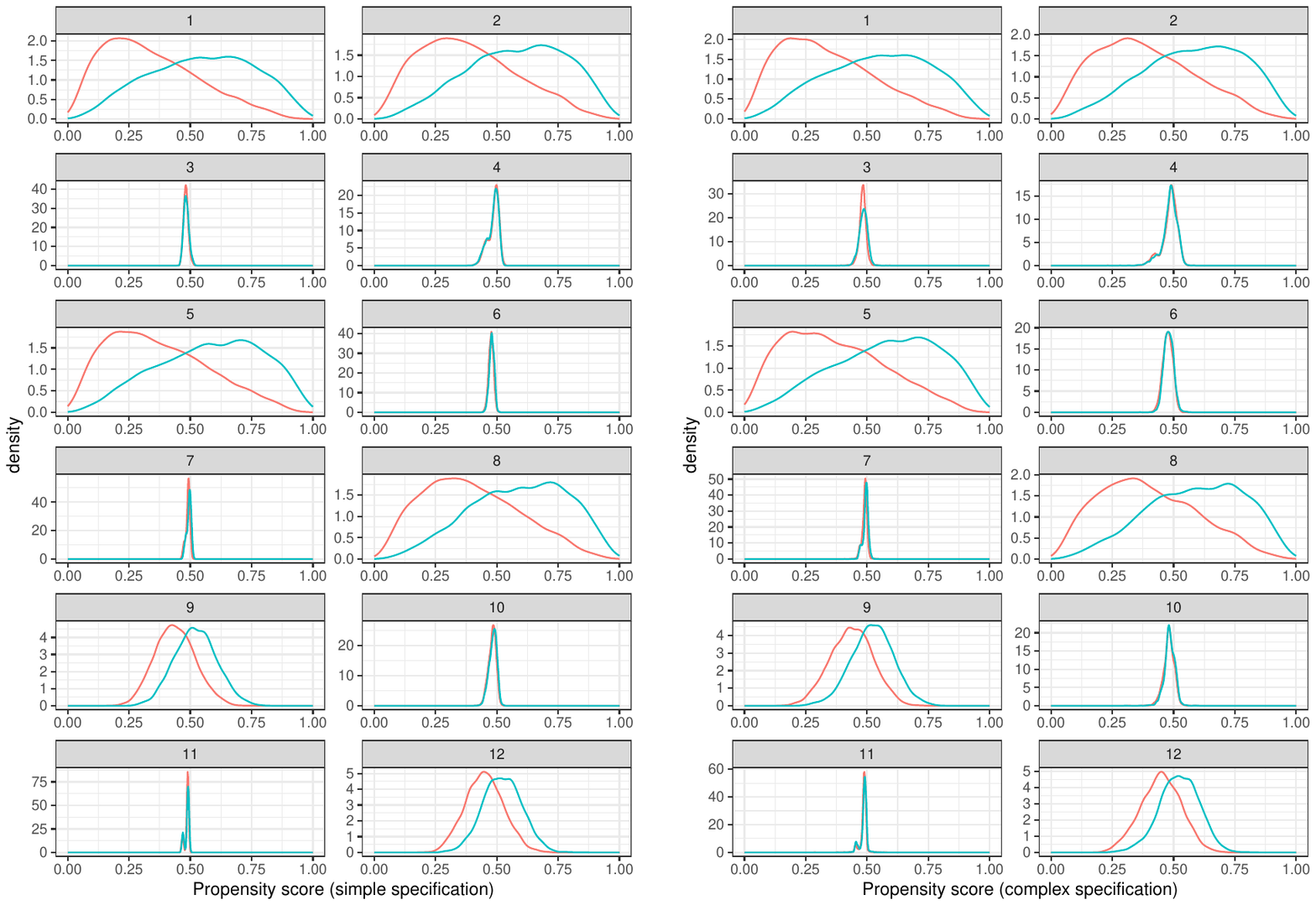}}
	\caption{Distribution of the propensity score in the base case scenario with censoring\label{fig5}}
\end{figure}

\end{document}